\documentclass[aps,prd,preprint,amsmath,amssymb,floatfix]{revtex4}
\usepackage{graphicx}
\usepackage{subfigure}
\usepackage{rotating}

\input{abbreviations.h}
\input{hyphenation.h}

\DeclareGraphicsExtensions{ .eps , .eps.gz ,
                            .ps  , .ps.gz }
\begin{document}

%
%
\title{Final Report of the Muon E821
Anomalous Magnetic Moment Measurement at BNL}

\author{
G.W.~Bennett$^{2}$, B.~Bousquet$^{10}$, H.N.~Brown$^2$, G.~Bunce$^2$,
R.M.~Carey$^1$, P.~Cushman$^{10}$, G.T.~Danby$^2$, P.T.~Debevec$^8$,
M.~Deile$^{13}$, H.~Deng$^{13}$, W.~Deninger$^8$, S.K.~Dhawan$^{13}$,
V.P.~Druzhinin$^3$, L.~Duong$^{10}$, E.~Efstathiadis$^1$, F.J.M.~Farley$^{13}$,
G.V.~Fedotovich$^3$, S.~Giron$^{10}$, F.E.~Gray$^8$, D.~Grigoriev$^3$,
M.~Grosse-Perdekamp$^{13}$, A.~Grossmann$^7$, M.F.~Hare$^1$, D.W.~Hertzog$^8$,
X.~Huang$^1$, V.W.~Hughes$^{13}$\footnote[2]{Deceased.}, M.~Iwasaki$^{12}$,
K.~Jungmann$^{6,7}$, D.~Kawall$^{13}$, M.~Kawamura$^{12}$, B.I.~Khazin$^3$,
J.~Kindem$^{10}$, F.~Krienen$^1$, I.~Kronkvist$^{10}$, A.~Lam$^1$,
R.~Larsen$^2$, Y.Y.~Lee$^2$, I.~Logashenko$^{1,3}$, R.~McNabb$^{10,8}$,
W.~Meng$^2$, J.~Mi$^2$, J.P.~Miller$^1$, Y. Mizumachi$^{11}$, W.M.~Morse$^2$,
D.~Nikas$^2$, C.J.G.~Onderwater$^{8,6}$, Y.~Orlov$^4$, C.S.~\"{O}zben$^{2,8}$,
J.M.~Paley$^1$, Q.~Peng$^1$, C.C.~Polly$^8$, J.~Pretz$^{13}$, R.~Prigl$^{2}$,
G.~zu~Putlitz$^7$, T.~Qian$^{10}$, S.I.~Redin$^{3,13}$, O.~Rind$^1$,
B.L.~Roberts$^1$, N.~Ryskulov$^3$, S.~Sedykh$^8$, Y.K.~Semertzidis$^2$,
P.~Shagin$^{10}$, Yu.M.~Shatunov$^3$, E.P.~Sichtermann$^{13}$, E.~Solodov$^3$,
M.~Sossong$^8$, A.~Steinmetz$^{13}$, L.R.~Sulak$^{1}$, C.~Timmermans$^{10}$,
A.~Trofimov$^1$, D.~Urner$^8$, P.~von~Walter$^7$, D.~Warburton$^2$,
D.~Winn$^5$, A.~Yamamoto$^9$ and
D.~Zimmerman$^{10}$ \\
(Muon $(g-2)$ Collaboration) }

\affiliation{
\mbox{$\,^1$Department of Physics, Boston University, Boston, MA 02215}\\
\mbox{$\,^2$Brookhaven National Laboratory, Upton, NY 11973}\\
\mbox{$\,^3$Budker Institute of Nuclear Physics, 630090 Novosibirsk, Russia}\\
\mbox{$\,^4$Newman Laboratory, Cornell University, Ithaca, NY 14853}\\
\mbox{$\,^5$Fairfield University, Fairfield, CT 06430}\\
\mbox{$\,^6$ Kernfysisch Versneller Instituut, Rijksuniversiteit Groningen,} \\
\mbox{NL-9747 AA, Groningen, The Netherlands}\\
\mbox{$\,^7$ Physikalisches Institut der Universit\"at Heidelberg, 69120 Heidelberg, Germany}\\
\mbox{$\,^8$ Department of Physics, University of Illinois at Urbana-Champaign, Urbana, IL 61801}\\
\mbox{$\,^9$ KEK, High Energy Accelerator Research Organization, Tsukuba, Ibaraki 305-0801, Japan}\\
\mbox{$\,^{10}$Department of Physics, University of Minnesota,
Minneapolis, MN 55455}\\
\mbox{$\,^{11}$ Science University of Tokyo, Tokyo, 153-8902, Japan}\\
\mbox{$\,^{12}$ Tokyo Institute of Technology, 2-12-1 Ookayama, Meguro-ku, Tokyo, 152-8551, Japan}\\
\mbox{$\,^{13}$ Department of Physics, Yale University, New Haven, CT 06520} }

\date{\today}

\begin{abstract}
  We present the final report from a series of precision measurements of
  the muon anomalous magnetic moment, $\amu=\gm/2$.
  The details of the experimental method, apparatus, data taking,
  and analysis are summarized.
  Data obtained at Brookhaven National Laboratory, using nearly equal samples of positive
  and negative muons, were used to deduce
  $\amu{\rm (Expt)} = 11\,659\,208.0(5.4)(3.3) \times 10^{-10}$, where the statistical
  and systematic uncertainties are given, respectively. The combined uncertainty
  of 0.54~ppm represents a 14-fold improvement compared to previous measurements at CERN.
  The standard model value for \amu\ includes contributions from virtual
  QED, weak, and hadronic processes.  While the QED processes account for most of the
  anomaly, the largest theoretical uncertainty, $\approx 0.55$~ppm,
  is associated with first-order hadronic vacuum polarization.
  Present standard model evaluations, based on $e^+e^-$ hadronic cross sections,
  lie 2.2 - 2.7 standard deviations below the experimental result.

\end{abstract}

\pacs{}

\maketitle
\tableofcontents
%
\newpage
\section{Introduction \label{sec:introduction}}

The muon magnetic moment is related to its intrinsic spin by the
gyromagnetic ratio $g_{\mu}$:
\begin{equation}
\vec{\mu}_{\mu} = g_{\mu}\left(\frac{q}{2m}\right) \vec{S},
\end{equation}
where $g_{\mu}=2$ is expected for a structureless,
spin-$\frac{1}{2}$ particle of mass $m$ and charge $q = \pm e$.
Radiative corrections (RC), which couple the muon spin to virtual
fields, introduce an anomalous magnetic moment defined by
\begin{equation}
a_\mu = \frac{1}{2}(g_\mu - 2).
\end{equation}
The leading RC is the lowest-order (LO) quantum electrodynamic
process involving the exchange of a virtual photon, the ``Schwinger
term,''~\cite{Schwinger:1948} giving $a_\mu({\rm QED;LO}) =
\alpha/2\pi \approx 1.16 \times 10^{-3}$. The complete standard
model value of \amu, currently evaluated to a precision of
approximately 0.6~ppm (parts per million), includes this first-order
term along with higher-order QED processes, electroweak loops,
hadronic vacuum polarization, and other higher-order hadronic loops.
The measurement of \amu, carried out to a similar precision, is the
subject of this paper.
The difference between experimental and theoretical values for \amu\
is a valuable test of the completeness of the standard model. At
sub-ppm precision, such a test explores physics well above the
100~GeV scale for many standard model extensions.

The muon anomalous magnetic moment was measured in a series of three
experiments at CERN and, most recently in our E821 experiment at
Brookhaven National Laboratory (BNL). In the first CERN
measurement~\cite{Charpak:1962}  muons were injected into a 6-m long
straight magnet where they followed a drifting spiral path, slowly
traversing the magnet because of a small gradient introduced in the
field. The muons were stopped in a polarimeter outside the magnet
and a measurement of their net spin precession determined \amu\ with
an uncertainty of 4300~ppm.  The result agreed with the prediction
of QED for a structureless particle. The second CERN
experiment~\cite{Bailey:1972} used a magnetic ring to extend the
muon storage time. A primary proton beam was injected directly onto
a target inside the storage ring where produced pions decayed to
muons, a small fraction of which fell onto stable orbits. The muon
precession frequency was determined by a sinusoidal modulation in
the time distribution of decay positrons, measured by detectors on
the opposite side of the ring from the injection point. The result
to 270~ppm agreed with QED only after the theory had been
recalculated~\cite{Aldins}. The CERN-III
experiment~\cite{Bailey:1977mm} used a uniform-field storage ring
and electric quadrupoles to provide vertical containment for the
muons having the ``magic'' momentum of 3.1~GeV/$c$.  At this
momentum, the muon spin precession is not affected by the electric
field from the focusing quadrupoles. Additionally, pions were
injected directly into the ring, which resulted in a higher stored
muon fraction and less background than proton injection. The
CERN-III experiment achieved a precision of 10~ppm for each muon
polarity.  CPT symmetry was assumed, and the results were combined
to give a 7.3~ppm measurement, which agreed with theory. The result
served as the first confirmation of the predicted 60~ppm
contribution to \amu\ from hadronic vacuum polarization.

The present BNL experiment follows the general technique pioneered
by CERN-III, but features many innovative improvements. A continuous
superconducting magnet, having high field uniformity, is used
instead of a lattice of discrete resistive magnets. A direct
current, rather than pulsed, inflector magnet permits the ring to be
filled at 33~ms intervals, the bunch extraction interval from the
AGS. Muons are injected directly into the storage ring, which
increases the storage efficiency and reduces the intense
hadron-induced background ``flash.'' A pulsed kicker places the
muons onto stable orbits and centers them in the storage region. The
electrostatic quadrupoles permit operation at about twice the field
gradient of the CERN experiment. The transverse aperture of the
storage region is circular rather than rectangular, in order to
reduce the dependence of the average field seen by a muon on its
trajectory. The magnetic field is mapped using an array of NMR
probes, mounted on a trolley that can be pulled through the vacuum
chamber. Waveform digitizers provide a time record of energy
deposition in calorimeters.  The records are used to determine
electron energies and times and to correct for multi-particle
overlap--``pileup." (Note: In this manuscript, we use {\em electron}
to represent either the positron or electron in the generic $\mu
\rightarrow e \nu \bar{\nu}$ decay chain.)

Combining the results of four positive muon runs and a final run
using negative muons, \amu\ was determined to a precision of
0.54~ppm. A summary of the CERN and BNL measurements is given in
Table~\ref{tab:aMuSummary}. This paper reviews the BNL E821 results,
all reported in Letters~\cite{Carey:1999dd,Brown:2001mg,
Bennett:2002jb,Bennett:2004xx} or Brief Reports~\cite{Brown:2000sj}.
Many of the key experimental components have been described in
separate papers; brief summaries are given here. The paper begins
with the basic principle of the experimental method including
subsections on the apparatus. The dynamics of muon storage, which
shape the observed decay electron distribution, are discussed next.
Then the data analysis is described and the paper concludes with a
discussion of the theoretical standard model value for \amu\ and its
comparison to the final result.

\begin{table}
\caption{Summary of $a_\mu$ results from CERN and BNL, showing the
evolution of experimental precision over time.  The average is
obtained from the 1999, 2000 and 2001 data sets only.}
\label{tab:aMuSummary}
\begin{center}
\begin{tabular}{ccclcc}
\toprule
Experiment & Years & Polarity &~~~ $a_\mu \times 10^{10}$ ~~~& Precision [ppm] & Reference \\
\colrule
CERN I   & 1961 & $\mu^+$ & 11\,450\,000(220\,000) & 4300 & \cite{Charpak:1962} \\
CERN II  & 1962-1968 & $\mu^+$ &   11\,661\,600(3100)& 270 & \cite{Bailey:1972} \\
CERN III & 1974-1976 & $\mu^+$ &   11\,659\,100(110) & 10 & \cite{Bailey:1977mm} \\
CERN III & 1975-1976 & $\mu^-$ &   11\,659\,360(120) & 10 & \cite{Bailey:1977mm} \\
BNL      & 1997 & $\mu^+$ &   11\,659\,251(150) & 13 & \cite{Carey:1999dd} \\
BNL      & 1998 & $\mu^+$ &   11\,659\,191(59) & 5 & \cite{Brown:2000sj} \\
BNL      & 1999 & $\mu^+$ &   11\,659\,202(15) & 1.3 & \cite{Brown:2001mg} \\
BNL      & 2000 & $\mu^+$ & 11\,659\,204(9) & 0.73 & \cite{Bennett:2002jb} \\
BNL      & 2001 & $\mu^-$ & 11\,659\,214(9) & 0.72 &
\cite{Bennett:2004xx} \\ \hline
Average  &  & & 11\,659\,208.0(6.3) & 0.54 & \cite{Bennett:2004xx} \\
\botrule
\end{tabular}
\end{center}
\end{table}

%
\section{Experimental Method \label{sec:exptmethod}}
\subsection{Overview \label{ssec:exptoverview}}

%
%

The cyclotron $\omega_c$ and spin precession $\omega_s$ frequencies
for a muon moving in the horizontal plane of a magnetic storage ring
are given by:
\begin{equation}
{\vec \omega_c} = -{q {\vec B} \over m \gamma},\qquad {\vec
\omega_s} = -{gq{\vec B} \over 2 m} - (1-\gamma) {q {\vec B} \over
\gamma m}. \label{eq:basiceq}
\end{equation}
The anomalous precession frequency $\wa$ is determined from the
difference
\begin{equation}
{\vec \omega_a} = {\vec \omega_s} - {\vec \omega_c} = -\left( {g-2
\over 2} \right) {q{\vec B} \over m} = -\amu {q{\vec B} \over m}.
\end{equation}
Because electric quadrupoles are used to provide vertical focusing in the storage ring, their electric
field is seen in the muon rest frame as a motional magnetic field that can affect the spin precession
frequency.  In the presence of both ${\vec E}$ and ${\vec B}$ fields, and in the case that ${\vec
\beta}$ is perpendicular to both ${\vec E}$ and ${\vec B}$, the expression for the anomalous
precession frequency becomes
\begin{equation}
   \vec \omega_a =  -{q \over m }\left[ a_{\mu} \vec B -
   \left( a_{\mu}- {1 \over \gamma^2 - 1}\right)
   \frac{\vec \beta  \times \vec E}{c} \right].
   \label{eq:omega}
\end{equation}
The coefficient  of the $\vec{\beta} \times \vec{E}$ term vanishes
at the ``magic'' momentum of 3.094~GeV/$c,$ where $\gamma = 29.3$.
Thus \amu\ can be determined by a precision measurement of \wa\ and
$B$. At this magic momentum, the electric field is used only for
muon storage and the magnetic field alone determines the precession
frequency. The finite spread in beam momentum and vertical betatron
oscillations introduce small (sub ppm) corrections to the precession
frequency.

The longitudinally polarized muons, which are injected into the
storage ring at the magic momentum, have a time-dilated muon
lifetime of $64.4~\mu$s.  A measurement period of typically
$700~\mu$s follows each injection or ``fill.''  The net spin
precession depends on the integrated field seen by a muon along its
trajectory. The magnetic field used in Eq.~\ref{eq:omega} refers to
an average over muon trajectories during the course of the
experiment. The trajectories of the muons must be weighted with the
magnetic field distribution. To minimize the precision with which
the average particle trajectories must be known, the field should be
made as uniform as possible.

\begin{figure}
\begin{centering}
\subfigure[Center-of-mass frame]
{\includegraphics[width=0.45\textwidth]{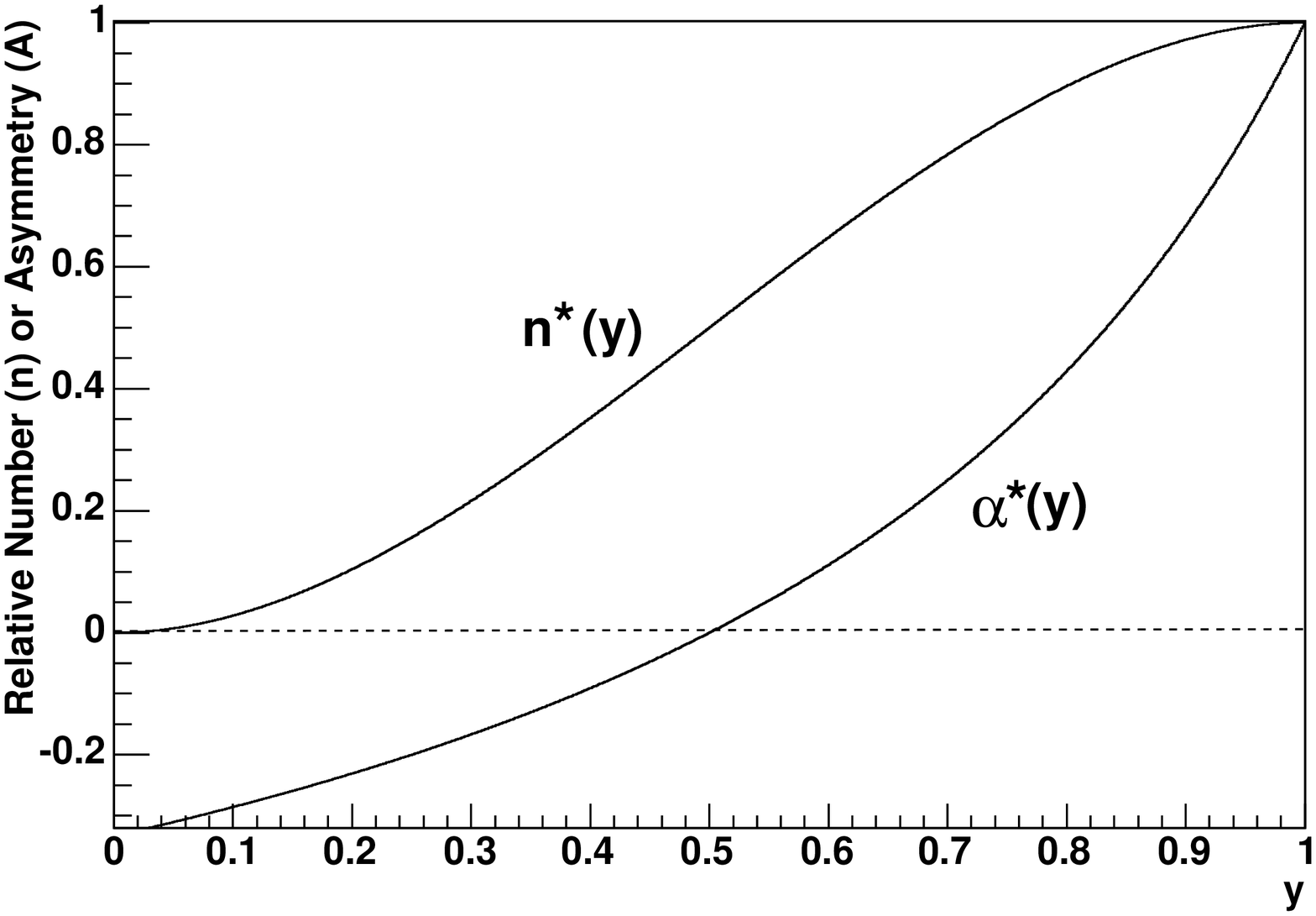}}
\subfigure[Lab frame]
{\includegraphics[width=0.45\textwidth]{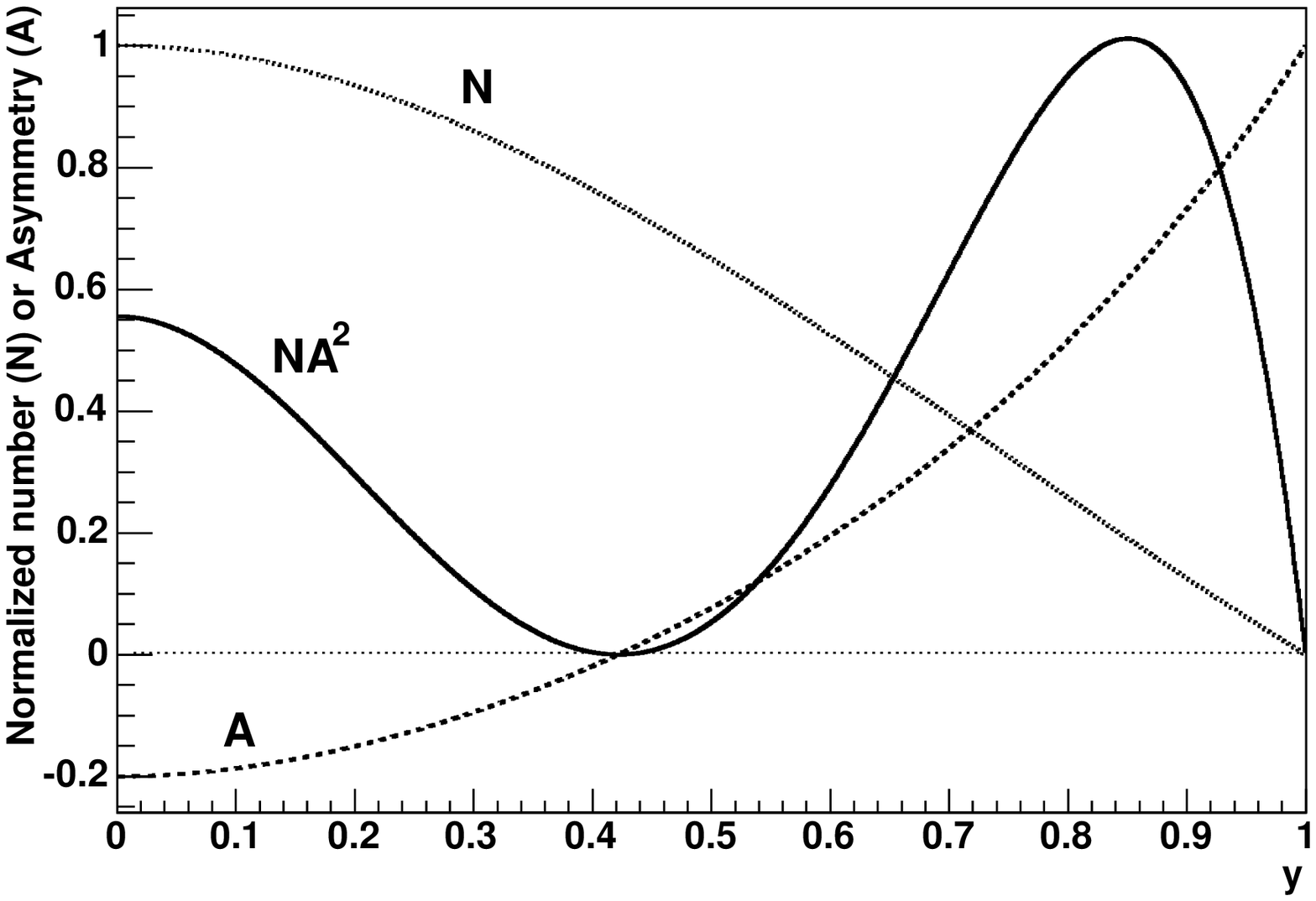}}
\end{centering}
\caption{Relative number and asymmetry distributions versus electron
fractional energy $y$ in the muon rest frame (left panel) and in the
laboratory frame (right panel).  The differential figure-of-merit
product $NA^2$ in the laboratory frame illustrates the importance of
the higher-energy electrons in reducing the measurement statistical
uncertainty.} \label{fg:nA}
\end{figure}

Because of parity violation in the weak decay of the muon, a correlation
existsbetween the muon spin and decay electron direction.  This correlation
allows the spin direction to be measured as a function of time. In the rest
frame of the muon---indicated by starred quantities---the differential
probability for the electron to emerge with a normalized energy $y =
E^*/E_{max}$ ($E_{max}=52.8$~MeV) at an angle $\theta^*$ with respect to the
muon spin is~\cite{kono}
\begin{eqnarray}
\frac{dP(y, \theta^*)}{dy~d\Omega} = (1/2\pi) n^*(y)[1-\alpha^*(y) \cos \theta^*] & ~{\rm with} \\
n^*(y) = y^2(3-2y) & ~{\rm and} \label{eq:n(y)}\\
\alpha^*(y) = \frac{q}{e}\frac{2y-1}{3-2y} ~. \label{eq:A(y)}
\end{eqnarray}
Figure~\ref{fg:nA}a shows the quantities $n^*(y)$ and $\alpha^*(y)$.
Electrons with $y < 0.5$ are emitted preferentially along the
(negative) muon spin direction and those with $y > 0.5$ are more
likely emitted opposite to the spin. Because both $n^*$ and
$\alpha^*$ are larger for $y
> 0.5$, decay electrons tend to emerge in the direction opposite to
the muon spin. Like the muon spin, the angular distribution of the electrons in
the muon rest frame rotates at the angular frequency \wa. Figure~\ref{fg:nA}b
shows the same differential quantities in the boosted laboratory frame ($n^*
\rightarrow N, \alpha^* \rightarrow A$) (here, $E_{max}\approx 3.1$~GeV and $A$
is the laboratory asymmetry). As discussed later, the statistical uncertainty
on the measurement of \wa\ is inversely proportional to the ensemble-averaged
figure-of-merit (FOM) $NA^2$. The differential quantity $NA^2$, shown in the
Fig.~\ref{fg:nA}b, illustrates the relative weight by electron energy to the
ensemble average FOM.

\begin{figure}
\begin{centering}
\includegraphics*[angle=-90,width=0.9\textwidth]{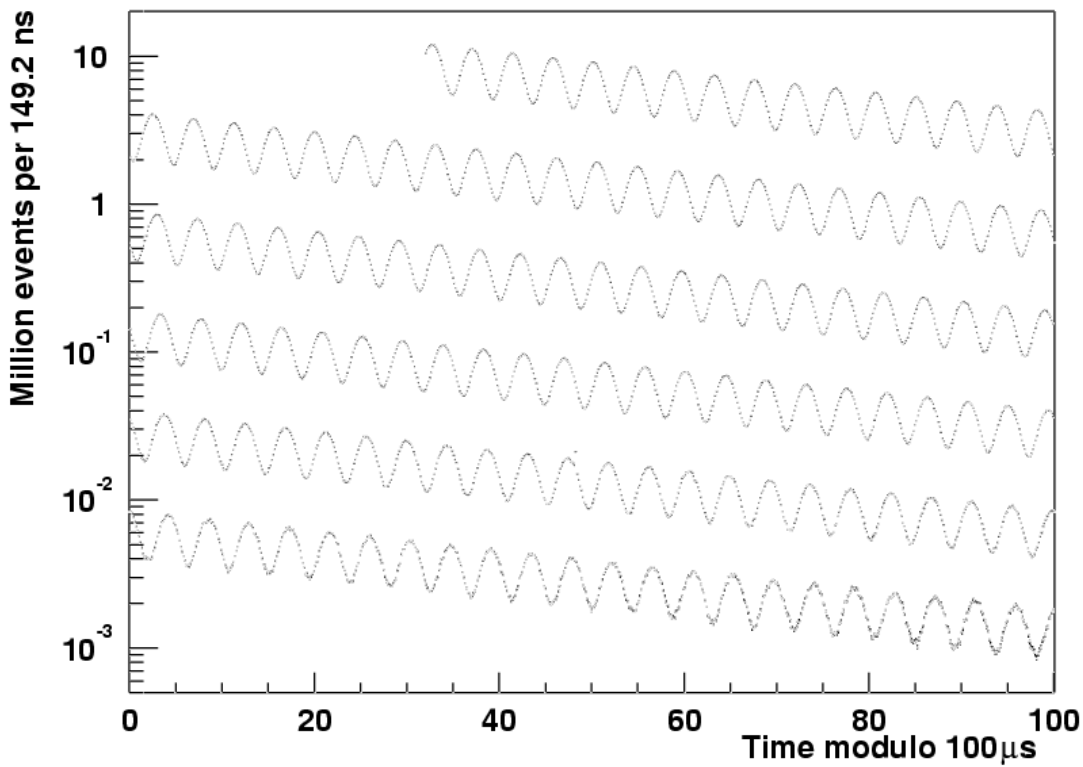} \caption{Distribution
of electron counts versus time for the 3.6 billion muon decays in
the R01 $\mum$ data-taking period. The data is wrapped around modulo
$100~\mu$s.  \label{fig:wiggles}}
\end{centering}
\end{figure}

Because the stored muons are highly relativistic, the decay angles
observed in the laboratory frame are greatly compressed into the
direction of the muon momenta.  The lab energy of the relativistic
electrons is given by
\begin{equation}
E_{lab} = \gamma(E^{*} + \beta p^{*}c \cos\theta^{*}) \approx
\gamma E^{*} (1 + \cos\theta^{*}).
\end{equation}
Because the laboratory energy depends strongly on the decay angle
$\theta^{*}$, setting a laboratory threshold $E_{th}$ selects a
range of angles in the muon rest frame. Consequently, the
integrated number of electrons above $E_{th}$ is modulated at
frequency $\wa$ with a threshold-dependent asymmetry. The
integrated decay electron distribution in the lab frame has the
form
\begin{equation}
N_{ideal}(t) = N_{0}\exp(-t/\gamma\tau_{\mu})\left[1 - A\cos(\wa t
+ \phi)\right], \label{eq:fivepar}
\end{equation}
where $N_{0}$, $A$ and $\phi$ are all implicitly dependent on $E_{th}$. For a threshold energy of
1.8~GeV ($y \approx 0.58$ in Fig.~\ref{fg:nA}b), the asymmetry is $\approx0.4$ and the average FOM is
maximized. A representative electron decay time histogram is shown in Fig.~\ref{fig:wiggles}.

To determine \amu, we divide $\wa$ by \wpt, where \wpt\ is the
measure of the average magnetic field seen by the muons. The
magnetic field, measured using NMR, is proportional to the free
proton precession frequency, $\omega_{p}$.
The muon anomaly is given by:
\begin{equation}
\amu =  \frac{\wa}{\omega_L - \wa}
      =  \frac{\wa/\wpt}{\omega_{L}/\wpt - \wa/\wpt}
      =  \frac{\mathcal{R}}{\lambda - \mathcal{R}},
      \label{eq:amuR}
\end{equation}
where $\omega_L$ is the Larmor precession frequency of the muon. The ratio $\mathcal{R} = \wa/\wpt$ is
measured in our experiment and the muon-to-proton magnetic moment ratio
\begin{equation}
\lambda = \omega_L/\omega_p = 3.18334539(10) \label{eq:lambda}
\end{equation}
is determined from muonium
hyperfine level structure measurements~\cite{Liu:1999,PDG:2000}.

The BNL experiment was commissioned in 1997 using the same pion
injection technique employed by the CERN~III experiment. Starting in
1998, muons were injected directly into the ring, resulting in many
more stored muons with much less background. Data were obtained in
typically 3-4 month annual runs through 2001. In this paper, we
indicate the running periods by the labels R97 - R01. Some facts
about each of the runs are included in Table~\ref{tb:runs}.

\begin{table}
  \caption{Running periods, total number of electrons recorded $30~\mu$s or more after
  injection having $E > 1.8$~GeV.  Separate systematic uncertainties are given for the field
  ($\omega_p$) and precession (\wa) final uncertainties. \label{tb:runs} }
\begin{tabular}{ccccccc}
\toprule
  Run  & Polarity & Electrons &  Systematic & Systematic & Final Relative \\
  Period  & & [millions] &  \wp\ [ppm] & \wa\ [ppm] & Precision [ppm] \\
  \colrule
  R97   &$\mu^+$  & 0.8  &  1.4 & 2.5   & 13 \\
  R98   &$\mu^+$  & 84  &  0.5 & 0.8  & 5 \\
  R99   &$\mu^+$  & 950  &  0.4 & 0.3  & 1.3 \\
  R00   &$\mu^+$  & 4000  &  0.24 & 0.31  & 0.73 \\
  R01   &$\mu^-$  & 3600  &  0.17 & 0.21 & 0.72\\
  \botrule
  \end{tabular}
\end{table}

\subsection{Beamline \label{ssec:beamline}}

Production of the muon beam begins with the extraction of a bunch of 24~GeV/$c$
protons from the AGS. The protons are focused to a 1~mm spot on a 1-interaction
length target, which is designed to withstand the very high stresses associated
with the impact of up to $7 \times 10^{12}$ protons per bunch. The target is
composed of twenty-four 150-mm diameter nickel plates, 6.4-mm thick and
separated by 1.6~mm. To facilitate cooling, the disks rotate at approximately
0.83~Hz through a water bath. The axis of rotation is parallel to the beam.

Nickel is used because, as demonstrated in studies for the Fermilab
antiproton source~\cite{fnal_pbar}, it can withstand the shock of
the instantaneous heating from the interaction of the fast beam. The
longitudinal divisions of the target reduce the differential
heating. The beam strikes the outer radius of the large-diameter
disks. The only constraint on the target transverse size is that a
mis-steered proton beam does not allow production from a part of the
target that would result in a high flux of pions entering the
storage ring during muon injection running.  This region corresponds
to the outer edge of the disks. Otherwise, the production target
transverse size is defined by the beam size. With the large radius
disks, shock damage of the target is distributed over the disk
circumference as the disks rotate. Still, it was necessary to
replace the target after each running period, typically following an
exposure of $5 \times 10^{19}$ protons, or when nickel dust was
observed in the target water cooling basin.

\begin{figure}
  \begin{center}
    \includegraphics*[width=0.75\textwidth]{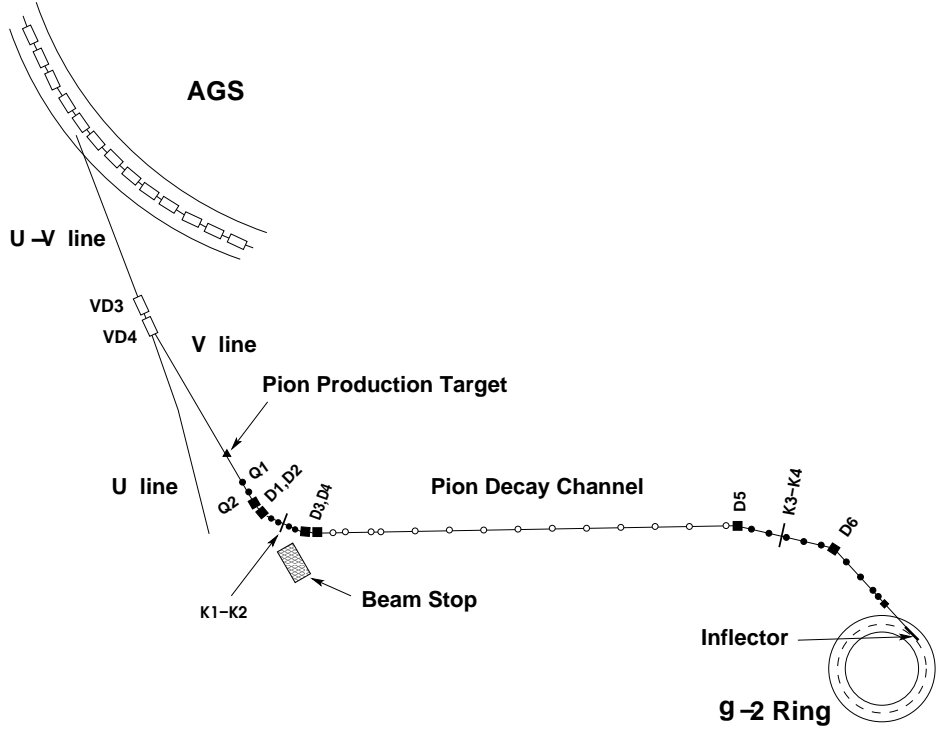}
    \caption{Plan view of the pion/muon beamline. The pion decay channel is 80~m
    and the ring diameter is 14.1~m.\label{fig:beamline}}
  \end{center}
\end{figure}

Pions are collected from the primary target at zero angle and
transferred into a secondary pion-muon decay channel, designed to
maximize the flux of polarized muons while minimizing pion
contamination.
A schematic representation of the beamline is shown in
Fig.~\ref{fig:beamline} and selected proton beam and pion beamline
parameters are given in Table~\ref{tab:beams}. Downstream of the
target, pions produced in the forward direction within a solid angle
of 32~mrad (horizontal) and 60~mrad (vertical) are collected by
quadrupoles Q1 and Q2. A momentum-dispersed image of the target is
created at the K1-K2 slits. The momentum dispersion is $0.04\%/$mm
and under typical running conditions the pion momentum width was
$\delta p/p \sim \pm 0.5\% $. The momentum-recombined beam then
traverses an 80~m quadrupole magnet FODO (alternating focusing and
defocusing) straight section. A momentum dispersed image of the
target is created at the K3-K4 slits. Here the momentum dispersion
is $0.06\%/$mm. The momentum-recombined beam is then focused to
allow passage through the hole in the back leg of the ring magnet,
through the strong vertically focusing storage ring fringing field,
the inflector, and into the storage volume. For pion injection, the
K1-K2 slits select the momentum of the pion beam and the K3-K4 slits
are effectively not used. For muon injection, the K3-K4 slits reject
most of the pion beam and select particles having the storage ring
momentum.

The beamline and ring were operated in different modes for the
five running periods.  In the R97 period, pions of central
momentum 0.5~percent above the magic momentum were injected
directly into the storage ring. With typically $5 \times 10^{12}$
protons on target, $10^8$ particles were injected into the muon
storage ring per fill. The muon capture efficiency was 25~ppm,
consistent with Monte Carlo estimates.  The majority of the pions
strike obstructions in the ring or decay to muons that are not
stored. This initial particle ``flash'' induces a considerable
background in the electron detectors.  All subsequent running
periods used direct muon injection,  where muons of the magic
momentum were injected into the ring and then kicked transversely
onto stable trajectories.

\begin{table}
\begin{center}
  \caption{Selected AGS proton beam and secondary pion beamline
  characteristics \label{tab:beams}}
  \begin{tabular}{lr|lr}
  \toprule
  Proton Beam & Value~ & ~Pion Beamline & Value \\
  \colrule
  Protons per AGS cycle & $5\times 10^{13}$ ~& ~Horizontal emittance &  $42~\pi $mm-mrad \\
  Cycle repetition rate & 0.37 Hz ~&~ Vertical emittance &  $56~\pi $mm-mrad \\
  Proton momentum & 24 GeV/$c$ ~&~ Inflector horizontal aperture        &  $\pm 9$ mm \\
  Bunches per cycle & 6 to 12 ~&~ Inflector vertical aperture &  $\pm 28$ mm \\
  Bunch width ($\sigma $) & 25 ns ~&~ Pions per proton$^*$ &  $10^{-5}$ \\
  Bunch spacing    & 33 ms ~&~ Muons per pion decay$^{**}$ &  0.012 \\
  \botrule
  \end{tabular}
  \end{center}
  {\small $^*$Captured by the beamline channel;
  $^{**}$Measured at the inflector entrance}
\end{table}

While the momentum of the downstream section of the beamline---after
the K3-K4 slits---was always set to the magic momentum of
3.094~GeV/$c$, the upstream capture and decay sections were adjusted
to meet the competing demands of high muon flux and low pion
contamination. The number of stored muons is maximized when the
upstream beamline is tuned 0.5~percent above the magic momentum.
However, this small momentum difference does not provide adequate
pion rejection at the K3-K4 slits.

The muon transmission to the storage ring entrance, the pion
survival fraction $F_\pi$ past K3-K4 (a figure of merit for pion
contamination), and the muon polarization, were calculated using the
{\tt BTRAF} beamline transport program~\cite{btraf}.  The results
were compared to measurements (See Table~\ref{tab:calc}). The muon
transmission efficiency is determined by counting the number of
high-energy electrons in the calorimeters in a given time period
well after the hadronic flash has dissipated. The pion survival
fraction is determined from beamline \v{C}erenkov measurements (see
below). The asymmetry in the observed \g2\ oscillations is
proportional to the beam polarization, as well as to the asymmetry
of the weak decay. It is also affected by the detector acceptance
and resolution. In particular, Monte Carlo simulations using $E_{th}
= 1.5$~GeV predict $A = 0.30\pm 0.01$ for a 100~percent polarized
beam. The measured asymmetry, obtained from fits of the data to
Eq.~\ref{eq:fivepar}, is found to be lower than the prediction when
$p_\pi /p_\mu$ is in the range $1.005 - 1.010$. The dilution is
caused by muons born from pion decays between the target and the
pion momentum selection at K1-K2, which for this small momentum
difference, are transported and stored in the ring. The ratio $p_\pi
/p_\mu = 1.017$ was chosen as the optimal running condition; it
features a high asymmetry and storage fraction, and an acceptably
low pion contamination.

\begin{table}
\begin{center}
\caption{As a function of the ratio of central-pion to muon momentum
$p_\pi /p_\mu$: From left to right, calculated relative muon
transmission fraction; measured relative stored muon flux;
calculated and measured pion transmission fraction into the ring;
calculated muon polarization; measured asymmetry $A$ using
Eq.~\ref{eq:fivepar} and $E_{th} = 1.5$~GeV.  The absolute fraction
of muons per pion decay is obtained by multiplying column~2 by
0.018.\label{tab:calc}}
\begin{tabular}{ccccccc}
\toprule $p_\pi /p_\mu $& ~~$N_{\mu}$(Calc)~~ &
~~$N_{\mu}$(Meas)~~
& ~~$F_{\pi}$(Calc)~~& ~~$F_{\pi}$(Meas)~~ & ~~$P_\mu$(Calc)~~ & ~~$A$(Meas)~~ \\
\colrule
1.005      & \underline{1} & \underline{1} & 0.78   &  0.80  & 0.99 & 0.22 \\
1.010      &   0.5   &  0.43 & 0.29   &  0.30  & 0.98 & 0.26 \\
1.015      &   0.29  &  0.21 & 0.04   &  0.065 & 0.96 & 0.30 \\
1.017      &   0.25  &  0.17 & 0.002  &  0.016 & 0.96 & 0.30 \\
1.020      &   0.18  &  0.12 &  -     &  0.009 & 0.95 & 0.30 \\
\botrule
\end{tabular}
\end{center}
\end{table}

Relative particle species fractions immediately downstream of the
K3-K4 slit were determined using a tunable threshold gas
\v{C}erenkov counter. Table~\ref{tab:cerenkov} lists the particle
fractions for a positive pion beam, measured during the R98 period.
The pion lifetime is 0.58~$\mu$s, much less than the detector
gate-on time. The fraction of pions transmitted to the storage ring
falls as $p_\pi / p_\mu $ is increased, which is expected from the
momentum selection at K3-K4. However, the increase in the beam
positron fraction was unanticipated. We believe it is due to
radiation in the beam monitoring elements and vacuum windows between
the momentum selection slits at K1-K2 and K3-K4. Positrons that are
stored in the ring lose energy by synchrotron radiation; they were
observed to spiral into the calorimeters during the first $5~\mus $
after injection. Protons, which could not be identified by the
\v{C}erenkov detector, were estimated to be about one third as
numerous as the pions. Stored protons are discussed in
Section~\ref{ssec:muloss}. The antiproton flux for the R01 running
period was negligible, typically suppressed by a factor of 500
compared to protons in the R00 running period.

\begin{table}
\begin{center}
\caption{The relative fraction of $e^+ $, $\mu^+ $ and $\pi^+$ versus the
pion-to-muon momentum ratio in the R98 period as determined by \v{C}erenkov
measurements after the K3-K4 slits. Protons are estimated to be 1/3 the pion
flux. \label{tab:cerenkov}}

\begin{tabular}{ccccc}
\toprule

  ~~~~$p_\pi /p_\mu $~~~~  &  ~~~~$e^+ $~~~~  &  ~~~~$\mu^+ $~~~~ & ~~~~$\pi^+$~~~~  & ~~~~$\pi /\mu $~~~~ \\
  \colrule
  1.000            &   0.16     &  0.014     & 0.83      & 59 \\
  1.009            &   0.14     &  0.05       & 0.81      & 16 \\
  1.014            &   0.17     &  0.06       & 0.77      & 13 \\
  1.017            &   0.34     & 0.34       & 0.32      &  1 \\
  \botrule
\end{tabular}
\end{center}
\end{table}

%
%
%
%
%

\subsection{Inflector \label{ssec:inflector}}
\begin{figure}
\includegraphics[angle=-90,width=\textwidth]{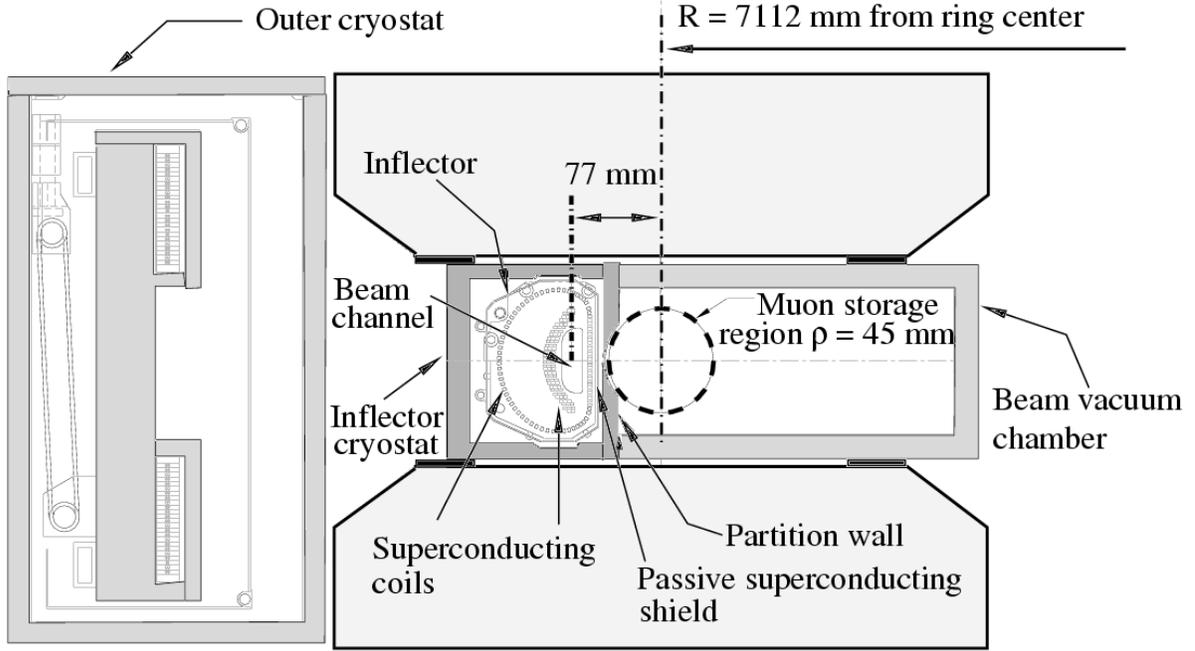}
\caption{The inflector/storage ring geometry.  The downstream end
of the inflector is shown, with the beam channel to the left of
the storage region (larger radius).  The ring center is to the
right.  Note the limited space between the pole pieces, which has
to contain the inflector and its cryostat along with the beam
vacuum chamber.  The current in the inflector flows into the page
in the ``C'' shaped arrangement of conductors just to the left of
the beam channel, and out of the page in the conductors that form
a backward ``D''.  The superconductor crosses over the beam
channel to connect the two coils. } \label{fig:inflectorgeometry}
\end{figure}

The muon beam passes through position and intensity instrumentation, through a thin vacuum window and
into a 1~m long hole in the backleg of the storage ring magnet, in air. After passing through an
opening in the outer coil cryostat and additional position instrumentation, it passes through a thin
vacuum window into the 1.7~m long superconducting inflector magnet~\cite{Yamamoto:2002}, whose 1.5~T
vertical field (field integral 2.55~T$\cdot$m) cancels the main storage ring field, permitting the
muons to pass largely undeflected into the storage ring. The current windings feature a unique
double-truncated cosine theta design~\cite{krienen:1989}, which minimizes the flux that leaks outside
the inflector volume. The geometry at the inflector exit is shown in Fig.~\ref{fig:inflectorgeometry}.
The inflector axis is approximately tangent to the storage ring, and it is adjustable by $\pm 4$~mrad.
The beam center in the inflector channel exit is 77~mm from the storage-ring center.

\begin{figure}
\includegraphics[width=\textwidth]{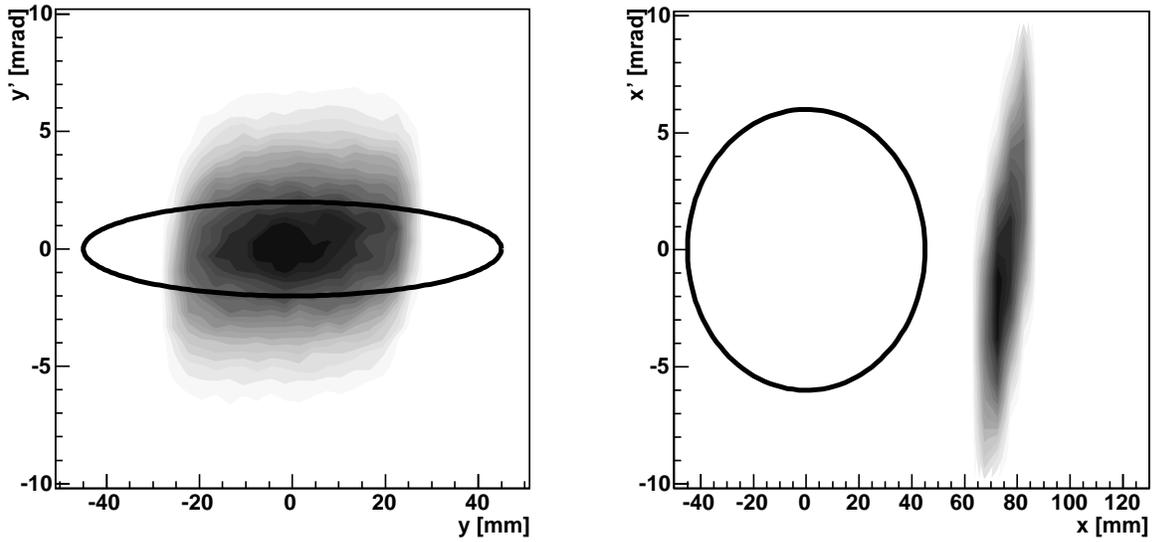}
\caption{The phase space plot for the inflector exit from a beam transport
simulation ($x$ is horizontal; $y$ is vertical).  Left plot: $p_y/p_z = y'$ vs.
$y$. Right plot: $p_x/p_z = x'$ vs. $x$.  The inflector center is displaced
from the storage ring central orbit by +77~mm. The ellipses represent the
storage ring acceptance.  After one quarter turn, the distribution in $x$ has
rotated through 90 degrees and lies below the ring acceptance.  It is then
kicked toward more positive $x'$, into the ring acceptance.}
\label{fig:inflectorphasespace}
\end{figure}

Placing the inflector cryostat in the limited space between the muon storage region and the outer main
magnet coil restricted the inflector aperture size to $18(w){\rm ~mm} \times 56(h) {\rm ~mm}$, which
is significantly smaller than the 90~mm diameter storage ring aperture.  The small size limits the
flux of incoming muons and introduces a mismatch in phase space with respect to the storage ring.
Figure~\ref{fig:inflectorphasespace} shows the vertical and horizontal muon beam phase space ($y, y'$
and $x,x'$) as simulated for the exit of the inflector.  Superimposed on the figures are the storage
ring acceptance ellipses. The muons undergo betatron harmonic motion in the storage ring, following
elliptical paths about the origin in phase space.

The precision magnetic field in the storage region is protected from the small
leakage flux from the end of the inflector by means of a passive
superconducting shield. The inflector is cooled down after the storage ring
magnet has been energized and the main field is stable. The superconducting
shield then pins the main field and traps the inflector fringe field as the
inflector is energized. The disturbance of the main storage ring field by the
inflector fringe field is negligible. However, in 1997 before installing it
into the ring, the first inflector required a repair, which could only be made
by cutting through the shield. The resulting fringe field reduced the storage
ring field by 600~ppm over a $~1^{\circ}$ azimuthal angle, resulting in
unacceptable magnetic field gradients for the NMR trolley probes closest to the
inflector body. The field in this region had to be mapped by a special
procedure following data taking. This introduced additional uncertainty into
the measurement of the average field, 0.20~ppm in the R99 result.

The damaged inflector was replaced before the 2000 running period.
In the new inflector, the superconducting shield was extended
further beyond the downstream end, and the lead geometry was changed
to reduce the fringe field due to the inflector leads. For both the
R00 and R01 running periods, the fringe field of the inflector was
negligible.



\subsection{Muon storage ring magnet \label{ssec:ring}}

\begin{figure}
\begin{center}
\includegraphics[width=.8\textwidth]{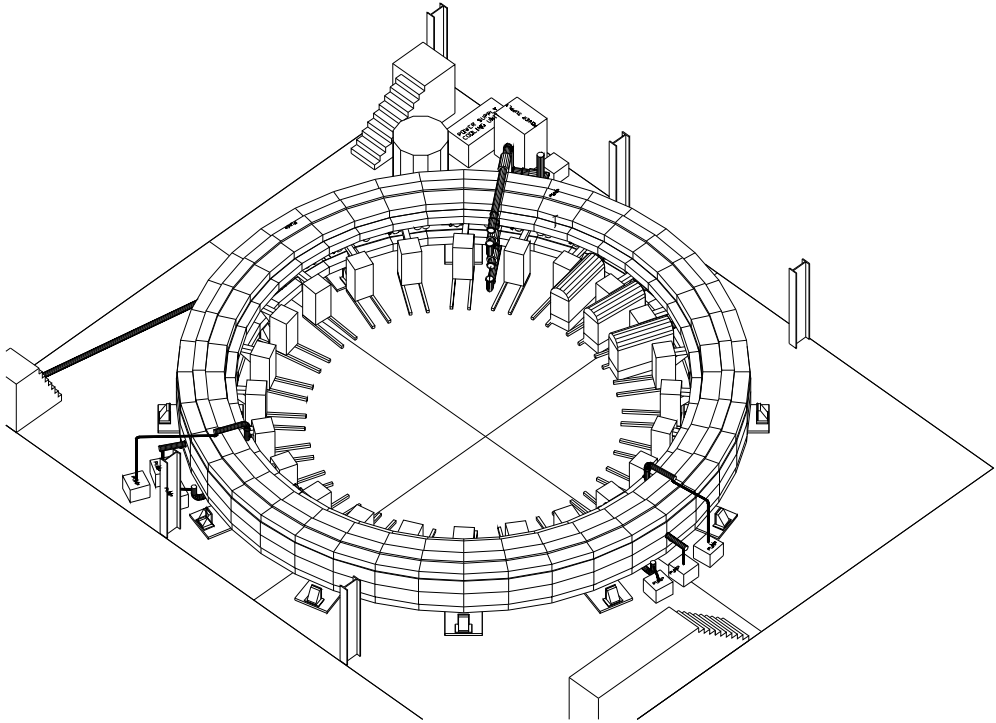}
\caption{A 3D engineering rendition of the E821 muon storage ring. Muons enter the back of the storage
ring through a field-free channel at approximately 10~o'clock in the figure.  The three kicker
modulators at approximately 2~o'clock provide the short current pulse, which gives the muon bunch a
transverse 10~mrad kick. The regularly spaced boxes on rails represent the electron detector systems.
\label{fig:3Dring}}
\end{center}
\end{figure}

\begin{figure}
\begin{center}
  \includegraphics[width=\textwidth,angle=0]{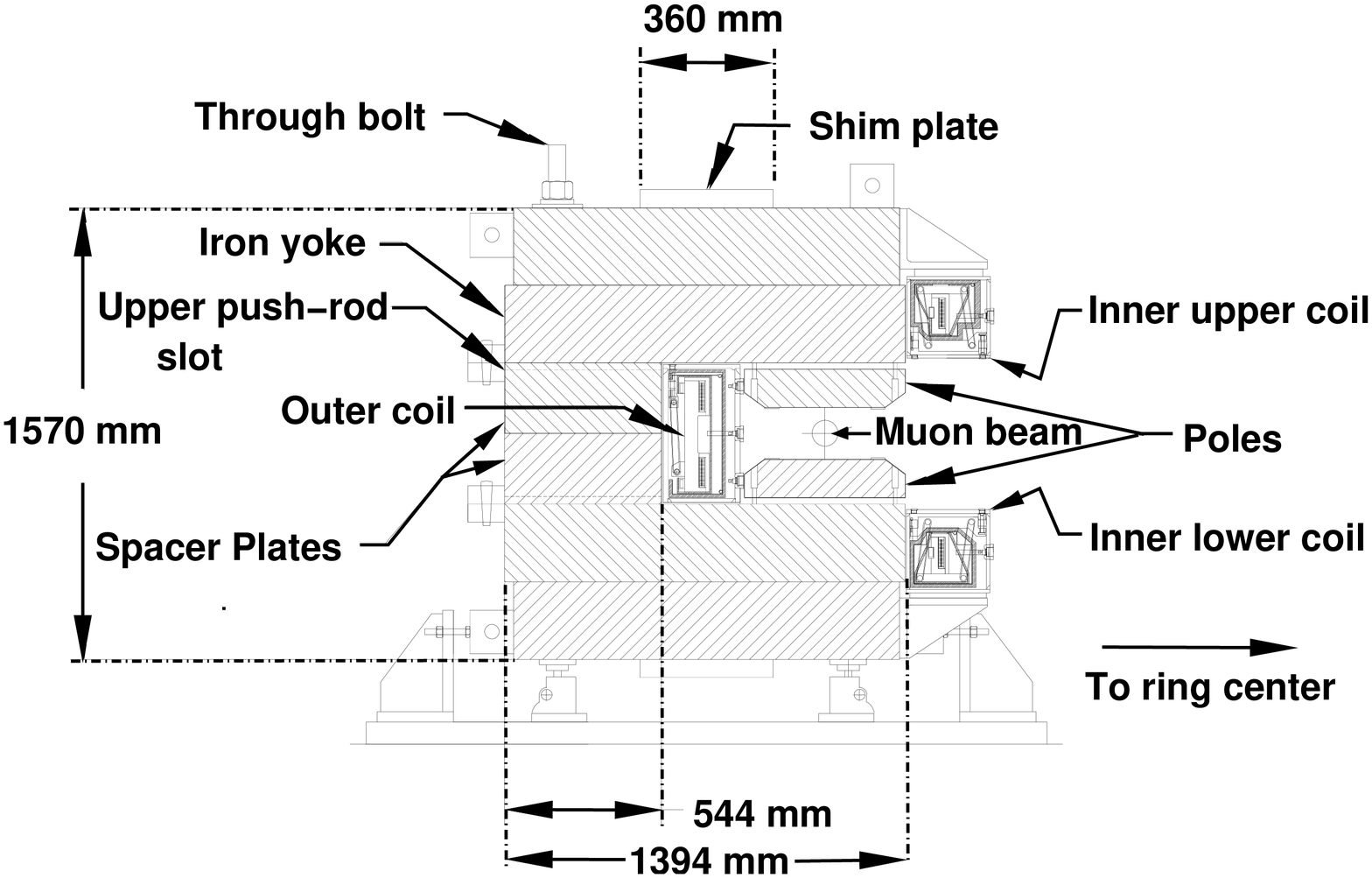}
  \caption{Cross sectional view of the ``C" magnet.}
  \label{fig:magnet}
  \end{center}
\end{figure}

The muon storage ring~\cite{Danby:2001} is a superferric ``C"-shaped magnet,
7.112~m in central orbit radius, and open on the inside to permit the decay
electrons to curl inward to the detectors (Fig.~\ref{fig:3Dring}). A 5~V power
supply drives a 5177~A current in the three NbTi/Cu superconducting coils.
Feedback to the power supply from the NMR field measurements maintains the
field stability to several ppm. The field is designed to be vertical and
uniform at a central value of 1.4513~T.   High-quality steel, having a maximum
of 0.08~percent carbon, is used in the yoke. Low-carbon steel is used for the
poles primarily because the fabrication process of continuous cast steel
greatly minimizes impurities such as inclusions of ferritic or other extraneous
material and air bubbles.  An air gap between the yoke and the higher quality
pole pieces decouples the field in the storage region from non-uniformities in
the yoke.    Steel wedge shims are placed in the air gap. Eighty low-current
surface correction coils go around the ring on the pole piece faces for active
trimming of the field. The opening between the pole faces is 180~mm and the
storage region is 90~mm in diameter. A vertical cross section of the storage
ring illustrating some of these key features is shown in Fig.~\ref{fig:magnet}.
Selected storage ring parameters are listed in Table~\ref{tab:StorageRing}.

Attaining high field uniformity requires a series of passive
shimming adjustments, starting far from and then proceeding towards
the storage region. First the twelve upper- and lower-yoke
adjustment plates are shimmed by placing precision spacers between
them and the yoke steel, modifying the air gap. Next the 1000 wedge
shims in the yoke pole-piece air gap are adjusted. With a wedge
angle of 50~mrad, adjusting the wedge position radially by 1~mm
changes the thickness of iron at the center of the storage aperture
by 50~$\mu$m. The wedge angle is set to compensate the quadrupole
component, and radial adjustments of the wedge and other changes to
the air gap are used to shim the local dipole field. The local
sextupole field is minimized by changing the thickness of the 144
edge shims, which sit on the inner and outer radial edges of the
pole faces. Higher moments, largely uniform around the ring, are
reduced by adjusting the 240 surface-correction coils, which run
azimuthally for 360 degrees along the surface of the pole faces.
They are controlled through 16 programmable current elements. With
adjustments made, the azimuthally averaged magnetic field in the
storage volume had a uniformity of $\simeq 1$~ppm during data-taking
runs.

The main temporal variation in the magnetic field uniformity is
associated with radial field changes from seasonal and diurnal drift
in the iron temperature. Because of the ``C'' magnet geometry,
increasing (or decreasing) the outside yoke temperature can tilt the
pole faces together (or apart), creating a radial gradient. The yoke
steel was insulated prior to the R98 run  with 150~mm of fiberglass
to reduce the magnetic-field variation with external temperature
changes to a negligible level.

\begin{table}
\begin{center}
  \caption{Selected muon storage ring parameters.\label{tab:StorageRing}}
  \begin{tabular}{lr}
  \toprule
  Parameter & Value \\
  \colrule
  Nominal magnetic field       &  1.4513 T \\
  Nominal current              &  5200 A \\
  Equilibrium orbit radius     &  7.112 m \\
  Muon storage region diameter &  90 mm \\
  Magnet gap                   &  180 mm \\
  Stored energy                &  6 MJ \\
  \botrule
  \end{tabular}
  \end{center}
\end{table}

\subsection{Electric quadrupoles \label{ssec:quads}}

\begin{figure}
\begin{center}
\includegraphics*[width=0.7\textwidth]{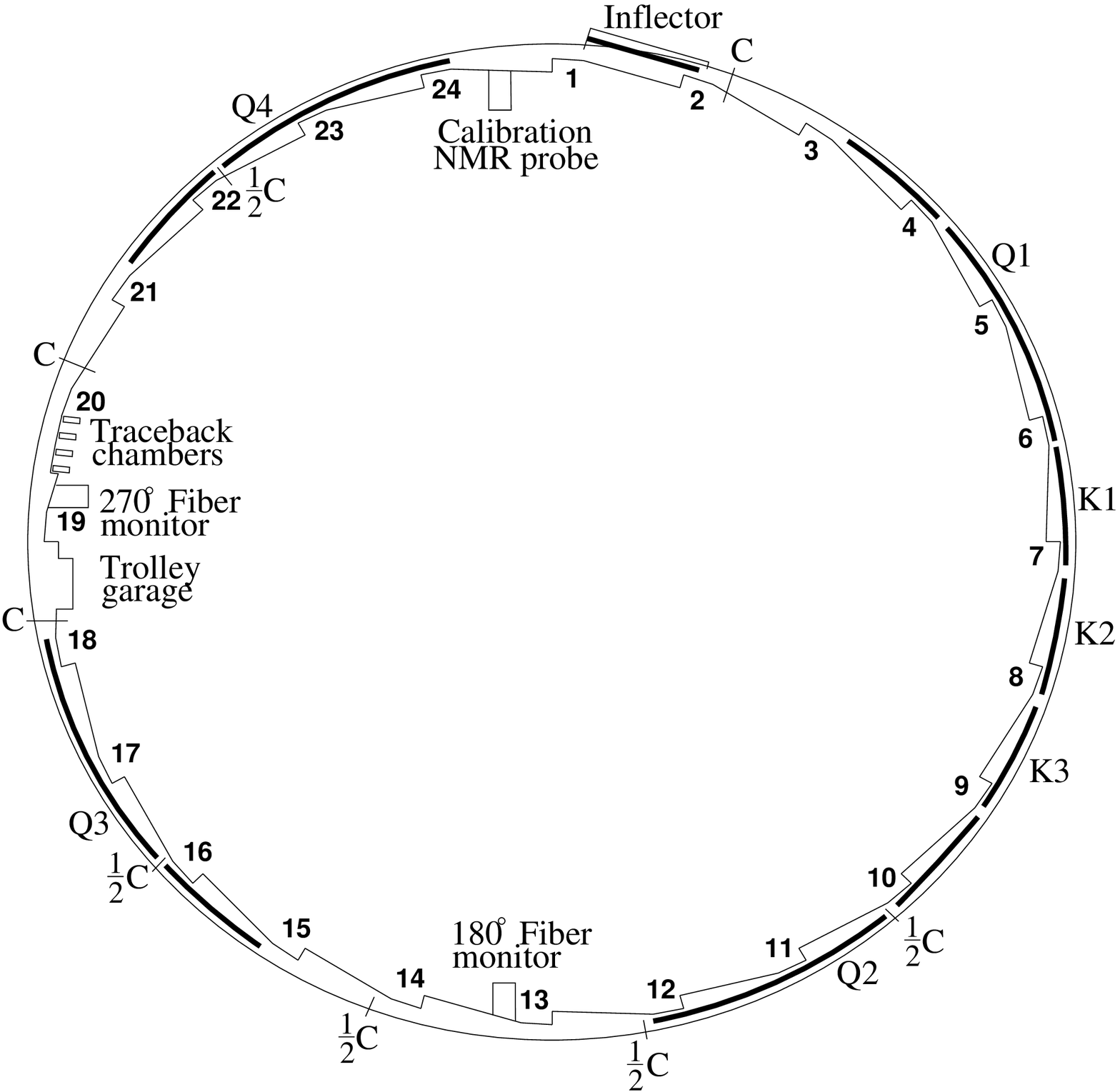}
 \caption{The ~\gm ~storage ring layout. The 24 numbers represent
 the locations of the calorimeters immediately downstream of
 the scalloped vacuum chamber subsections.  Inside the vacuum are four
 quadrupole sections  (Q1-Q4), three kicker plates (K1-K3) and
 full-aperture (C) and half-aperture ($\frac{1}{2}$C) collimators.
 The traceback chambers follow a truncated scalloped vacuum chamber
 subsection.  \label{fig:ringoverhead}}
\end{center}
\end{figure}

Electrostatic quadrupoles are used for vertical focussing of the
beam in the storage ring.  Ideally, the electrodes should fill as
much of the azimuth as possible; but, space is required for the
kicker magnet and the inflector. For symmetry reasons, the
quadrupoles are divided into four distinct regions: Q1-Q4 (see
Fig.~\ref{fig:ringoverhead}). Gaps at $0^\circ $ and $90^\circ $ for
the inflector and kicker magnets, along with empty gaps at
$180^\circ $ and $270^\circ $ provide a fourfold lattice symmetry.
Overall, the electrodes occupy 43 percent of the total
circumference. The fourfold symmetry keeps the variation in the beta
function small, $\sqrt {\beta_{\rm max} / \beta_{\rm min}}= 1.04$,
which minimizes beam ``breathing'' and improves the muon orbit
stability.

The quadrupole voltage should be set as high as possible to maximize muon
storage, subject to an upper limit of $\sim$25~kV (negative muons, can be
higher for positive muons) necessary for reliable operation in the vacuum
chamber, which has a typical vacuum of $\sim1.3 \times 10^{-5}$~Pa. The plates
are charged prior to each fill and the voltage is held constant through the
measuring period, which extends for at least $700~\mu$s. At injection, the
plates are charged asymmetrically to shift the beam horizontally and vertically
against a set of centered circular collimators---the {\it scraping} procedure.
Approximately $5 - 15~\mu$s later, the plate voltages are symmetrized to enable
long-term muon storage.
The
operating voltages, field indices (which are proportional to the
quadrupole gradient), the initial asymmetric scraping time, and the
total pulse length are given in Table~\ref{tb:quadstuff}.

\begin{table}[ht]
\begin{center}
\caption{Quadrupole high voltage, field index, scraping time and
the total pulse length for the different running
periods.\label{tb:quadstuff}}
\begin{tabular}{ c c c c c }
\hline \hline
Run &  ~~~HV~~~ & ~~~~$n$~~~~ & $T_{\rm scraping}$\   &  Pulse Length\  \\
Period & [kV]    &           & [$\mu$s]   &   [$\mu$s]   \\
\hline
R97   &  24.4 &  0.137  & 16 & 650 \\
R98   &  24.4 &  0.137  & 16 & 800 \\
R99   &  24.4 &  0.137  & 16 & 900 \\
R00   &  24.2 &  0.136  & 16 & 1400 \\ \hline
R01   &  21.7 &  0.122  & 7  & 700 \\
      &  25.3 &  0.142  & 7  &  700 \\
\hline \hline

\end{tabular}
\end{center}

\end{table}

\begin{figure}
\begin{center}
    \includegraphics*[width=0.5\textwidth]{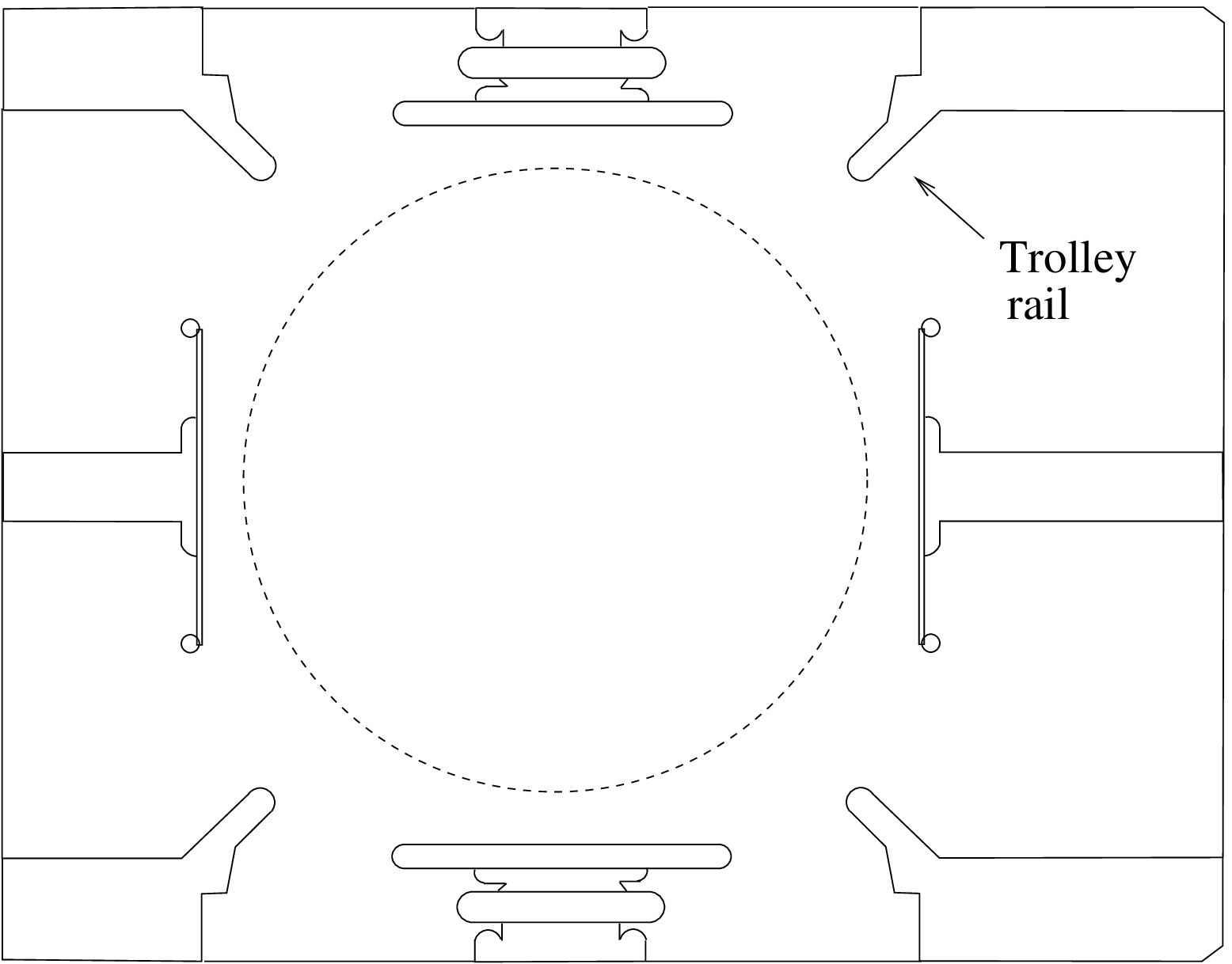}
 \caption{Schematic view of the electrostatic quadrupoles
 inside the vacuum chamber. For positive muons, the top and
 bottom electrodes are at $\sim +24$~kV; the side electrodes are
 at $\sim -24$~kV.
 The NMR trolley rails can be seen between  the electrodes in the
 $V = 0$ planes. The 90~mm  diameter storage region is depicted
 by the dashed  circle.\label{fig:quads}} \end{center} \end{figure}

A schematic representation of a cross section of the electrostatic quadrupole
electrodes~\cite{Semertzidis:2003} is shown in Fig.~\ref{fig:quads}. The four trolley rails are at
ground potential.  Flat, rather than hyperbolic, electrodes are used because they are easier to
fabricate. With flat electrodes, electric-field multipoles $ 8,\ 12,\ 16,\cdots$  in addition to the
quadrupole are allowed by the four-fold symmetry  (see Fig.~\ref{fig:quads}). Of these, the  12- and
20-pole components are the largest.   The ratio of the width (47~mm) of the electrode to the distance
between opposite plates (100~mm) is set to minimize the 12-pole component. Beam dynamics simulations
indicated that even the largest of the resulting multipoles, a 2 percent 20-pole component (at the
circular edge of the storage region), would not cause problems with muon losses or beam instabilities
at the chosen values of the field indices. The scalloped vacuum chamber introduces small 6- and
10-pole multipoles into the field shape.

The quadrupoles are charged for $\leq1.4$~ms of data taking during
each fill of the ring. Cycling the quadrupoles prevents the
excessive buildup of electrons around the electrodes, electrons
which are produced by field emission and gas ionization and
subsequently trapped in the electric and magnetic fields near the
quadrupoles. Trapping was particularly severe during the R01 running
period when negative muons were injected into the ring. The
continuous motion of the electrons---cyclotron motion in the dipole
magnetic field, magnetron motion along $\vec{E} \times \vec{B}$, and
axial oscillations along the vertical axis---ionizes the residual
gas and eventually produces a spark, which discharges the plates.
\par
Slight modifications of the magnetron motion were used to quench the electron trapping. In the
original design, electrons undergoing magnetron motion were trapped in horizontal orbits around the
vertical plates:  the natural symmetry of the electric field at the ends caused the circulating
electrons to re-enter the quadrupoles and return  to their starting point in approximately 50 (100)
$~\mu$s, for the short (long) sections. To prevent the recirculation of trapped electrons, the leads
were adjusted to rotate the field symmetry at the end of the plate by approximately $25^\circ$; the
resulting dipole field largely sweeps away the electrons. Also, field emission was minimized during
data-taking periods by conditioning the quadrupole electrodes and by returning them to full voltage
very slowly after every set of NMR trolley measurements.
\par
In addition to creating sparks, trapped electrons can also distort
the field created by the quadrupole electrodes. In a quadrupole
field with $E_y= ky$, the vertical oscillation frequency $\omega_A =
\sqrt{ek/m_e}$  provides a direct measure of the the actual gradient
$k$. Reducing the number of trapped electrons permitted the
application of a very large dc voltage to the electrodes and, in
turn, a measurement of the vertical resonance frequency of those
that remained. Thus we were able to set a sensitive limit for the
influence of trapped electrons on the electric-field configuration
as well as a limit on the magnetic fields that they must also
produce.

\subsection{Pulsed kicker magnet \label{ssec:kicker}}
Direct muon injection requires a pulsed kicker~\cite{kicker} to
place the muon bunch into the phase space acceptance of the storage
ring.  The center of the circular orbit for muons entering through
the inflector channel is offset from that of the storage ring and,
left alone, muons would strike the inflector exit after one
revolution and be lost.  A kick of approximately 10~mrad
($\sim0.1$~T$\cdot$m) applied one quarter of a betatron wavelength
downstream of the inflector exit is needed to place the injected
bunch on an orbit concentric with the ring center.  The ideal kick
would be applied during the first revolution of the bunch and then
turned off before the bunch returns on its next pass.

The E821 kicker makes use of two parallel current sheets with cross-overs at
each end so that the current runs in opposite directions in the two plates. The
80-mm high kicker plates are 0.75-mm thick aluminum, electron-beam welded to
aluminum rails at the top and bottom, which support the assembly and serve as
rails for the 2-kg NMR trolley.  The entire assembly is 94-mm high and 1760-mm
long. This plate-rail assembly is supported on Macor$^{\circledR}$ insulators
that are attached to Macor$^{\circledR}$ plates with plastic hardware forming a
rigid cage, which is placed inside of the vacuum chamber. {\tt
OPERA}~\cite{OPERA} calculations indicated that aluminum would minimize the
residual eddy currents following the kicker pulse, and measurements showed that
the presence of this aluminum assembly would have a negligible effect on the
storage ring precision magnetic field.

The kicker is energized by an LCR pulse-forming network (PFN),
which consists of a single capacitor, resistor, and the loop
formed by the kicker plates.  The capacitor is charged by a
resonant circuit and the current pulse is created when the
capacitor is shorted to ground at the firing of a deuterium
thyratron.  The total inductance of the PFN is $L= 1.6~\mu$H,
which effectively limits both the peak current and the minimum
achievable pulse width.  The resistance is $R=11.9~\Omega$ and the
capacitance is $C=10.1$~nF. For the damped LCR circuit, the
oscillation frequency is $f_d=1.08 \times 10^6$~Hz and the decay
time is $\tau_d=924$~ns.  The peak current is given by $I_0 =
{V_0/ ({2 \pi f_d L})}$, where $V_0$ is the initial voltage on the
capacitor.  The resulting current pulse has a base width of
$\sim400$~ns, which is long compared to the 149~ns cyclotron
period. Therefore, the positive kick acts on the muon bunch in the
first, second, and third turns of the storage ring.

\begin{figure}[htbp]
  \begin{center}
    \includegraphics*[width=0.5\textwidth]{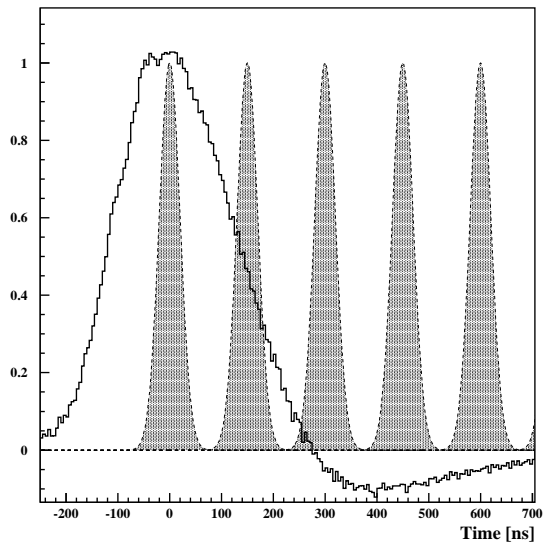}
    \caption{The trace is a sample kicker current pulse
    from one of the three kicker circuits. The periodic pulses
    provide a schematic representation of the unmodified muon
    bunch intensity during the first few turns. The vertical axis
    is in arbitrary units.
    \label{fg:kickwave}}
  \end{center}
\end{figure}

The kicker consists of three identical sections, each driven by a
separate PFN; this division keeps the inductance of a single
assembly at a reasonable value. In each circuit, an initial voltage
of $\sim90$~kV on the capacitor results in a current-pulse amplitude
of approximately 4200~A. Figure~\ref{fg:kickwave} shows the pulse
from one of the networks superimposed on a schematic representation
of the time and width of the muon bunch as it passes the location of
a single kicker section.  While a square-wave current
pulse---bracketing the injected bunch and turning off completely
before the next revolution---would be ideal, the actual pulse
waveform acts both positively and negatively on the bunch during the
first five turns in the ring. The injection efficiency is estimated
to be $3 - 5$ percent. Even at this modest efficiency, direct muon
injection is a significant improvement compared to pion injection,
not only in storage efficiency, but also because of the reduction of
the hadronic flash.



\begin{figure}
  \begin{center}
    \includegraphics*[width=9cm]{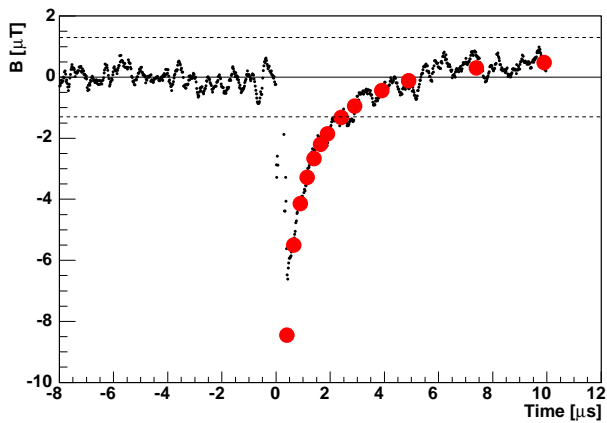}
    \caption{The magnetic field produced mainly by eddy currents on
             the kicker plates as measured by an optical polarimeter
             with the crystal at $(x,y) = (25~{\rm mm}, 0)$
             position.  The kicker is fired at 95~kV, producing
             4200~A at the kicker plates.  The band shows the
             range of $\pm 0.1$ ppm on integrated magnetic field.
             The large filled circles correspond to simulations with {\tt
             OPERA}, which used the measured current pulse as the
             input.  \label{fg:bfielde}}
  \end{center}
\end{figure}

The magnetic field produced by a prototype kicker was
measured~\cite{kicker} using a magnetometer based on the Faraday
effect. The main magnetic field, and the transient field following
the kicker pulse, were measured to a few percent. Excellent
agreement was obtained between {\tt OPERA} and measurement for the
field at the peak of the current pulse, and for the residual
magnetic field, see Fig.~\ref{fg:bfielde}.  The residual magnetic
field 30~$\mu$s after the main pulse contributes less than 0.1~ppm
to the integrated magnetic field seen by the muon beam.

\subsection{Field measurement instrumentation \label{ssec:NMR}}

\begin{figure}
\centering
\subfigure[~Absolute calibration probe]
{\label{fig:probes:a}\includegraphics*[width=4in]{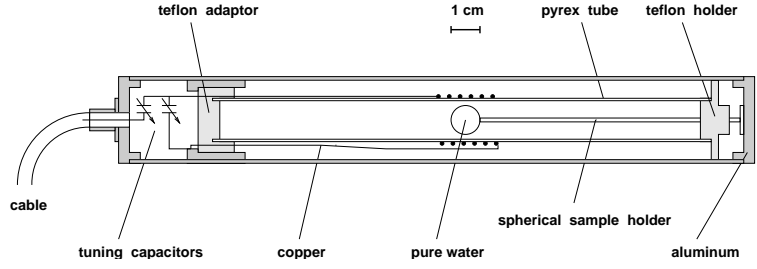}} \hspace{0.5in}
\subfigure[~Plunging probe]
{\label{fig:probes:b}\includegraphics*[width=4in]{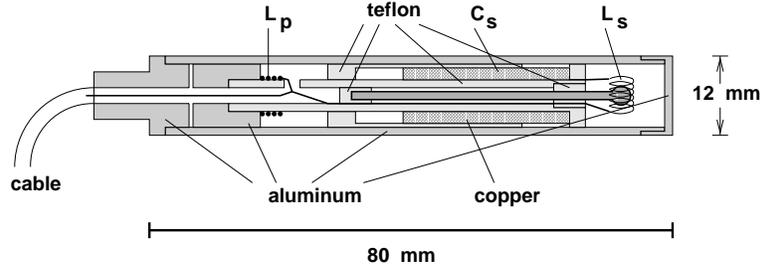}}
\hspace{0.5in}
\subfigure[~Trolley and fixed probe]
{\label{fig:probes:c}\includegraphics*[width=4in]{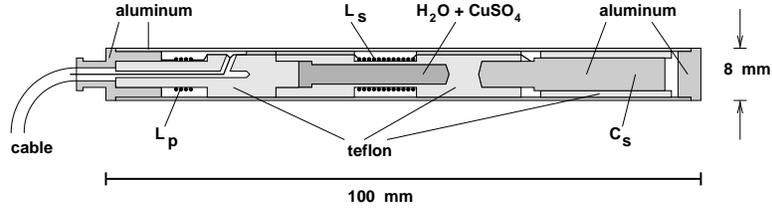}}

\caption{Schematics of different NMR probes. (a) Absolute probe
featuring a 10~mm diameter spherical sample of water.  This probe
was the same one used in Ref.~\cite{Liu:1999} to determine the
muon-to-proton magnetic moment ratio. (b) Plunging probe, which can
be inserted into the vacuum at a specially shimmed region of the
storage ring to transfer the calibration to the trolley probes. A
spherical water sample is enclosed inside the coil $L_s$. (c) The
standard probes used in the trolley and as fixed probes. The
resonant circuit is formed by the two coils with inductances $L_s$
and $L_p$ and a capacitance $C_s$ made by the Al housing and a metal
electrode. A cylindrical water plus CuSO$_4$ sample of approximately
10~mm length is used.} \label{fig:probes}
\end{figure}
\begin{figure}
\begin{center}
\subfigure[~NMR Trolley]{\includegraphics[width=2in]{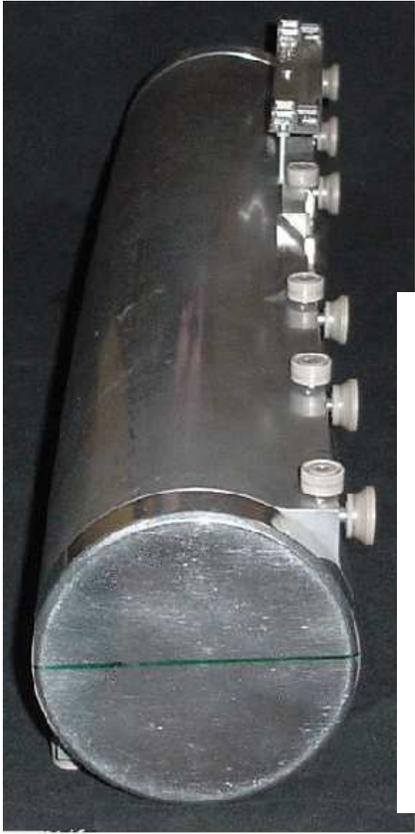}}
\subfigure[~Distribution of NMR probes over a cross section of the
trolley]{\includegraphics[width=4in]{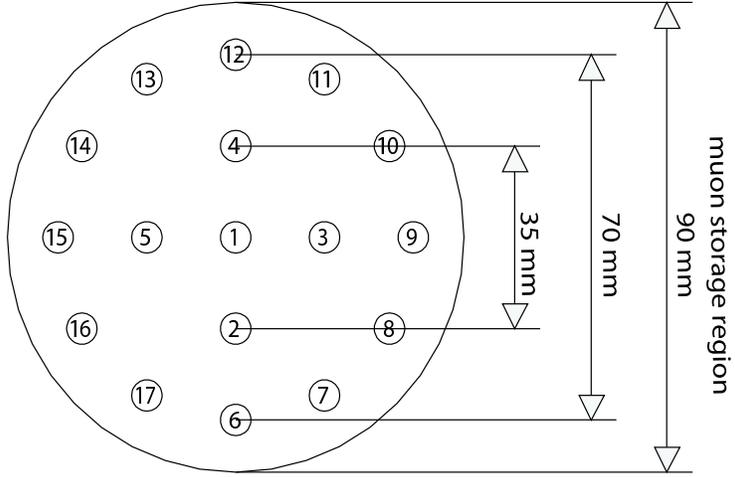}}
\end{center}
\caption{Photograph of the NMR trolley, which measures the magnetic
field in the storage ring.  The array of 17 NMR probes, which are
located inside the trolley housing, 82(1)~mm behind the front of the
trolley. Electronics occupies the back part of the device. At the
location of the probes, the field perturbation by these materials is
less than 2~ppm and is accounted for by the calibration method. The
probe numbers and placement are given by the schematic.}
\label{fig:trolleypos}
\end{figure}

Precision measurements of the magnetic field are made using the free-induction decay (FID) of nuclear
magnetic resonance (NMR) of protons in water~\cite{Prigl:1996}. The various NMR probes are calibrated
with respect to an absolute calibration probe, which is a spherical sample of water (see
Fig.~\ref{fig:probes:a}).  The functionality of the NMR measurement system and the reliability of the
absolute calibration probe were established in a wide bore superconducting magnet. The very same
equipment was used to calibrate the field in a muonium microwave experiment at Los Alamos National
Laboratory~\cite{Liu:1999}, where the Zeeman effect in the ground state was measured to obtain the
muon magnetic moment. The suite of NMR probes used in E821 includes:
\begin{itemize}

\item A calibration probe with a spherical water
sample (Fig.~\ref{fig:probes:a}),
 that provides an absolute
calibration between the NMR frequency for a proton in a water sample
to that of a free proton~\cite{phillips}.  This calibration probe is
employed at a  plunging station located at a region inside the
storage ring, where special emphasis was put on achieving high
homogeneity of the field.

\item A plunging probe (Fig.~\ref{fig:probes:b}),
which can be inserted into the vacuum in the plunging
station at positions matching those
of the trolley probe array. The plunging probe is used to transfer
calibration from the absolute calibration probe to the trolley
probes.

\item A set of 378 fixed probes placed above and below the storage
ring volume in the walls of the vacuum chamber. These probes have
a cylindrical water sample oriented with its axis along the
tangential direction (Fig.~\ref{fig:probes:c}).  They continuously
monitor the field during data taking.

\item Seventeen probes mounted inside a trolley that can be
pulled through the storage ring to measure the field
(Fig.~\ref{fig:trolleypos}).  The probes on board the trolley are
identical in design and shape to the fixed probes
(Fig.~\ref{fig:probes:c}).

\end{itemize}

Initially the trolley and fixed probes contained a water sample.
Over the course of the experiment, between run periods, the water
samples in many of the probes were replaced with petroleum jelly.
The jelly has several advantages over water: low evaporation,
favorable relaxation times at room temperature, a proton NMR signal
almost comparable to that from water, and a chemical shift (and thus
the NMR frequency) having a negligible temperature coefficient.

The free-induction decay signals are obtained after pulsed
excitation, using narrow-band duplexing, multiplexing, and filtering
electronics~\cite{Prigl:1996}. The signals from all probes are mixed
with a standard frequency $f_{ref} = 61.74$~MHz corresponding to a
reference magnetic field $B_{ref}$.  The reference frequency is
obtained from a synthesizer, which is phase-locked to the base clock
of the LORAN~C broadcast frequency standard~\cite{loran}, accurate
to $10^{-11}$. In a typical probe, the nuclear spins of the water
sample are excited by an rf pulse of 5~W and 10~$\mu$s length
applied to the resonance circuit.  The coil $L_s$ and the
capacitance $C_s$ form a resonant circuit at the NMR frequency with
a quality factor of typically 100.  The coil $L_p$ serves to match
the impedances of the probe assembly and the cable. The rf pulse
produces a linearly polarized rf field $\vec{H}$ in coil $L_s$,
orthogonal to the dipole field. The rf pulse rotates the macroscopic
magnetization in the probe by $90^{\circ}$.  The NMR signal from the
precessing magnetization at the frequency $f_{NMR}$ is picked up by
the coil $L_s$ of the same resonance circuit and transmitted back
through a duplexer to the input of a low-noise preamplifier. It is
then mixed with $f_{ref}$ to obtain the intermediate frequency
$f_{{\rm FID}}$.  We set $f_{ref}$ smaller than $f_{NMR}$ for all
probes so that $f_{{\rm FID}}$ is approximately 50~kHz.  The typical
FID signal decays exponentially and the time between the first and
last zero crossing---the latter defined by when the amplitude has
decayed to about $1/3$ of its initial value~\cite{Prigl:1996}---is
of order 1~ms. The interval is measured with a resolution of 50~ns
and the number of crossings in this interval is counted. The ratio
gives the frequency $f_{{\rm FID}}$ for a single measurement, which
can be converted to the magnetic field $B_{real}$ at the location of
the probe's active volume through the relation
\begin{equation}
B_{real} = B_{ref} \left(1+\frac{B_{real}-B_{ref}}{B_{ref}}\right) = B_{ref} \left(1 + \frac{f_{{\rm
FID}}}{f_{ref}}\right).
\end{equation}
The analysis procedure, which is used to determine the average
magnetic field from the raw NMR data, is discussed in
Section~\ref{sec:ana_field}.

\subsection{Detector systems, electronics and data acquisition
\label{ssec:det}}
\subsubsection{Electromagnetic calorimeters}
Twenty-four electromagnetic calorimeters are placed symmetrically
around the inside of the storage ring, adjacent to the vacuum
chamber, which has a scalloped shape to permit decay electrons to
exit the vacuum through a flat face cutout upstream of each
calorimeter (see Fig.~\ref{fig:ringoverhead}). The calorimeters are
used to measure the decay electron energy and time of arrival. They
are constrained in height by the magnet yoke gap.  The width and
depth were chosen to optimize the acceptance for high-energy
electrons, and minimize the low-energy electron acceptance. Each
calorimeter is 140~mm high by 230~mm wide and has a depth of 13
radiation lengths (150~mm).  The 24 calorimeters intercept
approximately 65 percent of the electrons having energy greater than
1.8~GeV. The acceptance falls with decreasing electron energy.

\begin{figure}
\begin{centering}
 \includegraphics*[angle=90,width=0.75\textwidth]{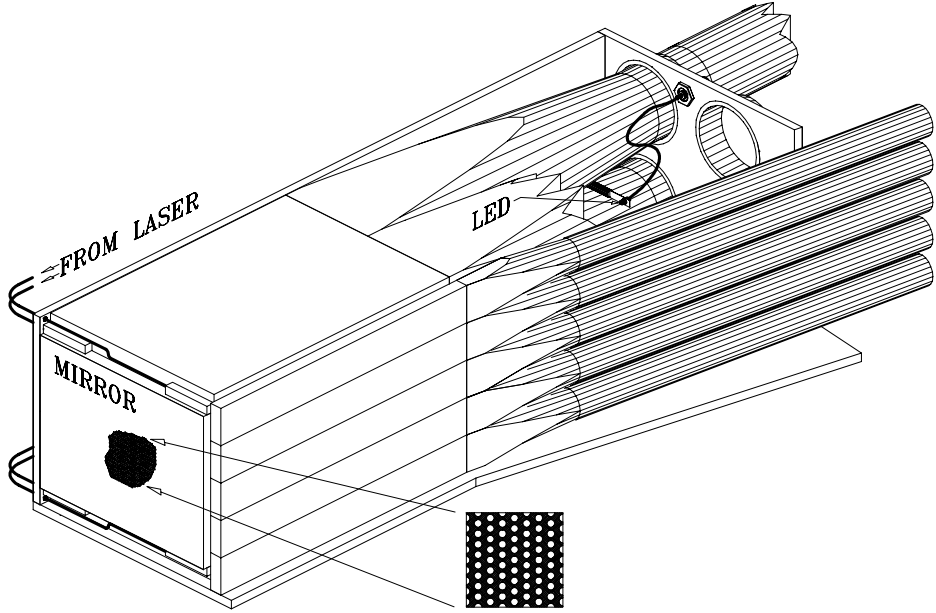}
\caption{Schematic of a Pb/SciFi calorimeter. The inset shows the close-pack
fiber-lead grid. The detector is subdivided into four quadrants, each viewed by
an independent PMT (not shown).  A pulsed nitrogen laser provides a time
calibration pulse into each of the four quadrants. The front scintillator
hodoscope, with its five-fold segmentation, is visible on the entrance face of
the calorimeter. \label{fig:caloblowup}}
 \end{centering}
\end{figure}

Because of the high rate (few MHz) at early times following
injection,fast readout and excellent pulse separation (in time) are
necessary characteristics of the design.  They are achieved by using
a plastic-scintillator-based sampling calorimeter read out by
photomultiplier tubes (PMTs). The calorimeter (Pb/SciFi) volume is
made of $52\%$ lead alloy, $38\%$ scintillating fiber, and $10\%$
epoxy.  The detector provides good light yield and, in the limited
space, adequate shower containment. The 1~mm scintillating fibers
are epoxied into grooved metal plates in a nearly close-packed
geometry. They are oriented radially, terminating on four
lightguides that pipe the light to Hamamatsu R1828 2-inch PMTs (see
Fig.~\ref{fig:caloblowup}).  The individual PMT gains and times were
carefully balanced because the four analog signals are added prior
to sampling by a waveform digitizer. The system is described in
detail in Ref.~\cite{Sedykh:2000}.

Prior to use in the experiment, each calorimeter was characterized
and calibrated at the AGS test beam.  The response deviates from a
linear form by less than 1~percent up to 3~GeV, falling short by
2.5~percent at 4~GeV. The detector fractional energy resolution is
approximately $7.0\%$ at 1.9~GeV and scales as $1/\sqrt{E}$. While
the initial absolute calibration and quadrant balancing were
performed at the test beam, final PMT balancing was established and
gain stability versus time was maintained by using the electron
energy spectra acquired during the data collection periods.  The
PMTs employed transistorized bases to maintain gain stability over
the very wide dynamic range of rates and background levels
encountered during the data acquisition period~\cite{Ouyang:1996}.

During normal data collection, an intense burst of particles strikes
the magnet pole pieces and calorimeters at the instant of injection.
The detectors immediately downstream of the inflector exit are
particularly affected. Because this ``prompt flash'' would
completely saturate the PMTs, they are gated off prior to injection
and turned back on some tens of microseconds later, depending on the
location of the calorimeter with respect to the inflector exit. The
output of a PMT was suppressed by a factor of $10^6$ by exchanging
the bias on dynodes 4 and 7.  When the dynodes are reset, the gain
returned to better than 99~percent of its steady-state value in
approximately $1~\mu$s.

The prompt flash creates an enormous number of neutrons, many of
which are thermalized and captured inside the Pb/SciFi calorimeters.
Capture gamma rays can knock out electrons, which traverse the
scintillating fibers.  The generated light gives an elevated
background pedestal that diminishes with the time dependence:
$\approx t^{-1.3}$ for $t>5~\mu$s.  The effective decay time was
minimized by doping detectors in the first half of the ring---where
the capture rate is highest---with a natural boron carbide powder.
The $^{10}$B component (20 percent) has a high thermal-neutron
capture cross section.

To monitor the detectors, a 300 ps uv ($\lambda = 337$~nm) pulse
from a nitrogen laser is directed through a splitter system into an
outside radial corner of each of the quadrants of all calorimeters.
The uv pulse is absorbed by a sample of scintillating fibers.  These
fibers promptly emit a light pulse having the same fluorescence
spectrum as produced by a passing charged particle. The calibration
pulse propagates through the entire optical and electronic readout
system. A reference photodiode and PMT---located well outside the
storage ring---also receive a fixed fraction of the light from the
laser pulse. The laser was fired every other fill, in parallel with
the beam data, during 20-minute runs scheduled once per 8-hour
shift. Firings alternated among four points in time with respect to
beam injection. These points were changed for each laser run,
providing a map of any possible gain or timing changes. Timing
shifts were found to be limited to less than 4~ps from early-to-late
times, corresponding to an upper limit of 0.02~ppm systematic
uncertainty in \wa. No overall trends were seen in the laser-based
gain change data. Observed gain changes were generally a few tenths
of a percent, without a preferred sign. Unfortunately, the scatter
in the these calibration measurements is much greater than is
allowed by statistics, and the mechanism responsible has not been
identified. Consequently, the laser-based gain stability (see
Section~\ref{sssec:gain}) could be established to no better than a
few tenths of a percent. Ultimately, monitoring the endpoints of the
electron energy spectra provided a better measurement of gain
stability.

\subsubsection{Special detector systems}\label{sssec:specdet}
Hodoscopes consisting of five horizontal bars of scintillator---the
front scintillating detectors (FSDs)---are attached to the front
face of the calorimeters (Fig.~\ref{fig:caloblowup}). The FSDs are
used to measure the rate of ``lost muons" from the storage ring and
to provide a vertical profile of electrons on the front face of the
calorimeters. The FSD signals are also used as a check on the pulse
times reconstructed from the calorimeter waveforms. The individual
scintillators are 235~mm long (radial), 28~mm high and 10~mm thick.
They are coupled adiabatically to 28-mm diameter Hamamatsu R6427
PMTs, located below the storage ring magnet midplane.  The PMT bases
are gated off at injection, following a scheme similar to that used
in the calorimeter PMT bases.  (Eventually, over the run periods,
about half of the FSD stations were instrumented with PMTs.) An FSD
signal is recognized by a leading-edge discriminator and is recorded
by a multi-hit time-to-digital converter (MTDC).

An $xy$ hodoscope with 7~mm segmentation is mounted on the front face of five calorimeters. These
position-sensitive detectors (PSDs) have 20 horizontal and 32 vertical scintillator sticks read out by
wavelength-shifting fibers and a Philips multi-anode phototube (later replaced by a multi-channel DEP
hybrid photodiode).  The MTDC recorded event time and custom electronics coded the $xy$ profile,
providing information on albedo and multiplicity versus time. In some cases, a calorimeter was
equipped with both PSD and FSD, providing efficiency checks of each. The vertical profiles of both FSD
and PSD provide sensitivity to the presence of an electric dipole moment, which would tilt the
precession plane of the muon spin.

Depending on the average AGS intensity for each running period,
either a thin scintillator or \v{C}erenkov counter is located in the
beamline, just outside the muon storage ring. This ``$T_0$'' counter
records the arrival time and intensity (time) profile of a muon
bunch from the AGS. Injected pulses not exceeding a set integrated
current in $T_0$ are rejected in the offline analysis because they
do not provide a good reference start time for the fill and because
they are generally associated with bad AGS extraction.

Scintillating-fiber beam-monitors (FBM), which are rotated into the
storage region under vacuum, were used to observe directly the beam
motion in the storage ring during special systematic study runs.
Separate FBMs measure the horizontal and vertical beam
distributions. Each FBM is composed of seven fibers centered within
the muon storage region to $\pm0.5$~mm. The 90-mm long fibers are
separated by 13~mm. One end is mated to a clear fiber, which carries
the light out of the vacuum chamber to PMTs mounted on top of the
storage ring magnet. The PMTs are sampled continuously for about
$10~\mu$s using 200~MHz waveform digitizers. The beam lifetime with
the fiber beam monitors in the storage region is about one half the
$64~\mu$s muon decay lifetime.

The scalloped vacuum chamber is truncated at one location and a
360~$\mu$m-thick mylar window (72~mm wide by 110~mm high) replaces
the aluminum flat exit face to allow electrons to pass through a
minimum of scattering material. A set of four rectangular drift
chambers---the traceback system~\cite{traceback}---is positioned
between this window and calorimeter station $\#20$. Each chamber
consists of three layers of 8-mm diameter straw drift tubes oriented
vertically and radially. An electron that exits the window and
passes through the four chambers is tracked by up to 12 vertical and
12 radial measurements.  Although decay electrons are not always
emitted tangentially, their radial momenta are sufficiently small so
that an extrapolation of the track back to the point of tangency
with the central muon orbit yields a measurement of the actual muon
decay position, as well as its vertical decay angle. The former is
helpful in establishing the average positions and therefore the
average field felt by the muons.

\subsubsection{Waveform digitizers and special electronics}
Waveform digitizers (WFDs) were used to collect raw data from the
PMT signals from the calorimeter and fiber monitors, and to monitor
voltages in the electrostatic quadrupoles and fast kicker as a
function of time.   Each WFD consists of a motherboard and an
application-specific daughtercard.  The motherboard features four
independent 8-bit digitizers with 64~kbytes of memory per channel.
For the calorimeter, signals from the four phototubes are summed on
the daughtercard and the resulting signal is directed into two
adjacent channels on the motherbaord, which sample the waveform on
alternate phases of a 200~MHz clock, thus effectively providing a
400~MHz sampling rate. Data from each clock phase is written into
separate memory buffers every 20~ns. Zero-suppressed operation, if
invoked, is provided by comparators located on the daughtercard. For
the calorimeter pulses, roughly 16~ns of the baseline are recorded
prior to the time of the trigger and about 64~ns worth of samples
are recorded after the trigger. For the fiber beam monitors and the
quadrupole and kicker voltage readout, four different versions of
the daughtercards were used. Sampling rates varied: 200~MHz for the
fiber monitors and kicker, 2~MHz for the quad voltage monitor. These
WFDs were operated without zero suppression.

\begin{figure}
  \begin{center}
  \includegraphics*[width=\textwidth]{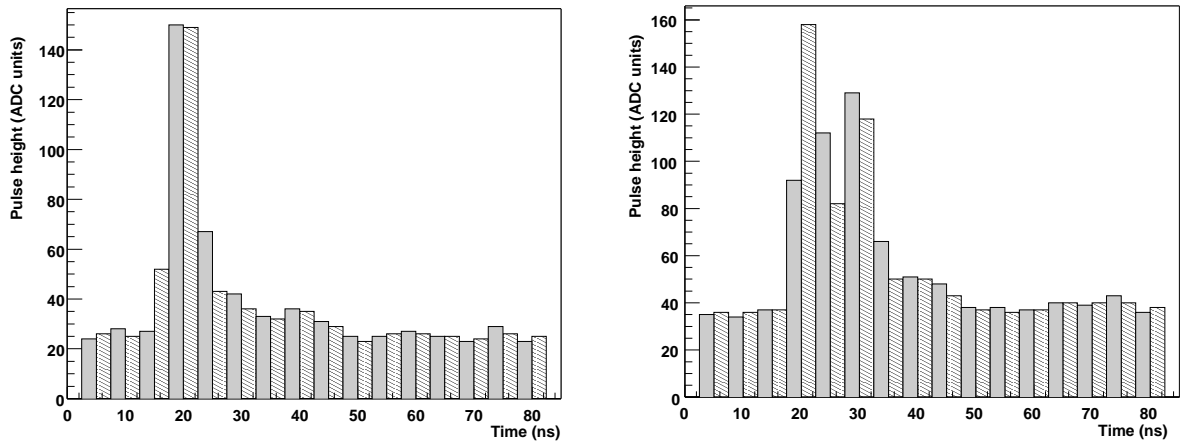}
  \caption{Examples of waveform digitizer samples from a calorimeter.
  The two WFD phases are alternately shaded.  The left panel shows a
  simple, single pulse, while the right panel has two overlapping pulses
  separated by 7.5~ns. \label{fig:samples}}
  \end{center}
\end{figure}

The time words written to memory in each of the two phases have a fixed but arbitrary offset in any
data-collection cycle. To resolve the ambiguity, a 150~ns triangular pulse is directed into a fifth
analog input on each daughtercard immediately prior to each fill. By reconstructing this ``marker
pulse,'' the unknown offset is determined unambiguously, and the data streams can then be combined in
the offline reconstruction program. Typical calorimeter pulses, with samples from the two phases
interleaved, are shown in Fig.~\ref{fig:samples}.

The clock signals for the WFD, MTDC and NMR are all derived from the
same frequency synthesizer, which is synchronized to the Loran~C
time standard~\cite{loran}. There is no correlation between the
experiment clock and the AGS clock that determines the time of
injection. This effectively randomizes the electronic sampling times
relative to the injection times at the $\sim5$~ns time scale,
greatly reducing possible systematic errors associated with odd-even
effects in the WFDs.

\section{Beam Dynamics \label{sec:beamdynamics}}
\subsection{Overview \label{ssec:beamoverview}}

\begin{figure}
\begin{center}
 \includegraphics*[width=0.8\textwidth]{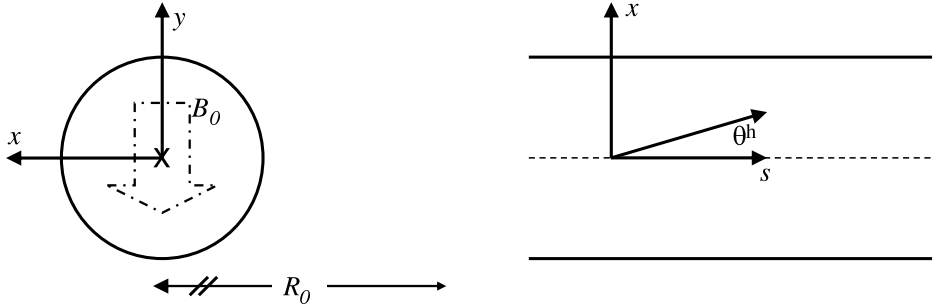}
 \caption{Coordinate system used to define the beam dynamics
 expressions.  An end view (left panel) with the negative
 muon beam into the page; the center of the storage ring is to the
 right, off the scale of this figure. The magnetic dipole field is oriented down.
 A top view (right panel) with the beam travelling in the $+s$ direction.
\label{fig:coordinate}}
\end{center}
\end{figure}

The muon storage ring uses the electrostatic quadrupoles to obtain
weak vertical focussing~\cite{wied,cp}. The field index is given by:
\begin{equation}
  n = \frac{R_0}{v B_0}\frac{\partial E_y}{\partial y},
  \label{eq:neff}
\end{equation}
where $R_0$ is the central orbit radius, $B_0$ is the dipole
magnetic field, and $v$ is the muon speed. The coordinate system is
shown in Fig.~\ref{fig:coordinate}. High voltage of $\pm 24$~kV
applied to the quadrupole plates gives an $n$ value of 0.137, which
is the field index that was used in the R97-99 periods. The R00
period used $n = 0.135$ and the R01 running was split into ``low-''
and ``high-$n$'' sub-periods with $n = 0.122$ and 0.142,
respectively. The field index value determines the stored muon beam
betatron motion, which in turn can perturb the electron time
distribution, as discussed in Section~\ref{ssec:CBO}.

For an ideal weak-focusing ring, where the quadrupole field is
constant in time and uniform in azimuth, the horizontal and vertical
tunes are given by $\nu_x = \sqrt {(1-n)}$ and $\nu_y = \sqrt {n}.$
Consider the motion of a particular muon having momentum $p$
compared to the magic momentum $p_0$. Its horizontal and vertical
betatron oscillations are described by
\begin{eqnarray}
x & = & x_e + A_x \cos(\nu_x \phi+\phi_{0x}) \\
y & = & A_y \cos(\nu_y \phi+\phi_{0y}).
\end{eqnarray}
Here $\phi = s/R_0$, where $s$ is the azimuthal distance around the ring and $A_x$ and $A_y$ are
amplitudes of the oscillations about the equilibrium orbit ($x_e$) and the horizontal midplane,
respectively, with $x_e$ given by
\begin{equation}
 x_e = R_0 \left(\frac{p - p_0}{p_0(1-n)}\right). \label{eq:equilorbit}
\end{equation}
The maximum accepted horizontal and vertical angles are defined by
the 45~mm radius of the storage volume, $r_{max}$, giving
\begin{eqnarray}
\theta ^{h}_{max} & = & \frac{r_{max} \sqrt{1-n}}{R_0}\\
\theta ^{v}_{max} & = & \frac{r_{max} \sqrt{n}}{R_0}.
\end{eqnarray}

\begin{figure}
\begin{centering}
 \includegraphics*[angle=0,width=0.7\textwidth]{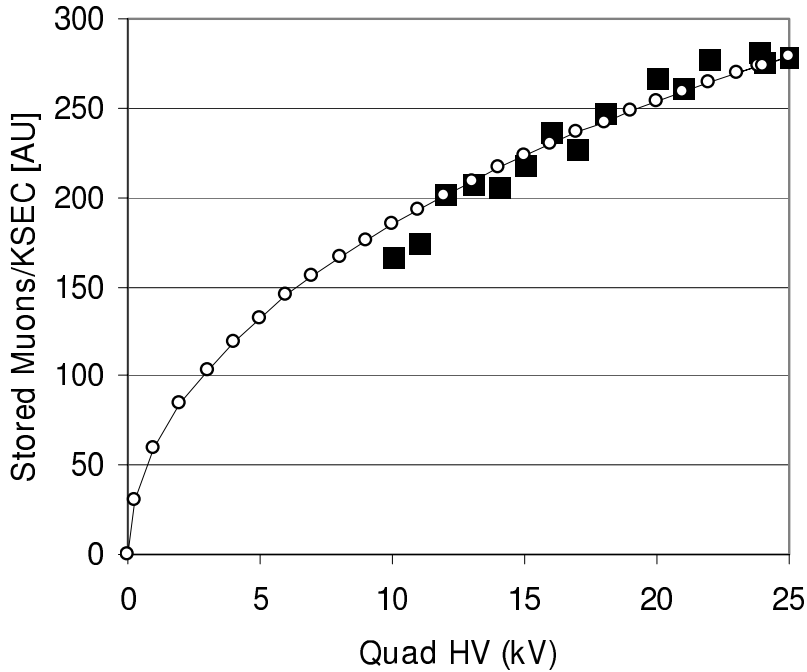}
 \caption{The normalized number of stored muons (squares) and
 the expected dependence (circles) for short runs varying the quadrupole high
 voltage. Here, KSEC is a relative measure of the proton intensity at the target
 as viewed by a secondary emission counter. \label{fig:quad_hv}}
 \end{centering}
\end{figure}

\begin{figure}
\begin{center}
    \includegraphics*[width=0.7\textwidth]{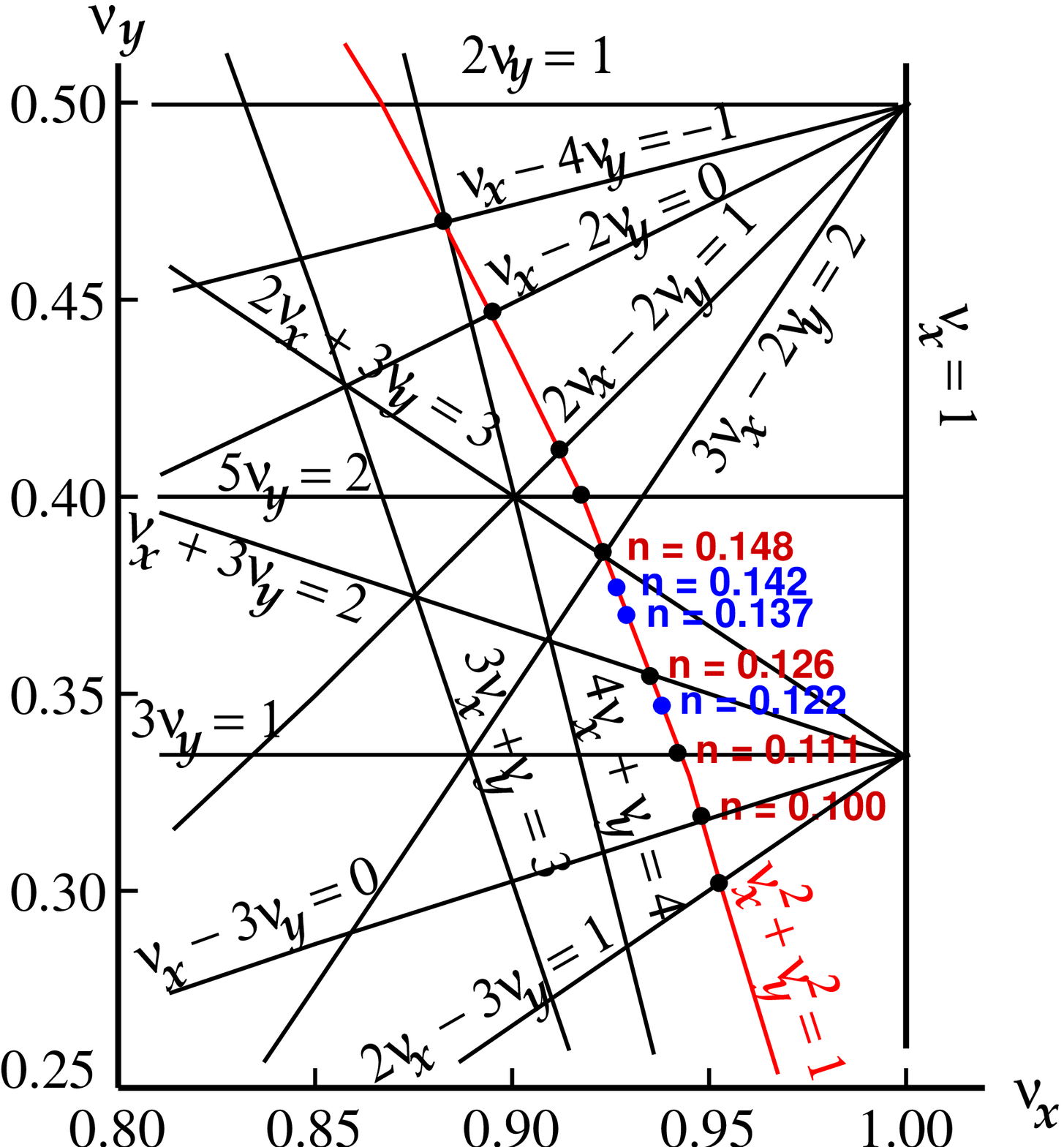}
 \caption{The tune plane showing resonance lines.
 Three of the $n$~values used to run the experiment, 0.122, 0.137, 0.142, are
 indicated on the arc of the circle defined as $v_{x}^2 + v_{y}^2 = 1$.
 They do not intersect any of the resonance lines, contrary to nearby tunes,
 which are also shown on the arc.
\label{fg:resonances}}
\end{center}
\end{figure}

In general, the muon storage fraction increases for higher
quadrupole voltage. The number of stored muons is plotted versus
high voltage in Fig.~\ref{fig:quad_hv}. As the high voltage is
increased, the vertical phase space increases, while the horizontal
phase space decreases. The operating quadrupole high voltage is
chosen to avoid the beam-dynamics resonances, which take the form $L
\nu_x + M \nu_y = N$, where $L$, $M$, and $N$ are integers. The
resonance lines and the storage ring working line given by $\nu^2_x
+ \nu^2_y = 1$ are shown in Fig.~\ref{fg:resonances}. Typical
working values are $\nu_x = 0.93$ and $\nu_y = 0.37$ for $n=0.137$,
giving $\theta^{h}_{max} = 5.9$~mrad and $\theta^{v}_{max} =
2.3$~mrad.

Numerical calculations, which include the finite extent of the
quadrupole electrodes, were used to select the exact high-voltage
values. Confirmation of resonance-induced muon losses was made
during special runs tuned to the resonances $\nu_x+3\nu_y=2$ and
$2\nu_x+3\nu_y=3$ ($n = 0.126$ and $n~=~0.148$).


\subsection{Fast rotation \label{ssec:fastrot}}
\begin{figure}
\begin{center}
\includegraphics*[width=0.5\textwidth]{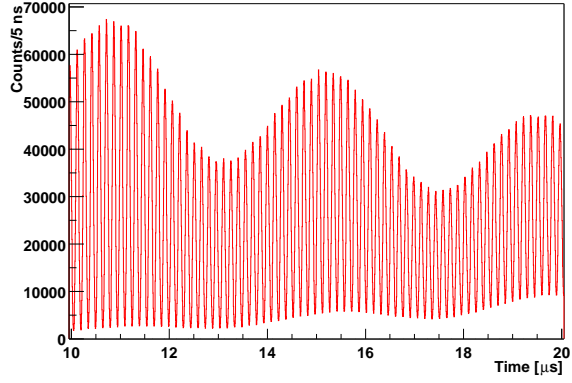}
 \caption{Intensity at a single detector station shortly after injection. The rapid
 modulation repeats at the cyclotron
 frequency as the muon bunch circles the ring.  The
 width of the bunch grows with time because of the finite $\delta p/p$
 of the stored muons.  The slow variation in the maximum amplitude
 is at the \gm\ frequency.
 \label{fig:fastrotation}}
\end{center}
\end{figure}

The stored muon momentum distribution is determined by analyzing the
debunching of the beam shortly after injection. This {\it fast
rotation} analysis is based on a few simple ideas. High-momentum
muons trace a larger average radius of curvature than low-momentum
muons. However, because all muons travel at the same speed
(0.9994$c$, constant to a part in $10^5$ over the aperture), the
higher (lower)-momentum muons have a smaller (larger) angular
frequency. A finite range in angular frequencies causes the initial
bunched beam to spread azimuthally over time.
Figure~\ref{fig:fastrotation} shows the decay electron rate at one
detector station from 10 to 20~$\mu$s following injection. The rapid
rate fluctuation reflects the 149~ns cyclotron period of the bunched
beam. The slower modulation is caused by the \g2\ spin precession.
The rapid bunch structure disappears over time, as the beam
uniformly fills the ring in azimuth.
\par
The initial bunched beam is modelled as an ensemble of particles
having an unknown frequency distribution and a narrow time spread
(rms $\sim25$~ns, occupying $\sim60$ degrees of the ring). The model
assumes that every time slice of the beam has the same frequency
profile but the time width is left as a fit parameter, as is the
exact injection time. The distribution of angular frequencies will
cause the bunched beam to spread out around the ring over time, in a
manner that depends uniquely on the momentum distribution. In
particular, the time evolution of any finite frequency slice is
readily specified. A given narrow bin of frequencies contributes
linearly to the time spectrum. The total time spectrum is a sum over
many of these frequency components, with amplitudes that can be
determined using $\chi^2$ minimization. The momentum distribution is
then determined from the frequency distribution (or equivalently,
from the radial distribution) by
\begin{equation}
\frac{p - p_0}{p_0} = (1-n)\left( \frac{R - R_0}{R_0} \right).
\end{equation}

\begin{figure}
\begin{center}
\includegraphics*[angle=-90,width=0.75\textwidth]{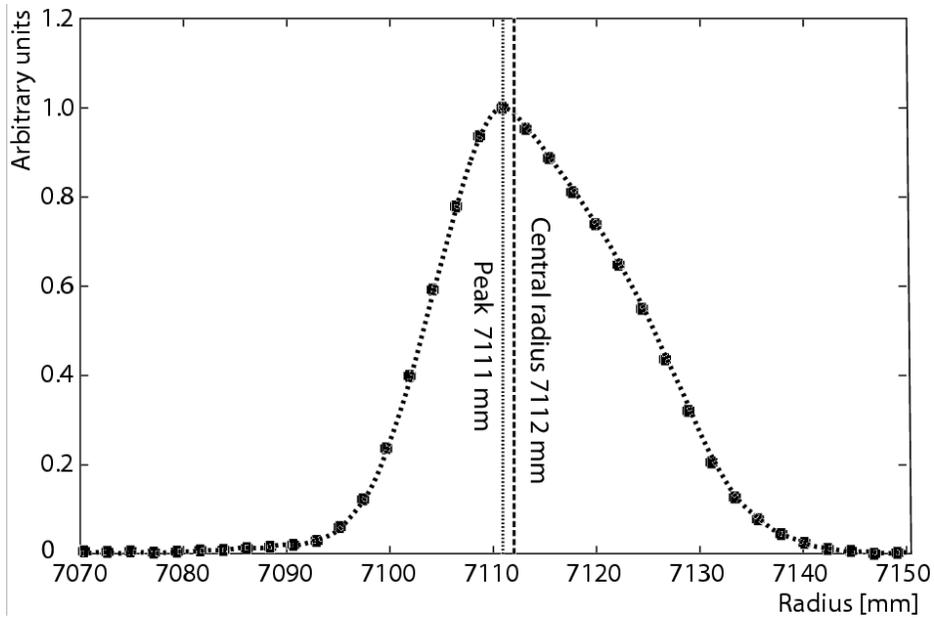}
 \caption{The distribution of equilibrium radii $dN/dx_e$, as determined from the
 fast-rotation analysis.  The dashed vertical line is
 at 7112~mm, the magic radius;  the dotted line is at 7111~mm.
 The solid circles are from a de-bunching model fit to
the data, and the dashed curve is obtained from a modified Fourier
analysis.
 \label{fig:radialdistribution}}
\end{center}
\end{figure}

The result of the fast-rotation analysis from the R00 period is
shown in a plot of the beam radius-of-curvature distribution shown
in Fig.~\ref{fig:radialdistribution}. The smooth curve is obtained
from a modified Fourier transform analysis. The peak of the
distribution lies below the nominal magic radius of 7112~mm but the
mean is somewhat larger, $7116 \pm 1$~mm for R00 and $7115 \pm 1$~mm
for R01.
The rms width is about 10~mm.

\subsection{Coherent betatron oscillations \label{ssec:CBO}}

\begin{figure}
\includegraphics*[width=\textwidth]{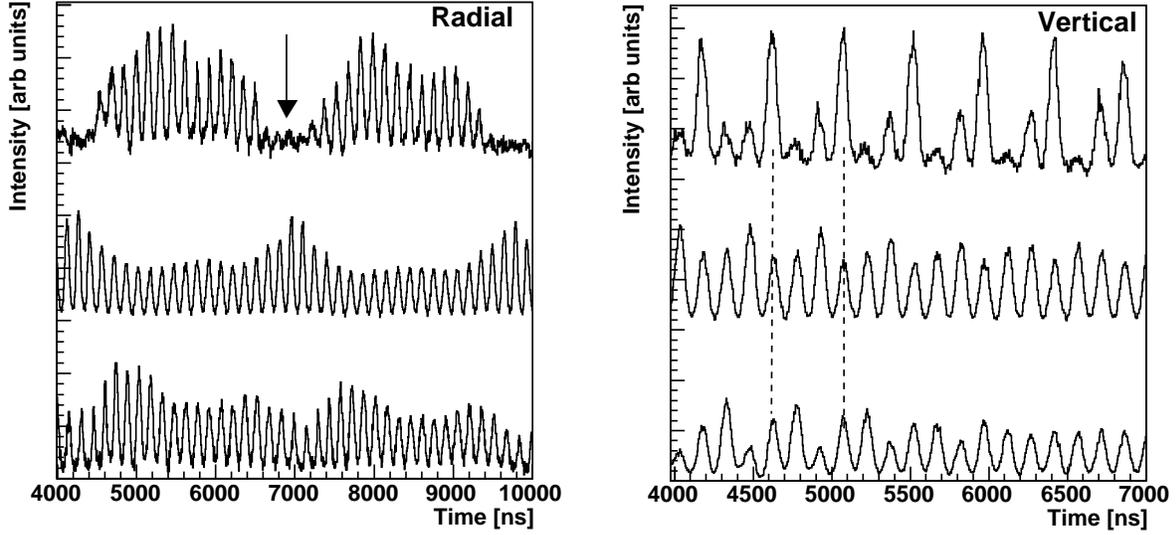}
\caption{Digitized waveform from the FBM located $180^{\circ}$ from the
inflector. The fast rotation period is observed in both the vertical and the
horizontal profiles. Left panel: Radial profile from an inner, middle, and
outer fiber trace (stacked top to bottom with an offset).  The width minimum is
indicated by the arrow.  The outer fibers do not peak at the same time because
the radial mean is oscillating at the same frequency as the width.  Right
panel: The vertical profile from a top, middle and bottom fiber. The two dashed
lines are placed at approximately 3.3~turns, the period of the vertical
betatron oscillation.  A vertical waist is apparent halfway between the dashed
lines.} \label{fg:harp180}
\end{figure}

Special short runs were devoted to observing the muon beam motion directly using the scintillating
fiber beam monitors (FBM) described in Section~\ref{sssec:specdet}. Figure~\ref{fg:harp180} shows
pedestal-subtracted data from the FBM, which sits $\sim180$ degrees around the ring from the inflector
exit. Data from fibers 2, 4, and 6 (where 4 is the central fiber) are shown.  The left (right) panel
shows the fiber intensity representing the radial (vertical) profile of the beam versus time shortly
after injection.
The average size of the signal across the aperture versus time reveals the longitudinal motion of the
bunched beam at one point in the storage ring. The rapid oscillation (period $=$ 149~ns) in
intensity, which is approximately in phase for every fiber, corresponds to the cyclotron motion of the
beam.
\par
The {\it relative} size of the signals determines the  moments of the transverse distributions, which
oscillate in time. Oscillations in the width arise from a mismatch between the inflector and storage
ring apertures. The former is 18~mm and the diameter of the latter is 90~mm. A narrow ``waist'' is
therefore imposed on the beam at injection, which then alternately widens and narrows as the beam
circulates. Because the deflection provided by the kick is smaller than that required to center the
beam in the storage aperture, the centroid of the beam oscillates. When the beam is observed at one
location (e.g., at a particular detector station), the width and centroid oscillate at frequency
$$f_{CBO} \approx f_c - f_x = f_c (1 - \sqrt{1-n}) \approx {\rm 476~kHz,}$$
the coherent betatron oscillation (CBO) frequency. The corresponding period is 14.1~turns.

Prominent oscillations are evident in both the mean and width of the radial distributions measured
with the FBM. In the left panel of Fig.~\ref{fg:harp180}, the traces are stacked from top to bottom
corresponding to the intensity of the muons at low, central, and high radii within the storage
aperture. The vertical arrow illustrates a time when the bunch width is narrow in radial profile. It
reaches a maximum width approximately $1.1~\mu$s later. The breathing frequency corresponds to the
horizontal CBO, which for this R01 low-$n$ period data, is $\sim425$~kHz.  The fact that the inner and
outer radial fiber intensities peak somewhat out of phase indicates that the mean radial position
oscillates, again, at the same CBO frequency.

\par
The interference of multiple frequencies is also apparent in the vertical profiles shown---with a
different time axis range---in the right panel of Fig.~\ref{fg:harp180}.  The top-to-bottom beam
intensity is given by the three shown fiber traces, which are stacked in the same sense.  While the
dominant feature is the repetitive cyclotron frequency, a slower, undulating pattern can also be seen.
This pattern---with a period roughly indicated by the two vertical dashed lines---is related to the
vertical oscillations of the beam. Recall that the phase space simulation plots in
Fig.~\ref{fig:inflectorphasespace} show that the beam initially fills only the central 56~mm of the
storage aperture because it is limited by the height of the inflector channel. However, after one
quarter of a vertical betatron oscillation, it fills the full 90~mm. The minimum width at 56~mm is
called the vertical waist (VW) (see, for example halfway between the dashed lines in the figure, where
the outer intensities are small). The vertical width is modulated at frequency $(1 - 2\sqrt{n})f_c
\approx 2.04$~MHz (low-$n$ R01 data). The period is $\sim3.3$~turns, or $0.49~\mu$s.
\par
The frequencies associated with the various beam motions are
revealed in Fourier analyses of both the fiber monitor and
muon-decay electron time spectra. However, the lifetime of the
oscillations can only be determined by the latter measurements,
which do not affect the beam directly. The observed CBO lifetime is
about $100-140~\mu$s. The VW lifetime is much shorter, about
$25~\mu$s. These lifetimes are determined primarily by the tune
spread. For the CBO, $\Delta \nu_x = \Delta n/(2\sqrt {1-n})$. For
the VW, $\Delta \nu_y = \Delta n/(N \sqrt {n})$ where $N$ is the
harmonic of the CBO frequency. From the FBM data, the VW has CBO
frequency harmonics with relative amplitudes: 0.6, 0.28, 0.08 and
0.04 for $N=1, 2, 3$, and 4, respectively. The observed lifetimes
imply an effective field index spread of $\Delta n \simeq \pm
10^{-3}$, in agreement with computer simulations.  A summary of
important frequencies for the case $n = 0.137$ is given in
Table~\ref{tb:freq}.

\begin{table}
\caption{Important frequencies and periods in the $(g-2)$ storage
ring for $n~=~0.137$. \label{tb:freq}}
\begin{tabular}{lllll} \hline \hline
Physical frequency & Variable & Expression & Frequency  & Period \\
\hline

Anomalous precession & $f_a$ &  ${e \over 2 \pi m} a_{\mu}  B$ & 0.23 MHz & 4.37 $\mu$s  \\
Cyclotron & $f_c$ & ${v \over 2 \pi R_0}$ & 6.71 MHz & 149 ns \\
Horizontal betatron & $f_x$ & $\sqrt{1-n}f_c$ & 6.23 MHz & 160 ns \\
Vertical betatron & $f_y$ & $\sqrt{n}f_c$ & 2.48 MHz & 402 ns \\
Horizontal CBO & $f_{\rm CBO}$ & $f_c - f_x$ & 0.48 MHz & 2.10  $\mu$s \\
Vertical waist & $f_{\rm VW}$  & $f_c - 2f_y$ & 1.74 MHz & 0.57  $\mu$s \\
\hline \hline
\end{tabular}
\end{table}

\begin{figure}
\begin{center} \includegraphics*[width=\textwidth]{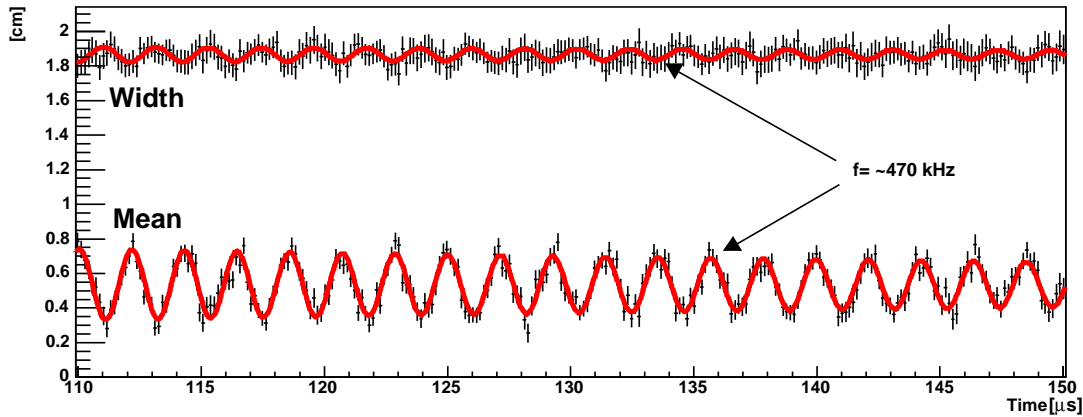}
\caption{The radial mean and rms width as determined by the traceback detector system in the R99
period.  Note that the width includes resolution smearing; the true width is $\approx 1.55$~ cm. Each
fit (solid lines) includes a damped exponential and an oscillatory term having the frequency of the
horizontal CBO effect. }
 \label{fg:CBO-Radial}
\end{center}
\end{figure}

The electron traceback system makes a more direct, but still non-destructive, measurement of the muon
phase space. Figure~\ref{fg:CBO-Radial} shows a representative distribution of the radial mean and
width of the muon population from 110 to 150~$\mu$s after injection during the R99 period. The
fundamental CBO frequency is the dominant feature of both plots. The amplitude is damped with an
exponential lifetime consistent with that found in other studies.

\subsection{Muon losses \label{ssec:muloss}}

Small perturbations in the magnetic or electric fields couple to the
horizontal and vertical betatron oscillations. If the forces arising
from these perturbations are periodic, the amplitude of the
oscillatory motion can grow without bound until the muon is lost.
The losses are minimized by choosing a field index that does not lie
on one of the resonance lines shown in Fig.~\ref{fg:resonances}, by
minimizing the electric and magnetic field perturbations, and by
``scraping'' the beam during the first $7 - 16~\mu$s after
injection.

Two perturbations of the field are of particular concern: a non-zero average radial component of the
magnetic field and the component of the main dipole field that varies as $\cos (s/R_0)$. The latter,
the first harmonic component, displaces the orbit horizontally.  It was measured with the NMR system
and shimmed at the beginning of each running period to be less than 30~ppm, well below the target
upper limit of 240~ppm. Monthly variations were kept under 20~ppm. A non-zero radial field displaces
the average orbit vertically. The radial magnetic field was adjusted using current shims so as to
maximize the number of stored muons. This centered the beam height to $\pm0.5$~mm with respect to the
collimators.

\begin{figure}
\begin{center}
\includegraphics*[width=0.5\textwidth]{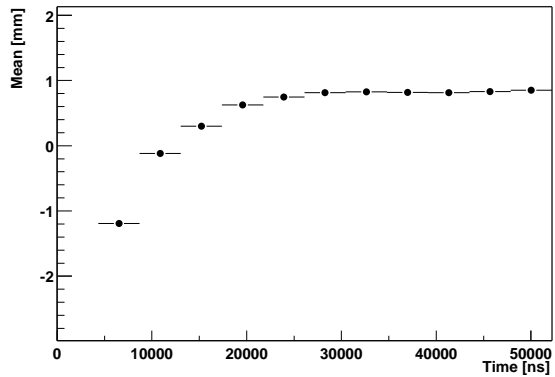}
 \caption{Mean vertical position of the beam as measured
 by FSD Station~22 during the
 R01 period.  The beam moves vertically with a time constant of $5~\mu$s
 after the end of scraping. \label{fg:muonVertical}}
\end{center}
\end{figure}

Beam scraping, which is discussed in detail in
Ref.~\cite{Semertzidis:2003}, refers to the application of
asymmetric voltages to the quadrupole plates shortly after
injection.  Scraping eliminates muons having trajectories close to
the collimators. The scraping procedure displaces the beam downward
by about $\sim$2~mm and re-centers the central orbit in the
horizontal plane, moving it $\sim$2~mm inward within two of the
quadrupole regions and $\sim$2~mm outward in the other two regions.
Muons at the extreme edges of the storage ring phase space then
strike a collimator (see below) and are removed from the ring. The
quadrupole plate voltages are restored to symmetric values (with a
5~$\mu$s time constant) after 7 to 15~$\mu$s (see
Table~\ref{tb:quadstuff}). The reduced effective aperture typically
results in a 10 percent loss of stored muons, in agreement with
analytical calculations and computer tracking simulations. The
effect of scraping on the average height of the muon distribution as
a function of time, as measured with the FSD counters, is shown in
Fig.~\ref{fg:muonVertical}. The mean beam position for Station~22 in
the R01 period starts a little over 1~mm low and rises as the
scraping voltages are turned off.

The collimators, which define the muon storage region, are made of
3-mm thick copper. They have an inner radius of 45~mm and an outer
radius of 55~mm.  Eight locations are available for the
collimators in the bellows adjoining the vacuum chambers (three
more are blocked by the kicker and one is unused due to proximity
to the inflector exit). The inner half circles of the four
collimators opposite the inflector are removed in order to avoid
scattering those low-momentum muons which, because of only a
two-thirds kick on the first turn, would otherwise strike the
collimators on the first turn and be lost.

\begin{figure}
\begin{center}
\includegraphics*[width=0.5\textwidth]{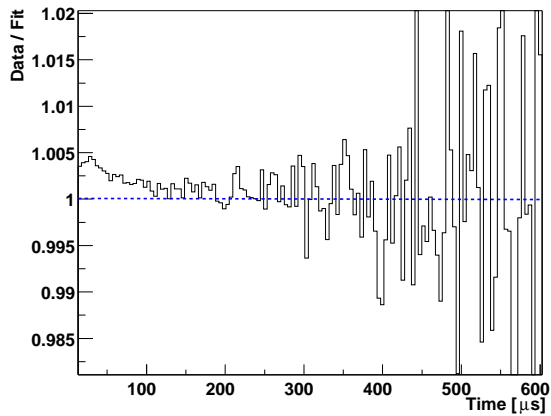}
 \caption{A fit to the simple function, $f=Ce^{-t/\gamma\tau}$ is made
 for the time period $300 - 600~\mu$s after injection for a subset
 of the pileup-subtracted R01 data.  The fit is extrapolated back
 to early times and the
 ratio data to fit is plotted for all times.  The excess of data compared to
 the fit at early times indicates the non-constant loss rate
 with time.  The bin width is the \gm\ period.
 \label{fg:mulossratio}}
\end{center}
\end{figure}

The muon decay fitting function $N(t)$ must account for muon losses.
A fit to the simple function $f=Ce^{-t/\gamma\tau}$, for data binned
in the \gm\ period and for times $t \geq 300~\mu$s, gives a
satisfactory description of the data. However, an extrapolation of
the fit back to early times reveals that the data lie above the fit
as shown in the ratio of data to fit in Fig.~\ref{fg:mulossratio},
demonstrating that the fractional loss of muons is greater early in
the measurement period.

The relative rate of muon losses can be measured directly using a coincidence of three FSD
scintillator hits in three successive detector stations. A 3.1~GeV muon loses approximately 100~MeV by
ionization in each calorimeter, well below the WFD hardware threshold. A muon loss event therefore
consists of a threefold FSD coincidence with no accompanying calorimeter signal. After
random-coincidence background subtractions, the time-dependent loss function is constructed, up to an
unknown efficiency factor, which is typically about 6 percent.

Beam protons were stored during the positive muon run periods R97-R00. Like muons, a fraction of these
protons exit the ring because of orbit perturbations. While proton losses have a negligible effect on
the decay positron distribution, they do form a significant background to the lost muon spectrum,
which must be removed. A proton loss event is defined as a muon loss candidate having large energy in
the third calorimeter, presumably from a hadronic shower. The proton loss signal was studied using
data acquired when the quadrupole voltages were dropped to zero, long after all muons had decayed in a
given fill. The uncertainty in the shape of the muon loss spectrum is dominated by uncertainty in the
proton-loss component---assuming, in addition, that the muon-loss monitor samples a constant fraction
of the actual ring losses versus time in the fill. The antiproton component for the R01 period with
negative muons was negligible.

\subsection{Electric-field and pitch correction
\label{ssec:pitch}}

The magnetic field is adjusted so that the magic momentum $p_0$ (see Eq.~\ref{eq:omega}) is at the
center of the aperture.  The \gm\ precession \wa\ of muons at this radius is not affected by the
electric field, which is, in any case, zero at the center of the quadrupoles.  However, muons of
higher or lower momentum are subject to a linearly rising field and deviate from the magic momentum,
so their \gm\ frequency is reduced.  We apply a correction for this effect.
As the muons oscillate about their equilibrium radii $x_e$, they experience a mean radial electric
field $\langle E_r\rangle = n (\beta B_0/R_0)x_e$. Differentiating Eq.~\ref{eq:omega} with respect to
$p$, and averaging over the muon ensemble gives
\begin{equation}
 \left\langle \frac{\delta\wa}{\wa}\right\rangle = -2\beta ^2 n (1 - n)
   \left\langle \left(\frac{x_e}{R_o}\right)^2 \right\rangle \label{eq:f4}
\end{equation}
where $\langle x_{e}^2\rangle$ is obtained from the fast rotation
analysis. While the radial distribution uncertainty is based on
data, it is checked against the simulation.  Both methods agree on
the mean-square width to a precision of 5 percent. This uncertainty
is added in quadrature with the uncertainty in the mean muon radial
position with respect to the electrostatic quadrupoles $\delta R =
\pm 0.5$~mm ($\pm $0.01~ppm in \amu), and with the uncertainty in
the mean vertical position of the beam $\pm $1~mm ($\pm $0.02~ppm in
\amu). As a quantitative example, the R01 electric-field correction
to the precession frequency for the low-$n$ subperiod is $(+0.47 \pm
$0.05)~ppm.

The pitch correction~\cite{farley:1972} accounts for the fact that
vertical betatron motion decreases slightly the magnetic field felt
by the stored muons in their rest frame. When the orbit is inclined
at angle $\psi$ to the horizontal, to sufficient accuracy $\omega_a$
is reduced by the factor $(1-{\frac{1}{2}}\psi^2)$. If $\psi_m$ is
the angular amplitude of the vertical oscillation, the average over
the ensemble of muons is $(1-{\frac{1}{4}}\langle \psi_m^2
\rangle)$. Here $\langle\psi_m^2\rangle = n \langle y^2\rangle/R_0^2
$, where $\langle y^2 \rangle$ is the mean-squared vertical spread
of the stored muons, which was measured by the traceback system,
confirming results from the tracking simulation.

The correction to the measured \wa\ due to vertical betatron
oscillations is calculated for each $n$ value setting. The
systematic error is estimated as the difference between the
correction calculated from the simulated and measured muon
distributions ($\pm 0.03$~ppm) added in quadrature with the
uncertainty in the mean muon radial position with respect to the
electrostatic quadrupoles and the uncertainty in the mean vertical
position of the beam.  For the R01 low-$n$ sub-period, the
correction is $(+0.27 \pm $0.04)~ppm.

A tracking program, simulating the muon spin precession in the
storage ring---including the correct azimuthal quadrupole electric
field, confirmed the validity of the analytic electric-field and
pitch corrections. The equations given in Ref.~\cite{Jackson} were
used for the orbit and the spin motion. The validity of the
electric-field and pitch-correction expressions are verified to
0.01~ppm in \amu.

\section{Data Analysis \label{sec:dataanalysis}}
The calculation of \amu\ requires the determination of the muon spin
precession frequency, \wa, and the average magnetic field seen by
the muons, \wpt. ``Precession" and ``field" data are recorded
separately and analyzed independently by different groups in the
collaboration.  To prevent bias, each analyzer adds a secret offset
to any frequency result presented publicly. When {\it all} the
analyses of a given type---precession or field---are deemed
internally consistent, the individual offsets are replaced by a
common offset, to facilitate comparisons between different analyses.
Only when all the analyses of each type are in agreement, and all
systematic investigations finished, are the final offsets removed.
Then \amu\ is calculated from the true \wa\ and \wpt\ results.

Two independent determinations of \wpt\ were made per running period. The NMR tools, described in
Section~\ref{ssec:NMR}, were used to provide the measurements as follows.  Every 3-4 days, the field
distribution in the storage volume is mapped by the NMR trolley. The time-dependence of the field is
measured using the fixed-probe NMR system, which samples and records the magnetic field around the
ring at approximately 0.1~Hz. Special calibration measurements are made before and after the
data-taking periods at a fixed location inside the ring vacuum, where the field uniformity is
especially good, to transfer the absolute calibration to the plunging probe and to the trolley probes.
The calibration of trolley probes against the plunging probe demonstrates the stability of the trolley
probe over the run period.

Four or five independent determinations of \wa\ were made per
running period.  The principal data are the waveform digitizer
records for each calorimeter station, from which decay electron
times and energies are extracted.  Events above an energy threshold
are included in electron time distributions, such as the one shown
in Fig.~\ref{fig:wiggles}.  The precession frequency is extracted
from a $\chi^2$ minimization fitting procedure, where the fit
function is based on Eq.~\ref{eq:fivepar}, plus small perturbations
from known physical effects.

The data-taking was organized in short ($\approx$~45~min) runs,
during which the macroscopic conditions remained stable. A database
was used to mark each run for special characteristics, such as laser
calibration, field index and scraping time, special radial
magnetic-field settings and other systematic tests. Approximately
1000 - 1300 good production runs remain per data-taking period after
complete commissioning of the apparatus and excluding special runs
for systematic studies. The vast majority of runs took place under
ideal operating conditions. However, whole runs were removed if they
failed a Kolmogorov-Smirnov test that was designed to compare each
run against a standard data set; this test was aimed specifically to
search for hardware problems that may have gone unnoticed. One
special virtue of the method is that it does not rely on any
information associated with the precession frequency. Runs were also
discarded if the power supplies for the surface correction coils
were not stable.  Before being included in the final data sample,
each fill within a run was checked against a set of quality
standards: injection bunch intensity and shape must not vary
substantially; no AGS extraction misfires; no quadrupole sparks or
missing quadrupole traces from the data stream; no kicker
misfirings; no detector electronics malfunctions; no memory
overflows in the onboard electronics; and, no missing marker pulses.

\subsection{Determination of \wp\ and \wpt\ \label{sec:ana_field}}
Two independent analyses of the magnetic field were made for each
of the data-taking periods. The methodology was similar for all
analyses.  Consistent results were obtained for all run periods.
The description below focuses on one of the analyses for the R00
data-taking period.

\subsubsection{Absolute calibration probe}
The determination of \amu\ requires a measurement of the free-proton
precession frequency, $\omega_p( = 2 \pi f_p)$, in the storage ring
magnetic field. However, the free induction decay (FID) signals
observed with the trolley NMR probes were not those from free
protons, but protons in water with a CuSO$_{4}$ additive that
shortened the relaxation time of the FID signal. In addition the
field is different from that in which the muons were precessing,
being perturbed by the magnetization of the materials in the probe
and trolley construction. To determine \wp\ required an absolute
calibration probe to relate the trolley probe FID precession
frequencies to that of a free proton in an unperturbed field.

The absolute calibration probe~\cite{fei} measured the precession
frequency of protons in a spherical water sample. It is
constructed to tight mechanical tolerances, using materials of low
magnetic susceptibility. The perturbation of the probe materials
on the field at the spherical water sample was minimized,
measured, and corrected for. This allowed the storage ring
magnetic field to be expressed in terms of the precession
frequency of protons in a spherical water sample
$\omega_{p}(\mathrm{sph,H}_{2}\mathrm{O},T)$ at temperature $T$.

A precession frequency measured with the absolute calibration
probe was then related to the spin precession frequency of a free
proton through~\cite{mohr}:
\begin{equation}
\omega_{p}(\mathrm{sph,H}_{2}\mathrm{O},T)=\left[1-\sigma(\mathrm{H}_{2}\mathrm{O},T)\right]
\omega_{p}(\mathrm{free}).
\end{equation}
Here $\sigma(\mathrm{H}_{2}\mathrm{O},T)$ accounts for the
internal diamagnetic shielding of a proton in a water molecule,
determined from~\cite{phillips}:
\begin{eqnarray*}
\sigma(\mathrm{H}_{2}\mathrm{O},307.85~\mathrm{K})&=&
1-\frac{g_{p}(\mathrm{H}_{2}\mathrm{0},307.85~\mathrm{K})}{g_{J}(\mathrm{H})}
\frac{g_{J}(\mathrm{H})}{g_{p}(\mathrm{H})}\frac{g_{p}(\mathrm{H})}{g_{p}(\mathrm{free})}\\
&=& 25.790(14)\times 10^{-6}.
\end{eqnarray*}
The ratio of the $g$~factor of the proton in water to that of the electron in
ground-state atomic hydrogen was measured to 10~ppb~\cite{phillips}. The ratio
of the electron to proton $g$~factors in hydrogen is known to
9~ppb~\cite{winkler}, and the bound-state corrections relating the $g$~factor
of a free proton to one in hydrogen were calculated in~\cite{lamb41,grotch}. We
also corrected for the measured temperature dependence of the shielding
constant, $d\sigma(\mathrm{H}_{2}\mathrm{O},T)/dT=10.36(30)\times
10^{-9}/\mathrm{K}$ ~\cite{british}.

The overall accuracy of the free-proton precession frequency
determined with the absolute calibration probe was estimated to be
0.05~ppm. This same probe was used in the muonium experiment in
which $\mu_{\mu}/\mu_{p}$ has been determined~\cite{Liu:1999}. Since
the magnets were different for the two experiments, the
perturbations of the absolute calibration probe materials on the
field at the spherical water sample were different. The change in
field is of the order of a few ppb, and is caused by magnetic images
of the probe in the g-2 magnet pole pieces.

A cross-check on the accuracy of the calibration probe comes from
measurements made in the muonium experiment where the ground state
hyperfine interval $\Delta\nu_{\mathrm{HFS}}$ was measured to
12~ppb, a result that is independent of errors in the calibration
probe. Because the theoretical prediction for the hyperfine
interval, $\Delta\nu_{\mathrm{HFS,th}}$, depends on
$\mu_{\mu}/\mu_{p}$ as a parameter, by equating
$\Delta\nu_{\mathrm{HFS}}$ and $\Delta\nu_{\mathrm{HFS,th}}$, we can
determine $\mu_{\mu}/\mu_{p}$ to 30~ppb \cite{Liu:1999,mohr,hughes}.
The value extracted agrees with $\mu_{\mu}/\mu_{p}$, determined to
120~ppb from the measurement of two Zeeman hyperfine transitions in
muonium in a strong magnetic field \cite{Liu:1999}. The latter
measurement of the magnetic moment ratio is strongly dependent on
the magnetic field measured using the absolute calibration probe,
and it is only weakly dependent on $\Delta\nu_{\mathrm{HFS,exp}}$.
If we assume the theoretical value is correct, the consistency of
the two determinations of $\mu_{\mu}/\mu_{p}$ suggests the absolute
calibration probe is accurate to better than 120~ppb.

\subsubsection{Trolley probe and plunging probe relative calibration}

\begin{figure}
\begin{center}
\subfigure[~Calibration position]
{\includegraphics[width=3in]{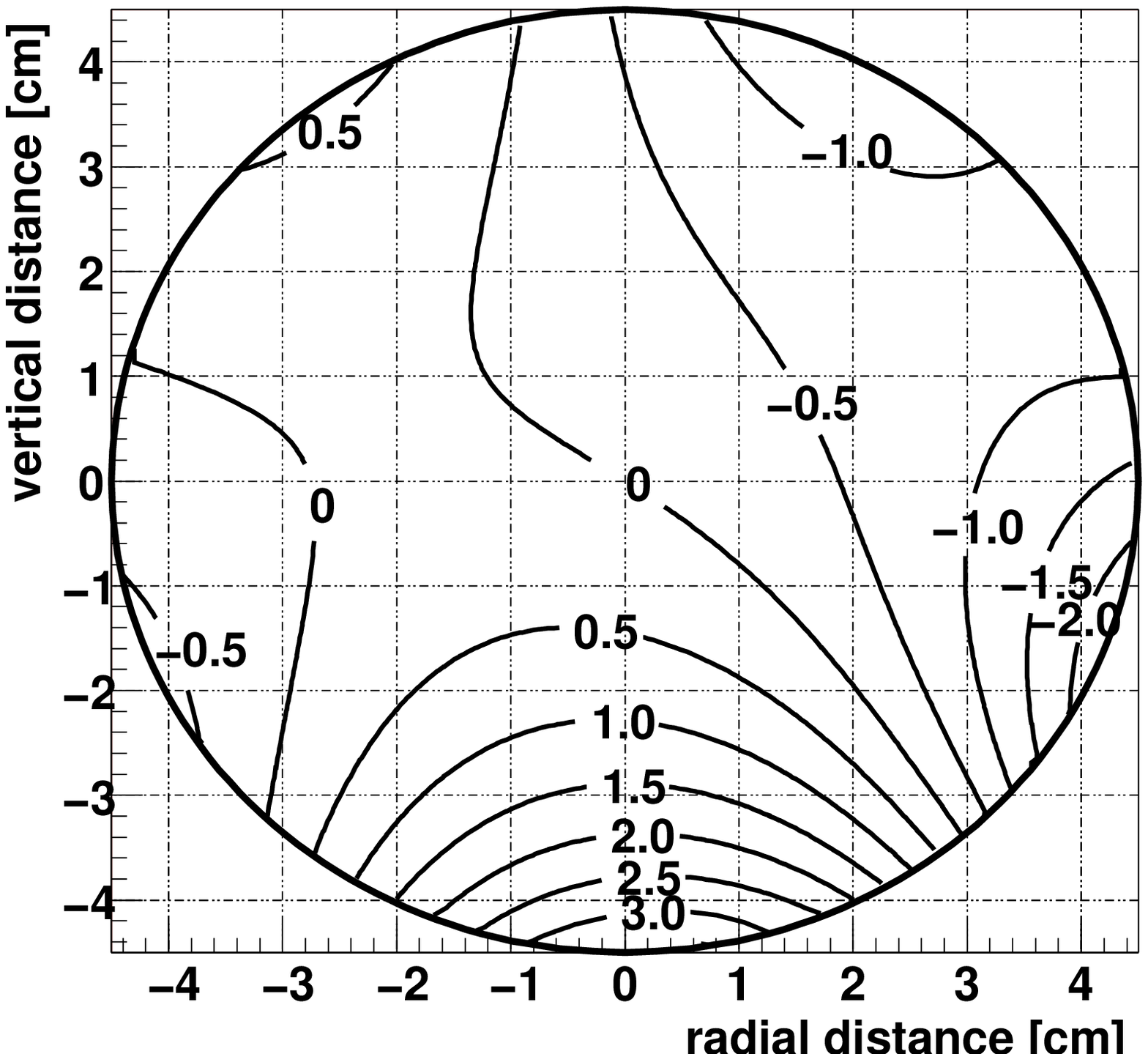}} \subfigure[~Azimuthal
average] {\includegraphics[width=3in]{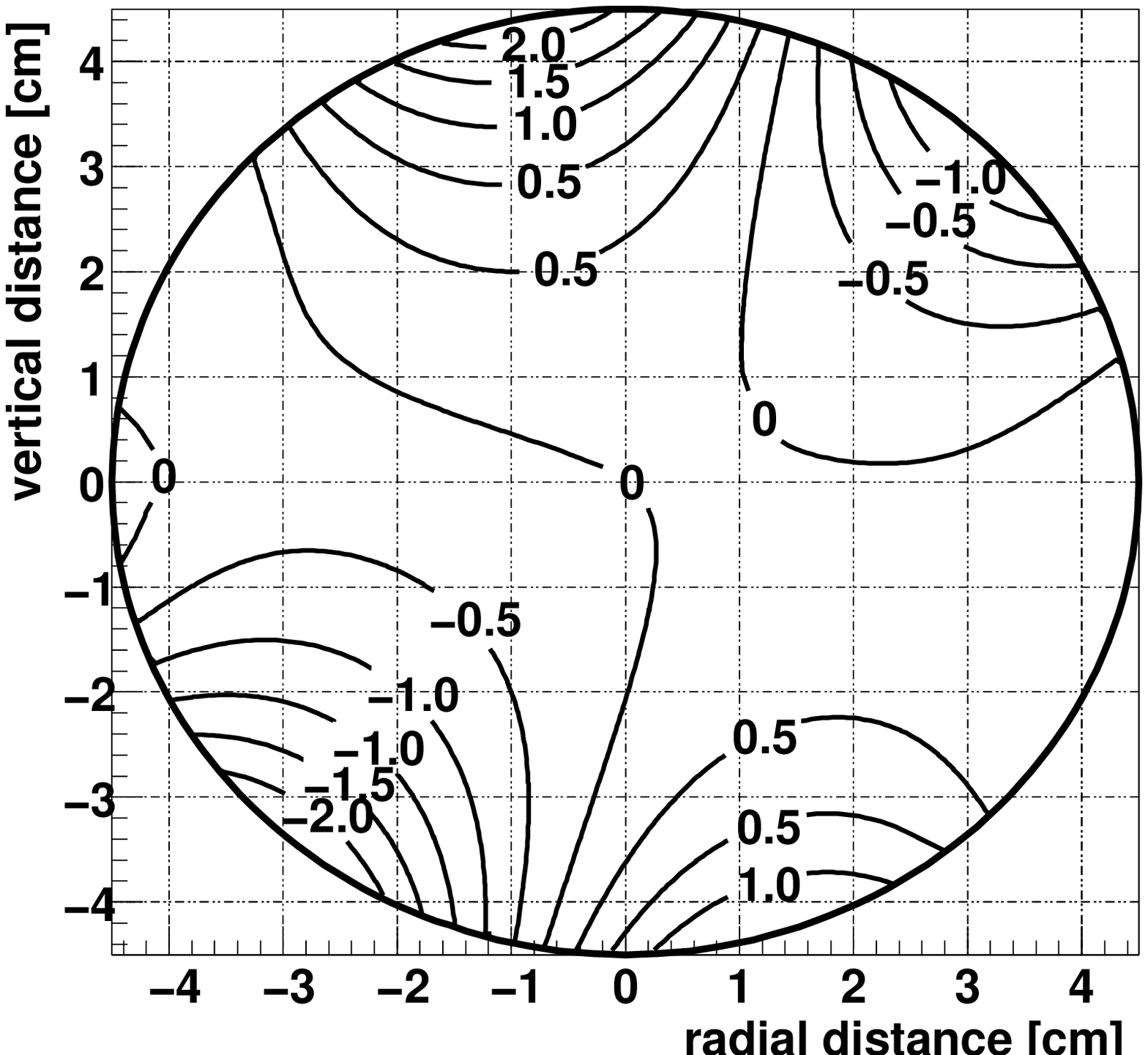}}
\end{center}
\caption{Homogeneity of the field at the (a) calibration position
and (b) in the azimuthal average for one trolley run during the
R00 period. The contour lines correspond to 0.5~ppm field
differences between adjacent pairs.} \label{fig:multipoles}
\end{figure}

Before and after each data-taking period, the absolute probe was
used to calibrate the plunging probe, the trolley center probe, and
several of the other trolley probes.  The procedure was carried out
at carefully shimmed locations in the storage ring having good field
homogeneity. Measurements from the 17 trolley probes, taken at such
an azimuthal location, were then compared to those from the plunging
probe and hence with respect to each other. The errors which arise
in the comparison are caused both by the uncertainty in the relative
positioning of the trolley probe and the plunging probe, and on the
local field inhomogeneity. The position of the trolley probes is
fixed with respect to the frame that holds them and to the rail
system on which the trolley rides. The vertical and radial positions
of the trolley probes with respect to the plunging probe are
determined by applying a sextupole field and comparing the change of
field measured by the two. The error caused by the relative position
uncertainty  is minimized by shimming the calibration field to be as
homogeneous as possible. In the vertical and radial directions, the
field inhomogeneity is less than 0.02 ppm/mm, as shown in
Fig.~\ref{fig:multipoles}, and the full multipole components at the
calibration position are given in Table~\ref{tb:multipoles}, along
with the multipole content of the full magnetic field averaged over
azimuth.   For the estimated 1~mm position uncertainty, the
uncertainty on the relative calibration is less than 0.02~ppm.

\begin{table}[b]
\begin{center} \caption{Multipoles at the outer edge of the storage
volume (radius = 45~mm).  The left-hand set are for the azimuthal
position where the plunging and calibration probes are inserted. The
right-hand set are the multipoles obtained by averaging over azimuth
for a representative trolley run during the R00 period.
\label{tb:multipoles}}
\begin{tabular}{lccccc}
\toprule Multipole & \multicolumn{2}{c}{Calibration Position}
          & ~~~~~~~~~
          & \multicolumn{2}{c}{Azimuthal Averaged} \\
~~$[{\rm ppm}]$     & Normal & Skew & ~~~~~~~~~~~~~& Normal & Skew \\
\colrule
Quadrupole~~~~~~~~& -0.71 & -1.04 & ~~~~~~~~~& 0.24 & 0.29 \\
Sextupole & -1.24 & -0.29 & ~~~~~~~~~& -0.53 & -1.06 \\
Octupole  & -0.03 & 1.06  & ~~~~~~~~~& -0.10 & -0.15 \\
Decupole  &  0.27 & 0.40  & ~~~~~~~~~&  0.82 &  0.54 \\
\botrule
\end{tabular}
\end{center}
\end{table}

Field variation along the azimuthal direction also leads to
calibration uncertainties because the precise azimuthal location of
the trolley probe's active volume is not known, {\it a priori}, to
better than a few~mm. Azimuthal field gradients were created at the
calibration location by powering correcting coils on the surface of
nearby magnet poles.  The gradients were measured with the trolley
positioned at various locations, typically 10~mm apart in azimuth.
The trolley measurements and the shift in field measured with the
plunging probe determine the azimuthal positioning of the active
volume within the trolley probes with respect to the plunging probe.
The corresponding contribution to the relative calibration of the
trolley probes amounts to 0.03~ppm.

The calibration of the NMR probes may vary with the measured NMR
frequency, since the frequency is determined by counting zero
crossings in a signal with a decaying baseline.  Other factors, such
as the temperature and power supply voltage, may have an effect as
well.  The effects were studied and an uncertainty contribution of
0.05~ppm in the field measurement was derived.

The absolute calibration of the trolley probes was made with the
storage ring at atmospheric pressure while the measurements used in
the analysis were made with the ring under vacuum. The paramagnetism
of O$_2$ in the air-filled trolley creates a small shift in the
measured field, which depends on the positions of the probes within
the trolley.  The size of the shift was estimated by taking a tube
of the same size and shape as the trolley, containing an NMR probe,
and filling the tube alternately with pure O$_2$ and N$_2$. The
resulting correction is 0.037~ppm.

\subsubsection{Measurement of azimuthal field average with trolley
probes}

\begin{figure}
\begin{center}
\includegraphics*[angle=-90,width=0.9\textwidth]{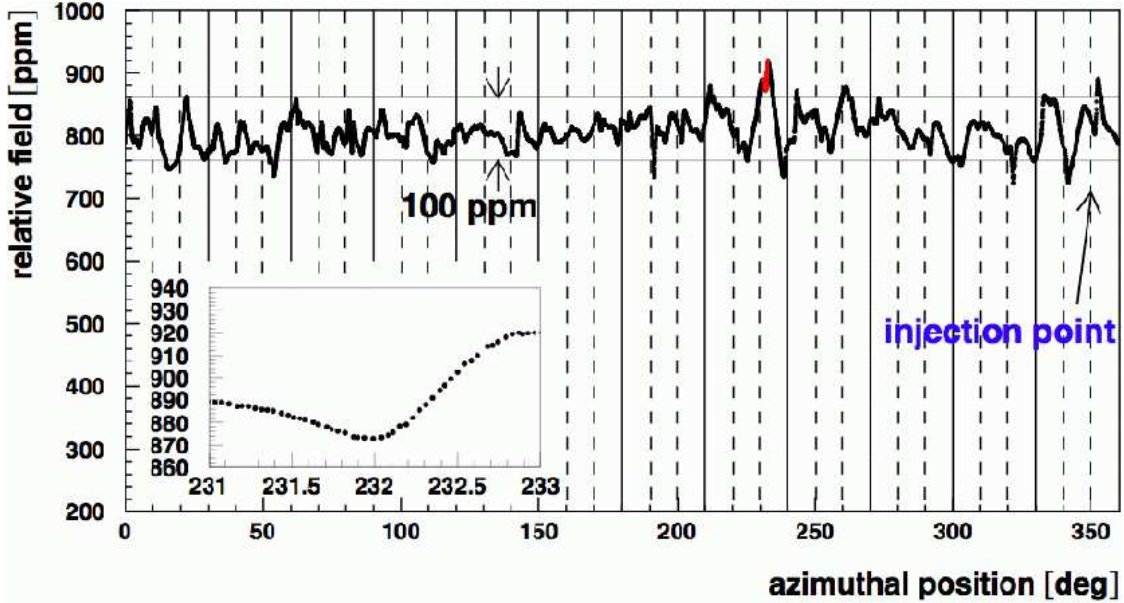}
\caption{The NMR frequency measured with the center trolley probe
relative to a 61.74~MHz reference versus the azimuthal position in
the storage ring for one of the  measurements with the field trolley
during the R01 period. The continuous vertical lines mark the
boundaries of the 12 yoke pieces of the storage ring. The dashed
vertical lines indicate the boundaries of the pole pieces.  The
inset focuses in on the measurements over an interval of two
degrees.  The point-to-point scatter in the measurements is seen to
be small. } \label{fig:fieldmap}
\end{center}
\end{figure}

The trolley is pulled by a cable around the inside of the storage
ring to measure the magnetic field integral in the same volume where
the muons are stored. Figure~\ref{fig:fieldmap} shows the field
measured by the center trolley probe for a typical measurement in
the R01 period. The trolley is pulled clockwise or counterclockwise
through the storage ring in about two hours, and the magnetic field
is measured at about 6000 locations.  Trolley measurements were made
two to three times per week at varying times during the day or
night. The azimuthal trolley positions were determined from the
perturbations to the fixed-probe readings when the trolley passes,
and from the readings of a pair of potentiometers whose resistances
were set by the drums that wind and unwind the trolley cables.  In
the final R01 running period, the use of an optical encoder system
further improved the trolley position measurement.

The trolley probe measurements are averaged over azimuth and a
multipole expansion including terms up to $4^{th}$ order for a
two-dimensional field without an azimuthal component is made
following
\begin{equation}
B(r,\theta)=B_0+\sum_{n=1}^{4}\left(\frac{r}{r_0}\right)^n[a_n \cos(n\theta)+b_n\sin(n\theta)].
\label{eq:multi_expension}
\end{equation}
Here $x$ and $y$ are the radial and vertical directions and the
corresponding cylindrical coordinates are $(r, \theta)$ with $r=0$
at the center of the storage region and $\theta=0$ defined to point
radially outward. The multipoles $a_i$ (normal) and $b_i$ (skew) are
normalized at $r_0=45~$mm; the contour plot obtained from one
trolley run is shown in Fig.~\ref{fig:multipoles}(b), which
corresponds to the multipole content in the right-hand list of
Table~\ref{tb:multipoles}.

The trolley measures the magnetic field in the muon storage region
up to a radius of 35~mm. The magnetic field beyond 35~mm radius is
obtained by extrapolating the measured moments up to and including
the decupoles. Data obtained in 1998 with a special shimming
trolley---having 25 NMR probes positioned to measure the field up to
45~mm radius---determined the field in the outer region and the
higher multipoles. The shimming trolley ran along the pole pieces
and could only be used with the vacuum chamber removed. The
multipole expansions of the shimming trolley data at six azimuthal
positions, both at the pole edge and at the pole center, all give
less than 8~ppm at 45~mm for multipoles higher than the decupoles.
By assuming an 8~ppm multipole at 45~mm and by multiplying its
radial dependence by the falling muon distribution, a maximum
uncertainty of 0.03~ppm is implied on the average field seen by the
muons, a negligibly small error.

The overall position uncertainty of the trolley, combined with the
field inhomogeneity, contribute to the error on the average field
$B_0$. The short-range jitter in azimuth is not correlated to the
field inhomogeneity and its effect, averaged out over many trolley
runs, is estimated to be smaller than 0.005~ppm. By contrast, the
long-range position error can be a systematic deviation. Combining
this uncertainty with the field inhomogeneity gives a 0.10~ppm
uncertainty in the worst case for runs prior to 2001. Improved
longitudinal position measurements reduced this uncertainty to
0.05~ppm for 2001. Both the normal and skew quadrupoles vary from
-0.5 to +0.5~ppm/mm over the azimuth. The vertical and radial
position of the trolley, over most parts of the ring are constrained
within $\pm0.5$~mm with respect to the center of the storage region.
The transverse position uncertainty should not be correlated with
the multipoles and hence there is no effect on average field due to
vertical and horizontal position uncertainties. There are some gaps,
about 100~mm in total, where the position uncertainty could be as
large as 3~mm and where the field inhomogeneity could be as large as
1.3~ppm/mm. The contribution to the average field uncertainty from
such an extreme case is 0.01~ppm, which is negligible. Reverse
direction trolley runs are also used in the analysis, which are
important to confirm the forward direction runs.

The NMR measurements provide only absolute field magnitudes,
$|\vec{B}|=\sqrt{B_x^2+B_y^2+B_z^2}$, where $B_x,~B_y$ and $B_z$
are the radial, vertical and longitudinal components,
respectively. The quantity $|\vec{B}|$ is used, rather than $B_y$
to calculate $a_\mu$.  This introduces an error
\begin{equation}
|\vec{B}|-|B_y|=\sqrt{B_x^2+B_y^2+B_z^2}-|B_y| \simeq \frac{B_x^2
+ B_z^2}{2|B_y|}, ~~{\rm for~} B_x \ll B_y,~B_z\ll B_y.
\end{equation}
The radial field $B_x$ is measured~\cite{nim_rad} in some locations
in the storage ring; $B_y$ and $B_z$ are both estimated using
Maxwell's equations. The difference between $B_y$ and $|\vec{B}|$ is
less than 0.01~ppm.

\subsubsection {Field tracking with the fixed probes}
The fixed probes are used to track the magnetic field during data
taking---the time between the direct measurements made by the NMR
trolley. Of the 378 fixed probes, which are placed in grooves in the
walls of the vacuum chamber, about half are useful for the analysis.
Some probes are simply too noisy. Other probes, located in regions
where the magnetic field gradients are large, have very short FID
times and therefore make measurements of limited precision. Still
other probes failed during the data-taking periods because of
mechanical and cable defects. The probes used in the analysis were
given a weight related to the region over which they are sensitive.
For example, the field measurements of a fixed probe at the junction
of magnet pole pieces are sensitive to a more restricted region
compared to a probe located at the center of the poles.  Therefore,
a higher weight in calculating the field average from the fixed
probes is used for probes near the pole centers than to probes near
the pole edges.  The values for the weights, about 0.3 and 0.7, were
chosen so as to optimize the agreement between the field average
determined with the trolley probes and that determined with the
fixed probes.  The weights were constant for all data-taking
periods.

\begin{figure}
\begin{center}
\includegraphics*[width=0.7\textwidth]{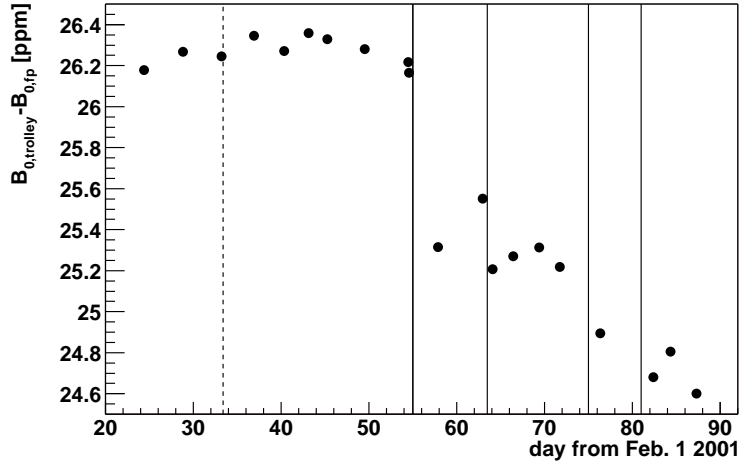}
\caption{The difference between the average field from the trolley
measurements and the one determined using a subset of the fixed
probes versus time. The solid vertical lines indicate times when the
difference is expected to change because the magnet power is cycled.
The vertical dashed line shows when the inflector was powered on.}
\label{fig:tro-nmr}
\end{center}
\end{figure}

\begin{figure}
\begin{center}
\includegraphics*[width=0.7\textwidth]{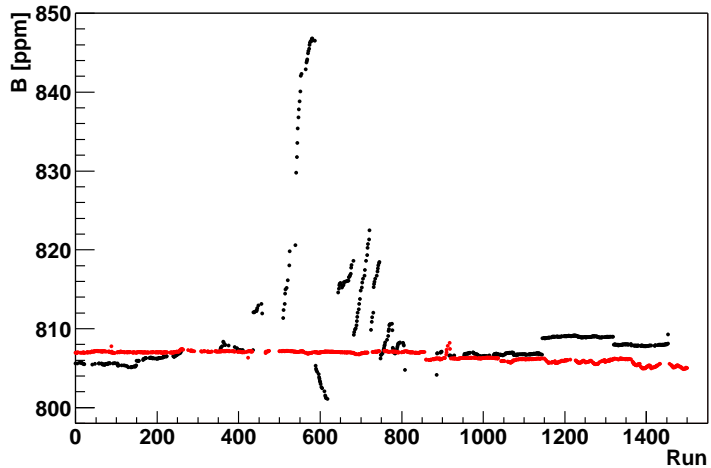}
\caption{The average magnetic field versus run number (relative to
the start of good data) for the R00 and R01 periods.  The relatively
flat distribution is from the R01 period, where the feedback system
kept the field stable at the few ppm level or better.  In R00, large
deviations are seen owing to a temporary fault in the feedback
system.} \label{fig:tracking}
\end{center}
\end{figure}

The differences between the average dipole field from the trolley
measurements and the weighted average field from the fixed probes,
made {\it during} the trolley measurements of the R01 running
period, are shown in Fig.~\ref{fig:tro-nmr}. Because of magnet
ramping or changes in the surface coil currents, the comparison
between pairs of adjacent points has meaning only when trolley
measurements were made during the same power cycle (the power cycles
are indicated by the vertical divisions). The rms distribution of
those differences is 0.10~ppm. The average magnetic field for the
R00 and R01 periods is shown in Fig.~\ref{fig:tracking}.  The
relatively flat distribution is from the R01 period, where the
feedback system kept the field stable at the few ppm level or
better. The large deviations near run 600 of the R00 data-taking
period were caused by a fault in the feedback system, but even then,
the average field was measured with sufficient accuracy.

\subsubsection{Average of the field over muon distribution and
time}

The stray magnetic field from the pulsed AGS magnets and from the
eddy currents caused by the kicker magnet has an effect on the
storage ring magnetic field. Fortunately, the field measured by the
fixed NMR probes is a time average, which is largely insensitive to
such transient effects. The field from the AGS magnets is determined
by plotting the fixed NMR probe readings obtained during the
$2\times10^5$ AGS cycles modulo the AGS period of 3~s. Its effect is
less than 0.01~ppm. The effect on the magnetic field in the storage
region from the kicker magnet eddy currents was measured in a
special setup of the prototype kicker magnet~\cite{kicker}, as this
transient magnetic field is largely shielded from the fixed NMR
probes by the aluminum vacuum chamber walls. At $30~\mu$s after
injection, when the \wa\ fits typically begin, the influence of
stray magnetic fields on the average magnetic field seen by the
muons is 0.02~ppm~\cite{kicker}.

A multipole expansion is used to express the average field felt by
the circulating muons. The azimuthally averaged magnetic field is
written
\begin{equation}
B(r,\theta) = \sum_{n=1}^{\infty} r^n(c_n \cos n\theta + s_n \sin
n\theta)
\end{equation}
where the coefficients $c_n$ and $s_n$ are the normal and skew
moments, respectively. The muon distribution moments $ I_n$, $J_n$
are defined:
\begin{equation*}
I_0 = \int_0^{r_0}\int_0^{2\pi}M(r,\theta)\,rdr d\theta \\
\end{equation*}
\begin{equation*}
I_n = \int_0^{r_0}\int_0^{2\pi} r^n M(r,\theta)\cos n\theta \,rdr d\theta \\
\end{equation*}
\begin{equation*}
J_n = \int_0^{r_0}\int_0^{2\pi} r^n M(r,\theta)\sin n\theta \,r dr
 d\theta.
\end{equation*}
The average field seen by the muons is then given by
\begin{equation*}
{\bar B} = c_0 + {\frac {1}{I_0}} \sum_{n=1}^{\infty} (c_n I_n + s_n
J_n).
\end{equation*}
For the circular storage ring aperture, only the first few
multipoles of the muon distribution are significant. Of those, the
normal moments ($I_n$) tend to be more important, except for the R00
period, when the muon beam was approximately 2~mm above the
mid-plane for much of the early data taking. The radial mean $I_1$
is deduced from the fast-rotation analysis. The higher moments
require the instantaneous radial distributions measured by the
traceback system, the most important of which is the radial RMS
width $\approx 15$~mm, consistent with the results of the full
tracking simulation.
\par
Combining the muon distribution and the field moments yields only very small
corrections to the lowest-order terms. In the R00 running period, the beam
radial centroid was $2.3\pm 1.0$~mm larger than the nominal aperture center and
the normal quadrupole moment of the magnetic field at the aperture of the muon
storage region was 0.06~ppm of 1.45~T.  Specifically,
$B_r(r,\theta)~=~f(r)\,\cos(\theta)$ and $B_y(r,\theta) = f(r)\,\sin(\theta)$,
where $f(r) = (0.06~$ppm$)\cdot(1.45~$T$)\cdot(r/45~$mm). Therefore, the
average magnetic field seen by the muons is 0.06~ppm~$ \times
~(2.3~$mm$/45$~mm) = 0.003~ppm higher than the field at the aperture center.
Because the beam position is not known on a run by run basis, the rms of the
normal quadrupole distribution is used to estimate the field error: 0.022~ppm.
In the vertical direction, the skew quadrupole from all trolley runs is
-0.11~ppm with an rms of 0.27~ppm. However, even when the beam was 2~mm high
during R00, the correction was negligible. The associated field error was taken
to be 0.27~ppm $\times ~(2~$mm$/45$~mm) = 0.012~ppm.
\par
The weighting of the sextupole and higher normal magnetic field
multipoles over the simulated muon distribution is shown in
Table~\ref{tab:avB}. The results were checked with full particle
tracking and the complete magnetic field map $B(x,y,s)$. All the
sextupole and higher skew multipoles of the muon distribution are
less than $10^{-3}$. No correction was made for moments higher
than the quadrupole. Instead, those values were incorporated into
the systematic error. The total systematic error in weighting the
magnetic field over the muon distribution is $\pm 0.03$~ppm.  To
compute the average field seen by the muons, the average magnetic
field is calculated for each run, and then averaged over all runs
with a weight corresponding to the number of decay electrons in
each run.

\begin{table}
\begin{center}\caption{Weighting of the magnetic field $B(x,y)$ from a
representative NMR trolley run during the R00 period over the muon
distribution $M(x,y)$ for the sextupole through decupole
components. \label{tab:avB}}
\begin{tabular}{lccc}
\toprule
Multipole~~~  & ~~~$M(x,y)$~~~ & ~~~$B(x,y)$~~~ & ~~~Product~~~ \\
           &          & [ppm] &  [ppb] \\
\colrule
Sextupole     &   0.02     &  -0.53& -11  \\
Octupole      &   -0.003    &  0.10& -0.3  \\
Decupole      &    0.005    &  0.82 & 4.0 \\
\botrule
\end{tabular}
\end{center}
\end{table}

\subsubsection{Results and systematic errors}
The weighted average field and systematic uncertainties for the
three running periods used in the final combined result are shown in
Table~\ref{tb:FinalFields}. A defect in the shield of the
superconductiong inflector resulted in imperfect shielding of the
inflector field for the R99 period. Over an azimuthal angle of $\sim
1^o$ degree, the residual fringe field lowered the magnetic field by
about 600 (3000)~ppm at the center (edge) of the storage aperture.
This made a separate measurement of the field necessary in this
region and led to an additional uncertainty of 0.2~ppm.  The
inflector was replaced between the R99 and R00 periods.

\begin{table}
\begin{center} \caption{Systematic errors for the magnetic field for the
different run periods. $^\dagger$Higher multipoles, trolley
temperature and its power supply voltage response, and eddy currents
from the kicker. \label{tb:FinalFields}}
\begin{tabular}{lccc} \toprule
Source of errors & R99 & R00 & R01 \\
                 & [ppm]&[ppm]&[ppm]\\
\colrule
Absolute calibration of standard probe~~~~~~~~~~ & 0.05 & 0.05 & 0.05\\
Calibration of trolley probes & 0.20 & 0.15 & 0.09\\
Trolley measurements of $B_0$ & 0.10 & 0.10 & 0.05\\
Interpolation with fixed probes & 0.15 & 0.10 & 0.07\\
Uncertainty from muon distribution & 0.12 & 0.03 & 0.03\\
Inflector fringe field uncertainty & 0.20 & -- & -- \\
Others $\dagger$ & 0.15 & 0.10 & 0.10 \\
\hline
Total systematic error on $\omega_p$ & 0.4 & 0.24 & 0.17\\
\hline Muon-averaged field [Hz]: $\wpt/2\pi$ & ~$61\,791\,256$~ &
~$61\,791\,595$~ & ~$61\,791\,400$~ \\
\botrule
\end{tabular}
\end{center}

\end{table}

\subsection{Analysis of \wa}
The anomalous muon spin precession at frequency $\omega_a$ leads to
a corresponding modulation in the number of electrons striking the
detectors. A general form of the electron time spectrum for events
having energy $E$ is
\begin{equation}
    N(E,t) = N_\circ(E,t)e^{-t/\gamma\tau_\mu}[1 - A(E,t)\cos(\wa t +
    \phi(E,t))],
\label{eq:spin_precess}
\end{equation}
where explicit energy and time dependencies for the normalization,
asymmetry and phase are included.  The replacement $E \rightarrow
E_{th}$ is a convenient simplification of Eq.~\ref{eq:spin_precess}
when the spectra include all events above energy threshold $E_{th}$.
An example of a histogram of the decay electron data from the R01
period was presented in Fig.~\ref{fig:wiggles}. The anomalous spin
precession frequency \wa, having period
$\tau_a=2\pi/\omega_a\approx4.365~\mu$s, is the prominent feature.
The asymmetry $A(E_{th},t)$ is $\approx0.4$, as the threshold energy
in this histogram is 1.8~GeV. The time-dilated muon lifetime is
$\gamma\tau_\mu\approx64.4~\mus$.

The only kinematic cut applied in the \wa\ analysis is the energy of a given
event; it is therefore important to establish the dependence of the fit
parameters on energy.  For example, $N(E)$ (integrated over all decay times,
$t$) is a gradually falling distribution in the ideal case, having an endpoint
at 3.1~GeV (see Fig.~\ref{fg:nA}b). Because of the size of the detectors and
their placement around the storage ring, their acceptance is a strong function
of energy. At 1.8~GeV, it is approximately 65 percent and the acceptance drops
with decreasing energy because low-energy electrons, having small bending
radii, can curl between detector stations. At higher energies, the acceptance
rises slowly until, close to 3.1~GeV, the bending radii are nearly equal to the
storage ring radius. These electrons strike the outer radial edge of a
calorimeter, where they are poorly measured because of incomplete shower
containment. Obstacles inside the storage aperture that are in the path of
decay electrons (quadrupole and kicker plates, the vacuum chamber wall) can
initiate partial showering, with the result that a higher-energy electron
appears in the detectors as a lower-energy event.  The observed energy spectrum
is shown in the inset to Fig.~\ref{fig:eSpectrum}. In the ideal case, the
asymmetry $A$ is also a monotonically increasing function of energy, starting
at -0.2 for $E = 0$ and rising to +1 at 3.1~GeV. It crosses 0 at approximately
1.1~GeV. Because of upstream showering and finite energy resolution, the
measured asymmetry differs from expectation.

\begin{figure}
\begin{center}
\includegraphics*[width=0.5\textwidth]{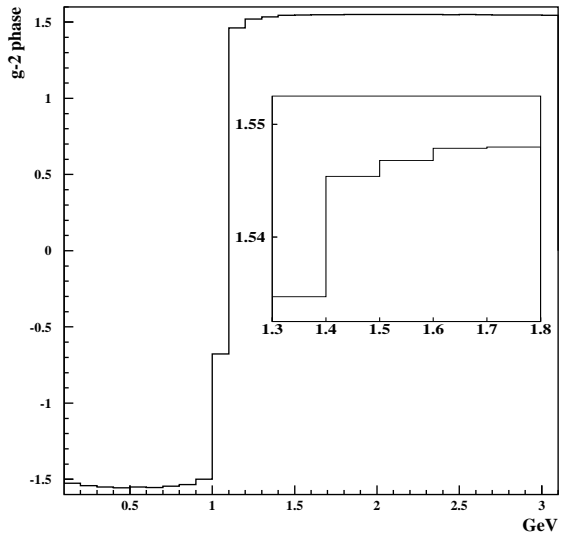}
\end{center}
\caption{Relative \gm\ phase versus energy from a tracking
simulation. There is an abrupt shift by $\pi$ radians at
approximately 1.05~GeV. Inset: an expanded view of the phase plot
from 1.3 to 1.8~GeV. The required gain stability with time is
proportional to the slope of the curve at the energy threshold, as
well as the differential asymmetry, $A(E_{th})$.}
\label{fig:PhaseVsE}
\end{figure}

The phase $\phi(E)$ represents the average muon spin angle as a
function of energy at $t=0$.  It is maximally correlated to
$\omega_a$ in the fitting procedure. In the ideal case, the energy
dependence is trivial. The phase angle is constant for $E$ above the
nominal threshold. The actual shape is explained by two geometrical
considerations, both of which depend on energy.  The time assigned
to a decay electron corresponds to the {\em arrival} at the
detector, not to the {\em decay}. High-energy electrons are assigned
times later after the decay than low-energy electrons because their
average path length to the detector is longer. In addition, the
detector acceptance is slightly greater when the electron decay
angle is inward---toward the center of the storage ring---compared
to when it is outward, and the magnitude of the difference scales
with energy.  A plot of phase versus energy, based on a tracking and
detector simulation, is shown in Fig.~\ref{fig:PhaseVsE}. The energy
dependences of $N$, $A$ and $\phi$ do not, in principle, affect the
anomalous precession frequency \wa.  Any superposition of decay
electrons having a time variation given by
Eq.~\ref{eq:spin_precess}, will be of the same form, having the same
modulation frequency, $\omega_a$. However, the time dependence of
$\phi(E)$ is an important concern. For example, an energy-scale
change as a function of time after injection is particularly
dangerous because $E_{th}$ is set at an energy where $d\phi/dE$ is
relatively large (see inset to Fig.~\ref{fig:PhaseVsE}).

Considerable effort was invested in developing a physically
motivated functional form that describes the electron decay
spectrum. For example, the coherent betatron oscillation described
in Sec.~\ref{ssec:CBO} modulates $N$, $A$ and $\phi$ with an
amplitude and decay time that must be determined from the data. Once
a suitable functional form is developed, the remaining task is to
show that the frequencies extracted from the resulting fits are
immune to a large number of possible systematic errors.

The \wa\ analyses all rely on non-linear $\chi^2$ minimization using
a fitting function $F(t,\vec{\alpha})$, having a range of free
parameters $\vec{\alpha}$ that depends on the specific fitting and
data-preparation strategy of the analyzer. The reduced chi-squared,
$\chi^{2}/{\rm dof}$ (dof~$\equiv$ degree of freedom), is used to
judge the goodness of fit with the condition that $\chi^{2}/{\rm
dof} = 1$ with a variance of $\sqrt{2/{\rm dof}}$, a property of the
$\chi^2$ distribution. For the typically 4000 degrees of freedom in
the fits described here, an acceptable reduced chi-squared is $1 \pm
0.022$, which is minimized by varying the free parameters. In the
limit of a large number of events in each bin (e.g., $N_k>30$) the
variance becomes Gaussian. The fit range in the different analyses
is always truncated well before the number of counts in the last bin
is fewer than this limit. In the sections that follow, we describe
the data preparation and fitting procedures.

\subsubsection{Data preparation and pulse-fitting procedure}
\label{ssec:production} 
\begin{figure}
\begin{center}
\includegraphics*[width=\textwidth]{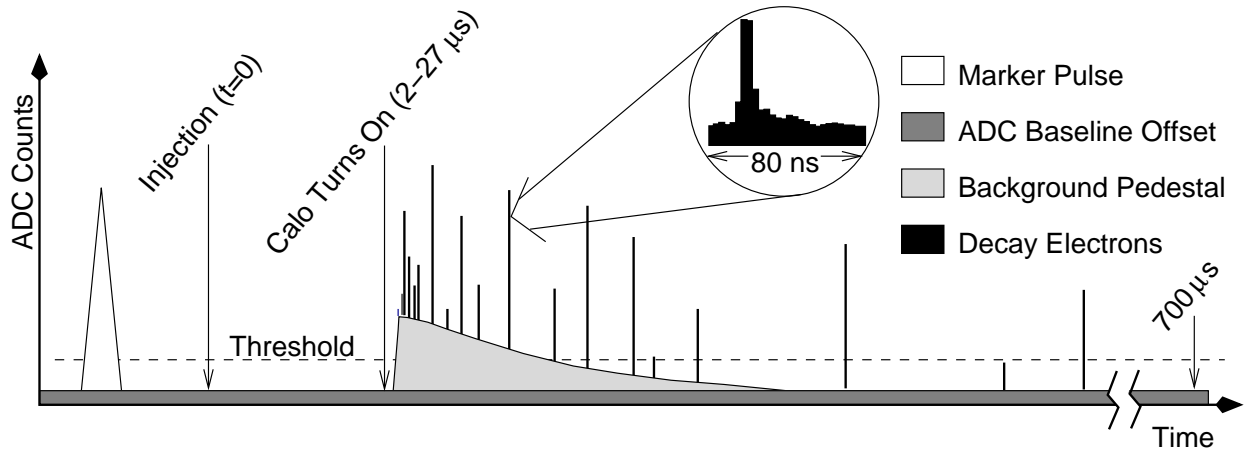}
\end{center}
\caption{Schematic representation of a WFD record obtained from a
single calorimeter in one fill of the storage ring. The dark grey
flat band represents an internal baseline added in hardware to
prevent underflows. The open triangle marker pulse is inserted
externally to establish the phase relation between the digitizers.
The light grey decaying pedestal corresponds to a flash-induced
background in the calorimeters.  The narrow spikes, see expansion,
represent the individual electron pulses of interest.}
\label{fig:fillcalo}
\end{figure}

Figure~\ref{fig:fillcalo} depicts a
schematic fill as recorded by the waveform digitizer for one
calorimeter station near the inflector. Prior to injection, the
calorimeters are gated off and a triangular-shaped marker pulse is
conveyed to each WFD to establish correct phasing of the two
independent 200~MHz digitizers. The calorimeters are gated on
$2-27~\mu$s after the muon bunch is injected.  A time-dependent
pedestal remains from the prompt hadronic flash. For detectors
immediately downstream of the inflector, the pedestal can be well
above the WFD trigger threshold for tens of microseconds, thus
initiating a continuous digitization of the analog signals. The
slowly decaying pedestal gives rise to an effective time change of
the hardware threshold. True electron pulses ride on top of this
background and must be identified and characterized as $(E,t)$
pairs.  The function of the pulse-finding algorithms is to identify
such events within the roughly $700~\mu$s measuring interval. This
task is particulary challenging because the initial instantaneous
rate of a few MHz implies a pileup rate above 1~percent. The rate
falls by a factor of $10^4$ over the measurement interval.  Because
background and electron pileup can introduce a time dependence in
the average phase, they can also bias the measurement of \wa.

The data are reconstructed by two different programs, {\tt g2Off}
and {\tt G2Too}. Their pulse-processing algorithms are conceptually
similar but the programs share no code. The two algorithms were
developed concurrently and the results obtained from the two were
compared. A Monte Carlo waveform generator was used to check that
each was immune to the difficulties outlined above. The fraction of
pulses that are found by only one of the two algorithms is small.
Slightly different fill-acceptance criteria, together with intrinsic
differences in the time and energy resolution of each method, lead
to a 1--2~percent difference in the reconstructed events. This
difference is accounted for in the comparison of \wa\ results
because it affects the statistical fluctuations.

%
%
%
%

The pulse-processing algorithm uses a standard pulse
shape---determined independently for each calorimeter---as a
template to extract the time and energy of the electron signals from
the raw waveform samples. These shapes are largely independent of
electron energy, which is proportional to the signal amplitude. The
algorithm finds this amplitude, along with the time of the true
maximum and the pedestal.

The average pulse shapes are built from sample pulses taken at times
long after injection, when the pedestal is constant, and background
and pileup are small. The sample pulses are aligned with respect to
the time of the true maximum, $t_m$.  A pseudotime is defined as
\begin{equation}
\tau=(2.5~{\rm{ns}})[i_{max}+(2/\pi) \tan^{-1}x_p],
\end{equation}
where
\begin{equation*}
x_p = \frac{S_{i_{max}}-S_{i_{max}-1}}{S_{i_{max}}-S_{i_{max}+1}},
\end{equation*}
and $i_{max}$ is the index of the maximum sample and $S_{i}$ is the value of
the $i$th sample. Here $x_p$ is a measure of where the samples lie with respect
to $t_m$. If the maximum sample corresponds to the amplitude of the pulse,
$x_p$ is nearly 1, because the pulse is fairly symmetric near its peak. When
the maximum sample moves earlier in time with respect to $t_m$, then $x_p$
decreases, whereas $x_p$ increases if it moves later. Note that $x_p$ is a
ratio of signal differences, in which the pedestal and scale cancel.
\par
The true time of the pulse is a complicated function of the
pseudotime. However, for any reasonable pulse shape, $t(\tau)$ is
a monotonically increasing function. Since the distribution $p(t)$
of true times is uniform, modulo 5~ns, the measured distribution
of pseudotimes is used to create a mapping between pseudotime and
true time
\begin{equation}
t(\tau
)=({\rm{5~ns}})\frac{\int\limits^{\tau }_{0}p(\tau ')d\tau '}
                           {\int\limits^{{\rm{5~ns}}}_{0}p(\tau ')d\tau '}
\end{equation}
to fix $t_m$ for the pulse within the 5~ns time bin. The time bin is
taken to be 5~ns rather than 2.5~ns because the pulse shapes are
slightly different as viewed by the two flash ADCs of a given input
channel. The pedestal is determined from the average signal far from
the maximum.
\par
\begin{figure}
\begin{center}
\includegraphics*[width=.7\textwidth]{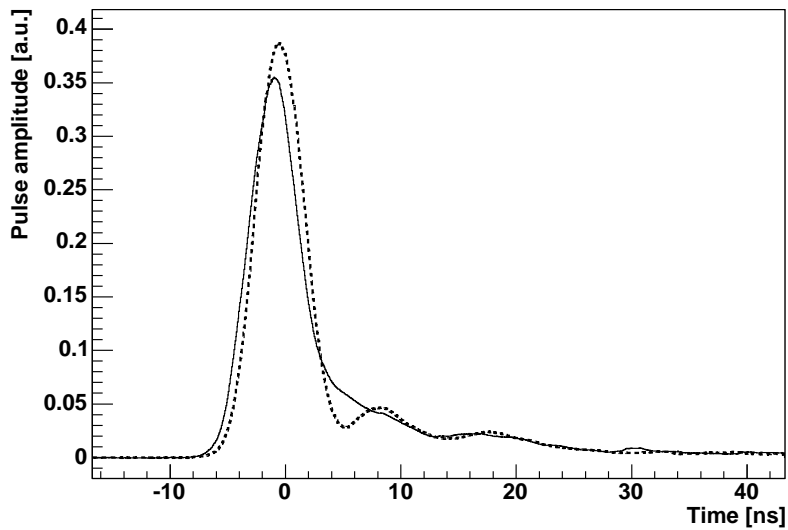}
\end{center}
\caption{Average pulse shapes for detector station 1. WFD phase~0
(dotted) has the more pronounced oscillatory behavior compared to
phase~1 (solid).} \label{fig:shapes1}
\end{figure}

Once the map between the pseudotime and true time has been
established, the average pulse shape is readily determined. The
pedestal-subtracted ADC samples of every pulse, time-aligned so that
the true maximum sits at 0, are added to a finely binned time
histogram. That histogram represents, by construction, the average
pulse shape. As an example, the pulse shapes from the two
independent WFD phases of one calorimeter are shown in
Fig.~\ref{fig:shapes1}. The shapes and amplitudes are slightly
different, as noted above.
\par
The main pulse processing routine fits each digitization interval
using the template average pulse shape for each detector. First, the
number of candidate pulses is determined by noting local peaks. The
expression
\begin{equation}
D = \sum_{i\,\in\, {\rm samples}}
  \left[S_{i}-P-\sum_{j\,\in\, {\rm pulses}} A_j f_i(t_j)\right]^2
\label{eqn:pulsefitting:DMinimized}
\end{equation}
is minimized. Here $f_{i}(t)$ is the average pulse shape and
variables $t_j$, $A_j$ and $P$ are fit parameters, representing the
times and amplitudes of each pulse and the pedestal. If the quality
of the fit is inadequate, a new model that includes additional
pulses is tested. For the typical digitization interval---having a
single pulse---only a subset of fifteen 2.5~ns ADC samples centered
on the assumed pulse is included in the fit. In practice, only the
times $t_j$ need to be included directly in the fit. The optimal
$A_j$ and $P$ may be calculated analytically.

\subsubsection{Energy calibration and stability}
\label{sssec:gain} 
\begin{figure}
\begin{center}
\includegraphics[width=4in]{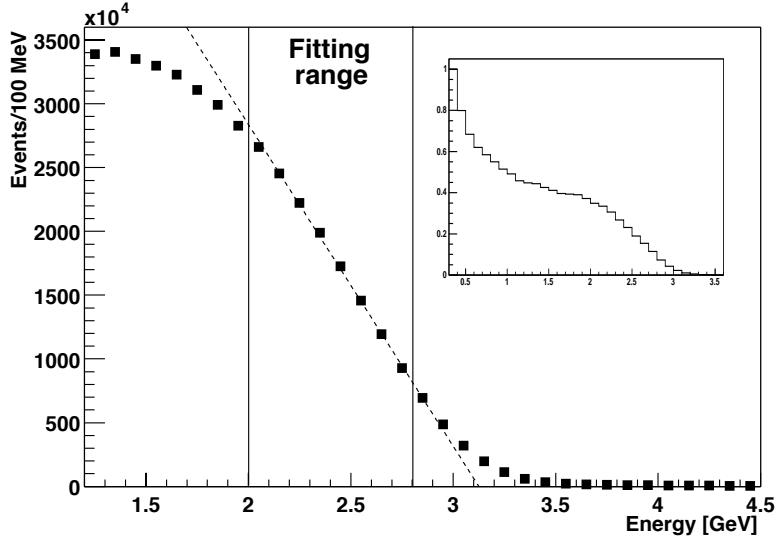}
\end{center}
\caption{Typical measured calorimeter energy distribution with an
endpoint fit superimposed. The inset shows the full reconstructed
energy spectrum from 0.3 to 3.5~GeV} \label{fig:eSpectrum}
\end{figure}

The calorimeter energy scale is determined for each detector from its measured
energy spectrum; a typical spectrum is shown in Fig.~\ref{fig:eSpectrum}. The indicated portion of the
falling edge, in which the number of counts declines from 80~percent to 20~percent of its peak value,
is fit by a straight line. The $x$-intercept of this line, the {\em endpoint}, corresponds to
approximately 3.1~GeV.

The absolute energy scale  needs to be known only approximately.
However, it is critical to measure variations in the energy scale
over the fill period. A time-changing gain---where the relevant time
scale is tens of microseconds following the fit start time---can
come from true hardware drifts in the amplification system and from
software reconstruction bias.  Both effects are associated with the
rate of events. To measure energy-scale stability, the mean energy
$\bar E(t)$ of electrons above threshold as a function of time is
constructed, suitably corrected for pileup (see next section) and
binned in full \gm\ cycles to avoid the natural dependence of the
average energy on the muon spin orientation. The gain stability
function $G(t)$ is related to the average energy through
\begin{equation}
\frac {G(t)-G(\infty)} {G(\infty)} = f\cdot\left[\frac {\bar{E}(t) -
\bar{E}(\infty)} {\bar{E}(\infty)} \right],
\end{equation}
where $f$ is a threshold-dependent constant, approximately equal
to 2 for $E_{th} = 1.8$~GeV and $t = \infty$ refers to late in the
fill. With the exception of a few detectors, $G(t)$ varies from
unity by at most 0.2 percent for the fitting period. Typically,
the electron time spectra are constructed applying $G(t)$ as a
correction to each electron's energy.

To determine the sensitivity of \wa\ on $G(t)$, electron time
spectra are reconstructed where the applied $G(t)$ is varied by
multiplying the optimal $G(t)$ by the multipliers $m_G = -1, 0, 1$
and 2. These spectra are then fit for \wa\ and the slope
$\delta\wa/\delta m_G$ is determined. Typically, the quoted result
from an individual analysis is based on $m_G = 1$ and the gain
systematic uncertainty is based on the change in \wa\ if $G(t)$ is
not applied. Because of differences in running conditions and fit
start times, the gain systematic uncertainty differs by running
period. The most complete studies of the gain systematic uncertainty
were carried out for the R00 and R01 analyses, giving 0.13 and
0.12~ppm, respectively.

\subsubsection{Correction for multi-particle pileup}
Two low-energy electrons arriving close together in time are
interpreted as one equivalent high-energy electron. Because
low-energy electrons have a shorter drift time and their acceptance
peaks at a different orientation  of the muon spin, the \gm\ phase
carried by these ``pseudo-electrons'' is different from a true
high-energy electron. The phase difference, by itself, would not be
of concern if the fraction of pileup pulses were constant over the
measurement period. But, the pileup rate is proportional to $\sim
e^{-2t/\gamma\tau}$, which has a time constant half that of the muon
lifetime. The pileup fraction therefore changes in time, which can
pull the \gm\ phase and thus bias \wa.  The general treatment of
pileup in the analyses is to correct for pileup, then fit for \wa.

Because the full raw waveforms are stored, most multi-particle
pileup events are recognized by the fitting algorithms. When two
pulses coexist on the same digitization interval and are separated
by more than 5~ns, the pulse-finding algorithms can identify the
individual pulses with nearly 100 percent reliability. The right
panel of Fig.~\ref{fig:samples} shows two pulses separated by
7.5~ns. The fitting algorithms distinguish these pulses and
correctly assign their times and energies.

Truly overlapped pulses cannot be separated, but they are accounted
for {\it on average} by subtracting an artificially constructed
pileup contribution from the spectra. Consider a digitization
interval having a ``trigger'' pulse and a ``shadow'' pulse that
trails the trigger pulse by a fixed time offset. The probability to
find a pulse at time $t$ is assumed to be the same as finding a
similar pulse at time $t+\delta$ for $\delta \ll T_c$, the cyclotron
period. The trigger pulse energy must exceed the $\approx 1~$GeV
hardware trigger threshold, but the shadow pulse can have any energy
above the fitting minimum of approximately 0.25~GeV. The
two-dimensional spectra $S_T(E,t)$ and $S_S(E,t)$ are obtained,
where $T$ and $S$ represent trigger and shadow pulses, respectively.
A ``double'' pulse having an energy equal to the sum of the trigger
and shadow is created at a time determined by the average of the two
pulses, weighted by their respective energies. Such events create
the doubles distribution $D(E,t)$. Finally, the pileup distribution
is
\begin{equation}
P(E,t) = D(E,t) - S_T(E,t)- S_S(E,t).
\end{equation}

The spectrum of counts can now be corrected for pileup by
subtracting this distribution. In doing so, care must be taken to
properly evaluate the errors because some of the events used to
construct the pileup spectrum are part of the uncorrected spectrum
already, depending on the energy cut. To sufficient precision, the
uncertainty for a bin in the pileup-subtracted histogram at time
$t$ is described by $\sigma^2(t) =
(1+k~\mbox{exp}(-t/\gamma\tau_{\mu}))\,F(t)$, where $F(t)$ is the
functional form describing the pileup-free data and $k$ is
determined from the data itself.

\begin{figure}
  \begin{center}
  \includegraphics*[width=0.75\textwidth]{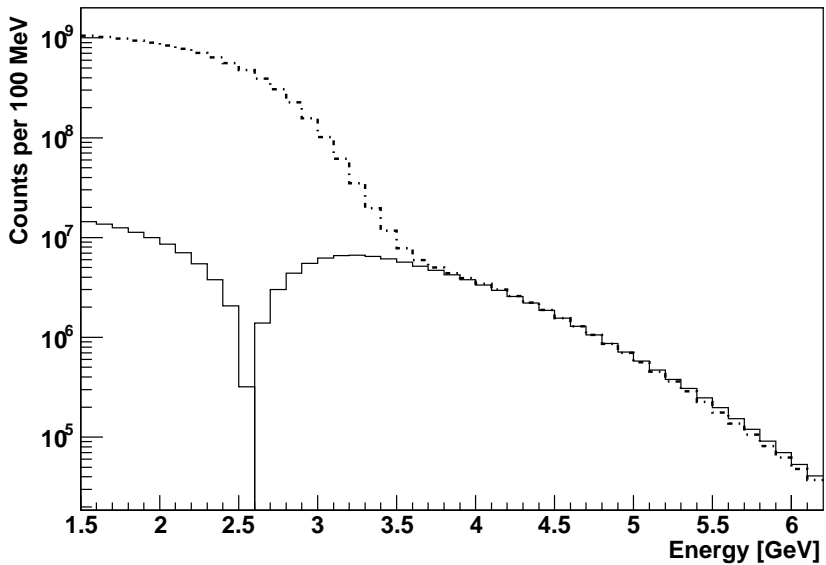}
  \caption{Uncorrected energy spectrum, including pileup events
  (dotted, top curve) and artificial pileup reconstruction spectrum
  (solid, bottom curve). The energy distributions smoothly coincide
  above about 3.5~GeV, where only true pileup events can exist.
  Note, below 2.5~GeV, the pileup spectrum is negative.
  The absolute value is plotted.
  \label{fig:pileupenergy}}
  \end{center}
\end{figure}

A pileup-free spectrum is created from normalized difference
spectra. Figure~\ref{fig:pileupenergy} illustrates the uncorrected
energy distribution (upper curve), which extends well beyond the
3.1~GeV natural maximum because of pileup events. The artificially
constructed pileup energy distribution is also shown (lower
curve).  It matches the raw distribution at the highest energies,
confirming that pileup has been correctly constructed from the raw
spectrum.  Below 2.5~GeV, the constructed pileup spectrum is
``negative'' because two lower-energy electrons are removed and
misinterpreted as a single higher-energy electron; in this figure,
the absolute value is plotted.

The pileup systematic uncertainty falls into three categories:
efficiency, phase, and unseen pileup.  The pileup {\em efficiency}
is established by creating a pileup event spectrum and adding it to
the raw spectrum.  A pileup multiplier, $m_{pu}$, is used to
construct modified electron time distributions with varying pileup
fractions.  These spectra are fit to determine $\delta\wa/\delta
m_{pu}$. Equality of the electron energy spectra early (high rate)
and late (low rate) in the fill indicates that pileup is corrected;
an uncertainty of 8~percent on this correction is assigned. The
systematic uncertainty on \wa\ from pileup subtraction efficiency is
0.036~ppm. The pileup {\em phase} reflects the error due to the
uncertainty in the phase of the constructed pileup spectrum.
Simulations determine the limits of the phase difference, and the
amplitude of pileup subtraction, combined with the phase difference,
yields an uncertainty in \wa\ of 0.038~ppm. Finally, a 0.026~ppm
uncertainty is assigned to the effect of those very-low-energy
pulses, unnoticed by the pulse-finding algorithm, which are  {\it
not} included in the constructed pileup spectra. The combined pileup
uncertainty on \wa\, is 0.08~ppm, where the efficiency and phase
uncertainties are correlated and add linearly and the unseen pileup
uncertainty is combined in quadrature.

\subsubsection{Elimination of fast rotation}

As described in Section~\ref{ssec:fastrot}, muons are injected into
the storage ring in approximately Gaussian bunches with RMS widths
of 25~ns. The momentum spread causes debunching with a time constant
of approximately $20~\mus$ and the leading and trailing edges begin
to overlap 5~$\mu$s after injection. Approximately $30~\mus$ after
injection---a typical fit start time---the underlying microstructure
remains, appearing as a rapid modulation of the electron decay
spectrum for a given detector. This fast-rotation signal is filtered
from the decay spectra by adding a random fraction of the cyclotron
period $T_c$ to the reference time $T_0$ that marks the arrival of
the bunch at the entrance to the storage ring, a procedure that
reduces the fast-rotation modulation by a factor of about 500.
Furthermore, if the calorimeter signals are aligned in time
according to their azimuthal location and their decay spectra are
combined, the fast-rotation structure is reduced by an additional
factor of 10.

In addition to the slow modulation caused by the \gm\ precession,
the actual rate in a detector station varies significantly over a
cyclotron period, from early times until the bunch structure has
disappeared. The corresponding modulation of the pileup rate is
handled automatically by the shadow pulse subtraction scheme.

\subsubsection{Multi-parameter fitting}
\begin{figure}
\begin{center}
\includegraphics*[width=.7\textwidth]{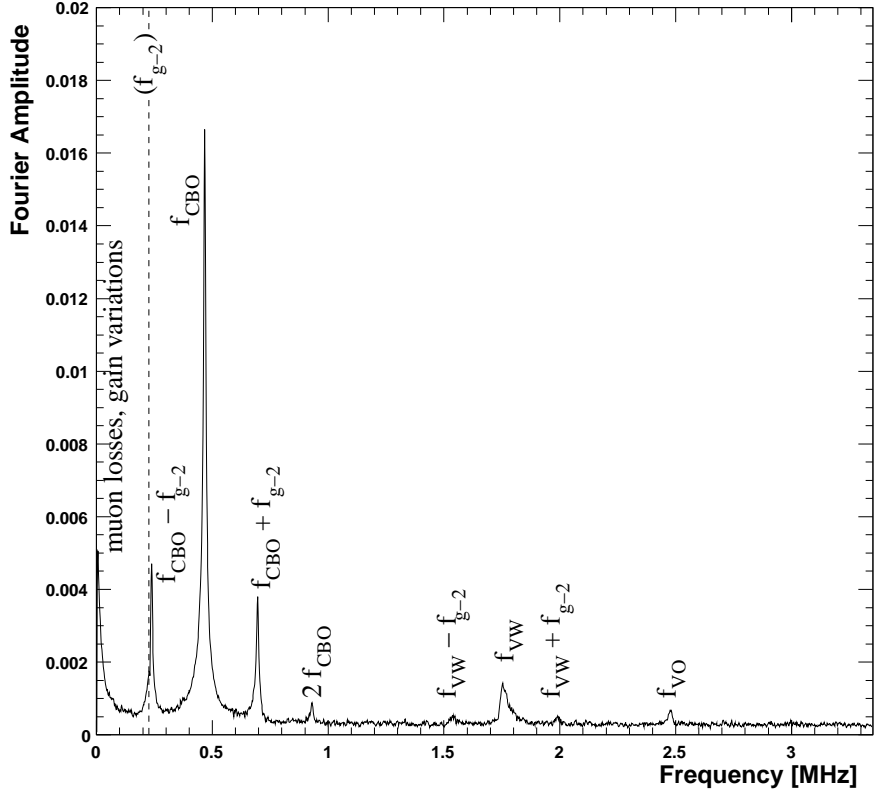}
\end{center}
\caption{The Fourier spectrum was obtained from residuals from a fit based on
the five-parameter, ideal muon decay and spin precession expression. The
horizontal coherent betatron oscillation (CBO) frequency at 466~kHz, its first
harmonic, and the difference frequency between CBO and the \gm\ frequency are
strong peaks. The vertical waist (VW) and CBO vertical oscillation (VO) produce
smaller, but still significant, effects at high frequencies. The low-frequency
rise stems from muon loss and gain distortions of the underlying decay
exponential.} \label{fig:fourier01-mario}
\end{figure}
The electron decay spectra prepared as described above, fit with the naive
five-parameter function in Eq.~\ref{eq:fivepar}, results in a very poor
$\chi^2/$dof. Fourier analysis of the residuals reveals identifiable
oscillatory features and slow changes to the overall spectral shape.
Figure~\ref{fig:fourier01-mario} shows the Fourier transform of the residuals
of such a fit to the R01 data. While the fit removes $\omega_a$ from the
residuals, strong peaks at the horizontal CBO frequency, its first harmonic,
and at the sum and difference between the CBO and \wa\ frequencies are evident.
Additionally, small peaks associated with the vertical CBO (VO) and the
vertical waist (VW) are seen at higher frequencies. The low-frequency rise is
ascribed to distortions to the exponential envelope from muon loss and gain
changes.  These physical terms motivate development of a multi-parameter
fitting function. A general form, which includes all known and relevant
physical perturbations, and assumes an energy threshold $E_{th}$, can be
written

\begin{equation}
  N(t) = \frac{N_0}{\gamma\tau_{\mu}}e^{-t/\gamma\tau_{\mu}}\cdot\Lambda(t)\cdot V(t)\cdot B(t)\cdot C(t)\cdot
  \left[ 1 - A(t)\cos(\omega_a\,t + \phi(t)  \right]
\end{equation}
with
\begin{eqnarray}
  \Lambda(t) & = & 1 - A_{loss} \int_{0}^{t} L(t') e^{-t'/\gamma\tau_{\mu}} dt'  \label{eq:lambda}\\
  V(t) & = & 1 - e^{-t/\tau_{VW}}A_{VW}\cos(\omega_{VW}\,t+\phi_{VW}) \label{eq:vertwaist}\\
  B(t) & = & 1 - A_{br}e^{-t/\tau_{br}}\label{eq:relaxation}  \\
  C(t) & = & 1 - e^{-t/\tau_{cbo}} A_1 \cos(\omega_{cbo}\,t + \phi_1) \label{eq:Ccbo} \\
  A(t) & = & A \,\left( 1 - e^{-t/\tau_{cbo}}A_2\cos(\omega_{cbo}\,t + \phi_2) \right) \label{eq:Acbo} \\
  \phi(t) & = & \phi_0 +  e^{-t/\tau_{cbo}}A_3\cos(\omega_{cbo}\,t + \phi_3)
   \label{eq:PHIcbo}.
\end{eqnarray}
While the additional terms are necessary to obtain an acceptable
$\chi^2$, they are not strongly correlated to \wa.

\begin{figure}
\begin{center}
\includegraphics*[width=0.5\textwidth]{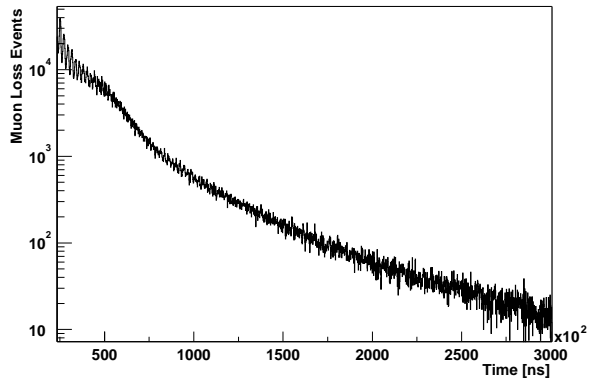}
 \caption{Muon loss rate vs. time from the
 R00 period.  Three consecutive and coincident FSD station signals
 form the muon loss signal.  The loss function $L(t)$ is proportional to this raw data plot.
 \label{fg:muonloss}}
\end{center}
\end{figure}

Equation~\ref{eq:lambda} describes $\Lambda(t)$, which is derived
from $L(t)$, which in turn is derived from the muon loss monitor
data, as illustrated in Fig.~\ref{fg:muonloss}. Muon loss introduces
a slowly changing modification to the normal exponential decay.  To
determine the absolute rate of muon losses, the acceptance of the
detection system must be established by the Monte Carlo simulation.
With an estimated acceptance of a few percent, results from the fits
indicate an approximate fractional loss rate of $10^{-3}$ per
lifetime.

Equation~\ref{eq:vertwaist} defines $V(t)$, which accounts for an
acceptance modulation owing to vertical betatron oscillations of the
stored muon beam---the vertical waist---having angular frequency
$\omega_{VW} \approx (1-2\sqrt{n})\omega_{c}$.  The vertical waist
dephases with a characteristic de-coherence time $\tau_{VW} \approx
25~\mu$s. The frequency and de-coherence time are determined in
dedicated early-start-time fits on isolated detectors---when the
vertical waist is large---and they are found to be consistent with
expectations from beam dynamics calculations. The phase is the only
free parameter in the physics fits. The term $V(t)$ is important in
fits that start before $30~\mu$s.

During scraping, which is completed $7 - 15~\mu$s after injection,
the muon beam is displaced both vertically and horizontally. As
the quadrupole voltages return to their nominal storage values,
the beam returns to the aperture's center. The relaxation time of
$5~\pm~1~\mu$s is observed directly in the mean position of decay
positrons on the FSD (see Fig.~\ref{fg:muonVertical}). The term
$B(t)$ modifies the normalization constant $N_0$ and accounts for
the corresponding change in acceptance. As was the case for
$V(t)$, $B(t)$ is important only for fits which start at very
early times.

The horizontal CBO describes the modulation of the normalization,
phase and asymmetry: $N$, $A$ and $\phi$. The modulation of $N$ and
$A$ results from the changing acceptance for detecting decay
positrons as the beam oscillates radially.  The modulation in $\phi$
results from changes in the drift time.
The amplitudes and phases in Eqs.~\ref{eq:Ccbo}~-~\ref{eq:PHIcbo}
are difficult to predict and are therefore determined from the fit.
Nonetheless, fit results for these parameters are consistent with
Monte Carlo predictions, including variations by detector. The
damping of the CBO is adequately described by an exponential having
the common decay time constant $\tau_{cbo}$ in the range $
90-130~\mu$s (dependent on quadrupole voltage). The horizontal CBO
frequency and lifetime are independently determined using the FSD
detectors. As the muons move in and out in the horizontal plane of
the storage ring, the average path length for the decay positrons to
strike the detectors grows shorter and longer. As a result, the rms
{\it vertical} width on the face of the detectors oscillates at the
horizontal CBO frequency with an amplitude that decays with
characteristic time $\tau_{cbo}$. The CBO frequency and lifetime,
measured using the FSD detectors, are consistent with those obtained
from the multi-parameter fits to the calorimeter data.

No CBO terms were used in the R98 fitting function. The small CBO
modulation was largely hidden, given the low statistical power of
the data set and the fact that the final histograms were formed from
sums of the individual detector time spectra. The effect of the CBO
would cancel exactly in the sum if the acceptance around the ring
were uniform. However, kicker and, to a lesser degree, quadrupole
electrodes---upstream of particular detector stations---lower the
acceptance of those detectors, breaking the symmetry. Still, the CBO
effect is reduced by an order of magnitude in the sum. However, fits
to the much larger R99 data set gave a poor $\chi^2 $/dof before
accounting for CBO modulation of the normalization, which was
incorporated by multiplying $N_{ideal}$ by the function
\begin{equation}
C'(t) = 1 + A_{cbo} e^{-t^{2}/\tau _{cbo} ^{2}} \cos(\omega _{cbo} t
+ \phi_{cbo}).
\end{equation}
With this modification, the $\chi ^2 $/dof was acceptable
(alternatively, Eq.~\ref{eq:Ccbo} was also used and, in subsequent
analyses, became the standard method to account for CBO). The fit
value $A_{cbo} \simeq 0.01$ was in good agreement with that found
from the full detector simulation. The conservative systematic error
of $\pm$0.05~ppm was assigned, based on changes to the \wa\
frequency when varying the CBO fit parameters over wide ranges.

\begin{figure}
\begin{center}
\includegraphics*[width=.7\textwidth]{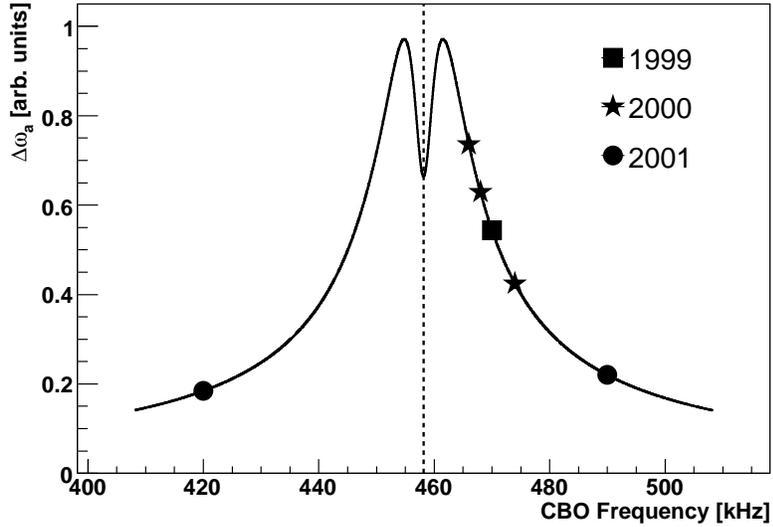}
\caption{The relative pull ($\Delta\omega$) versus the CBO modulation frequency
{\it if not} addressed by the fitting function. A typical full vertical scale
is several ppm; the actual scale depends on the specifics of the fit and the
data set used. The R00 data were acquired under run conditions in which \wa\
was very sensitive to CBO.  This sensitivity was minimized in the R01 period
where low- and high-$n$ subperiods, each having CBO frequencies well below or
above twice the \gm\ frequency, were employed.} \label{fig:yuriplot}
\end{center}
\end{figure}

After the R00 running was concluded, it was discovered that the CBO frequency
was unusually close to twice the \gm\ frequency; that is, $\omega_{cbo} - \wa
\approx \wa$.  Analytic and simulation studies indicate that in this case, the
fit value of $\omega_a$ is particularly sensitive to the CBO modulation.
Figure~\ref{fig:yuriplot} shows the relative pull ($\Delta\omega$) versus the
CBO modulation frequency {\it if not} addressed by the fitting function. As
evident in the figure, the R00 data were acquired under run conditions in which
\wa\ was very sensitive to CBO. Accordingly, the systematic uncertainty for
this period is larger than the R01 period, which featured low- and high-$n$
subperiods, each having CBO frequencies well below or above twice the \gm\
frequency. The CBO systematic uncertainties were conservatively estimated to be
$\pm0.21$~ppm and $\pm$0.07~ppm for the R00 and R01 periods, based on varying
the fixed CBO fit parameters over large ranges and comparing the \wa\ fit
values from fits to the single detector data and all detector data.

Table~\ref{tab_correlationmatrixA.physfull} shows a representative
correlation matrix from a fit to the low-$n$ data set in the R01
period.  The $V(t)$ and $B(t)$ terms are omitted because they are
negligible for the fit start time of 31.8\,$\mu$s in this fit.  Note
that none of the terms---apart from the \gm\ phase---correlates
strongly with \wa.

\begin{sidewaystable}[hp!]
\caption{Correlation matrix cov($p_i$p$_j$)/$\sigma_{i}\sigma_{j}$
from a fit to the low-$n$ data from R01 starting at 31.8\,$\mu$s.}
The $V(t)$ and $B(t)$ terms have been omitted because they are
negligible for this fit start time.
\label{tab_correlationmatrixA.physfull}
\begin{center}

\begin{tabular}{|c|r r r r r r r r r r r r r r|}
\hline
  &$N_{0}$ & $A$ & $\tau$ & $\phi$ & $\omega_{a}$ & $\omega_{\rm cbo}$
  & $\tau_{\rm cbo}$
  & $A_{1}$ & $\phi_{1}$ & $A_{2}$ & $\phi_{2}$
  & $A_{3}$ & $\phi_{3}$ & $A_{\rm loss}$ \\
\hline
$N_{0}$          & 1.000 &-0.040 &0.816 &-0.010 &\bf0.007 &0.017 &-0.022 &0.030 &-0.025 &-0.023 &0.028 &-0.013 &-0.071 &0.982\\
$A$              &-0.040 &1.000 &-0.025 &-0.008 &\bf0.004 &0.006 &0.011 &-0.014 &-0.010 &0.066 &0.038  &-0.068 &0.016 &-0.037\\
$\tau$           & 0.816 &-0.025 &1.000 &-0.007 &\bf0.005 &0.010 &-0.013 &0.018 &-0.015 &-0.014 &0.018 &-0.010 &-0.044 &0.873\\
$\phi$           &-0.010 &-0.008 &-0.007 &1.000 &\bf-0.834 &0.018 &-0.025 &0.035 &-0.024 &0.054 &-0.082  &0.004  &0.068 &-0.009\\
$\omega_{a}$     &0.007 &0.004 &0.005 &-0.834 &\bf1.000 &-0.013 &0.018 &-0.026 &0.017 &-0.040 &0.058 &-0.002 &-0.050 &0.007\\
$\omega_{\rm cbo}$ & 0.017 &0.006 &0.010 &0.018 &\bf-0.013 &1.000 &-0.004 &0.007 &-0.319 &0.002 &-0.049 &-0.002 &-0.016 &0.015\\
$\tau_{\rm cbo}$ &-0.022 &0.011 &-0.013 &-0.025 &\bf0.018 &-0.004 &1.000 &-0.903 &0.003 &0.132 &0.005 &0.016 &-0.007 &-0.020\\
$A_{1}$          & 0.030 &-0.014 &0.018 &0.035 &\bf-0.026 &0.007 &-0.903 &1.000 &-0.006 &-0.134 &-0.007 &-0.010 &0.008 &0.028\\
$\phi_{1}$       &-0.025 &-0.010 &-0.015 &-0.024 &\bf0.017 &-0.319 &0.003 &-0.006 &1.000 &-0.003 &0.046 &-0.001 &0.014 &-0.023\\
$A_{2}$          &-0.023 &0.066 &-0.014 &0.054 &\bf-0.040 &0.002 &0.132 &-0.134 &-0.003 &1.000 &-0.006 &0.009 &0.023 &-0.021\\
$\phi_{2}$       & 0.028 &0.038 &0.018 &-0.082 &\bf0.058 &-0.049 &0.005 &-0.007 &0.046 &-0.006 &1.000 &-0.007 &0.005 &0.026\\
$A_{3}$          &-0.013 &-0.068 &-0.010 &0.004 &\bf-0.002 &-0.002 &0.016 &-0.010 &-0.001 &0.009 &-0.007 &1.000 &-0.015 &-0.013\\
$\phi_{3}$       &-0.071 &0.016 &-0.044 &0.068 &\bf-0.050 &-0.016 &-0.007 &0.008 &0.014 &0.023 &0.005 &-0.015 &1.000 &-0.065\\
$A_{\rm loss}$   & 0.982 &-0.037 &0.873 &-0.009 &\bf0.007 &0.015 &-0.020 &0.028 &-0.023 &-0.021 &0.026 &-0.013 &-0.065 &1.000\\
\hline
\end{tabular}
\end{center}
\end{sidewaystable}

\subsubsection{The ratio-fitting method}
\label{sec:Ratio_Method}

\begin{figure}
\begin{center}
\includegraphics*[angle=-90,width=.7\textwidth]{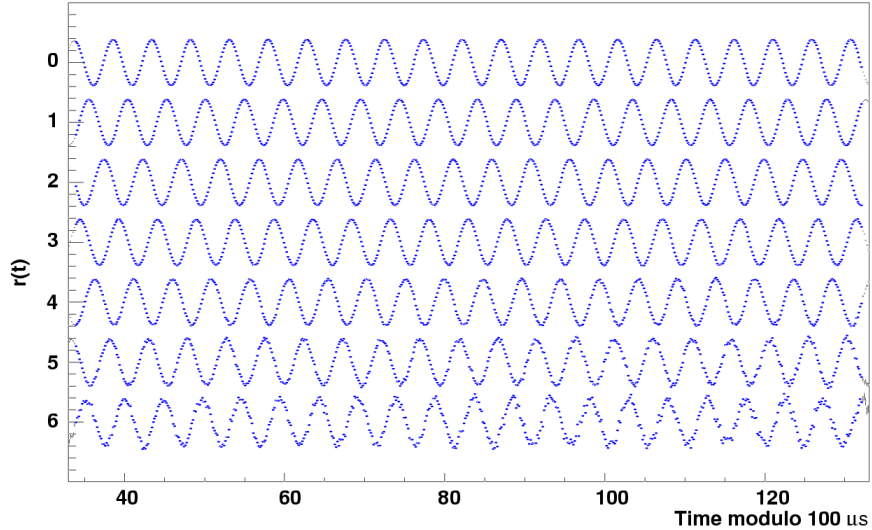}
\caption{Ratio spectrum from the R01 period. The top band begins
$27~\mu$s after injection.  For display purpose, each successive
100~$\mu$s band is displaced downward. } \label{fg:Ratio_Teach}
\end{center}
\end{figure}

An alternative analysis of \wa\ first randomly sorts the electron time spectra
into four equal subsets, labelled $n_1(t)$ through $n_4(t)$.  Each is similar
to the standard $N(t)$ distribution.  The four subsets are then recombined in
the ratio
\begin{equation}
    r(t) = \frac{n_1\left(t+\frac{1}{2}T\right) +
    n_2\left(t-\frac{1}{2}T\right) - n_3(t) - n_4(t)}
    {n_1\left(t+\frac{1}{2}T\right) +
    n_2\left(t-\frac{1}{2}T\right) + n_3(t) + n_4(t)}
\label{eq:ratio_definition}
\end{equation}
to produce a spectrum as shown in Fig.~\ref{fg:Ratio_Teach}. Here, $T =
2\pi/\omega_a$ is a (very good) estimate of the spin precession period.  In
preparing the ratio histograms, gain corrections are applied to the data and
pileup and fast rotation are removed. The ratio histograms can then be fit well
by a three-parameter function, which accounts for the amplitude, frequency and
phase of the spin precession:
\begin{equation}
  r_3(t) = A\cos(\omega_a\,t + \phi) +
\frac{1}{16}\,\left(\frac{T}{\gamma\tau_{\mu}}\right)^2.
\label{eq:r3}
\end{equation}
The final term arises from the exponential decay when shifting the
subsets forward or backward in time by $T/2$. Equation~\ref{eq:r3}
describes well the ratio histograms formed when data from all
detectors are combined. In the R01 analysis, it was used
successfully to fit the detector-combined data from both $n$-value
data sets.

However, Eq.~\ref{eq:r3} is inadequate in fitting individual
detector data subsets, or those of a given field index, as was done
in an independent analysis of the R01 data.  The ratio function must
be expanded to incorporate the CBO modulation of the normalization,
asymmetry and phase, as described for the multi-parameter fitting.
Therefore, Eq.~\ref{eq:r3} is replaced by the more general
nine-parameter function
\begin{equation}
r_9(t) = \frac{2f_0(t) - f_+(t) - f_-(t)}{2f_0(t) + f_+(t) +
f_-(t)}
\end{equation}
where
\begin{equation}
f_0(t) = C(t)\cdot(1 + A(t) \cos(\wa t + \phi(t))
\end{equation}
and
\begin{equation}
 f_{\pm}(t) = C(t')e^{\mp T/(2\gamma\tau_{\mu})}\cdot[1 + A(t') \cos(\wa t +
 \phi(t'))]
\end{equation}
with $C(t), A(t)$ and $\phi(t)$ defined in
Eqs.~\ref{eq:Ccbo},~\ref{eq:Acbo}, and \ref{eq:PHIcbo}, and with
$t'=t \pm T/2$.  This expression is sufficient to obtain good fits
to individual data subsets.

\subsubsection{The asymmetry-weighting method
\label{sssec:aweight}}
In the standard multi-parameter or ratio fitting described above,
the statistical error on the frequency \wa\ is
\begin{equation}\label{eq:fom1}
  \frac{1}{\sigma^2_{\wa}} = \frac{{\mathcal N} \langle A \rangle^2
  (\gamma\tau_{\mu})^2}{2},
\end{equation}
where $\langle A \rangle$ is the average of the asymmetry of the
${\mathcal N}$ events having energy greater than $E_{th}$. In the
asymmetry-weighting method, each decay is weighted by the
asymmetry, the weight being implied by the nominal $A(E)$
relationship. The resulting statistical error
\begin{equation}\label{eq:fom2}
  \frac{1}{\sigma^{2}_{\wa}} = \frac{{\mathcal N} \langle A^2 \rangle
  (\gamma\tau_{\mu})^2}{2}
\end{equation}
is optimized.
 The difference between Eqs.~\ref{eq:fom1} and \ref{eq:fom2} is
the asymmetry term, $\langle A^2 \rangle$ vs. $\langle A
\rangle^2$, which reduces the uncertainty by 10 percent. The
systematic sensitivities are different but of comparable size.
This method is equivalent to binning and analyzing the data in
discrete energy bins in the limit of an infinite number of bins.

\subsubsection{Internal and mutual consistency}

Before the individual offsets assigned to the analyses can be
replaced by a single common offset, each analysis must demonstrate a
high level of internal consistency. Then, before the common offset
is removed, the results of the different analyses are examined for
mutual consistency.

Internal consistency demands:
\begin{itemize}
\item a reduced $\chi^2$ equal to unity within the expected
statistical spread;
\item no pronounced structures in the fit residuals, both in the
time and frequency domains;
\item that the fit results must be independent of any specific
subset of data, within expected statistical and systematic errors.
\end{itemize}

The optimal set of parameters is obtained by minimizing $\chi^2$,
defined as
\begin{equation}
\chi^2 = \sum\limits_{\text{bins}}
\cfrac{[N_i-F(t_i,\vec{\alpha})]^2}{\sigma_{i}^2}.
\end{equation}

The error $\sigma_i$ must account for pileup subtraction. In some
analyses it was calculated during the pileup correction procedure,
as a combination of the contents of the bin and the amount
subtracted from that bin. In others, it was approximated as
$(1+k\,e^{-t/\gamma\tau})\,F(t_i,\vec{\alpha})$. The latter method
was inspired by the observation that without pileup subtraction, $\sigma_{i}^2 =
F(t_i,\vec{\alpha})$, that is the best estimate of the error squared at
any point is given by the fit value itself. The factor
$(1+k\,e^{-t/\gamma\tau})$ reflects a small correction because of
the pileup subtraction.

The quality of the fits is tested by splitting the data into
subsets, especially those that enhance or reduce particular
systematic uncertainties. In estimating the impact of statistical
fluctuations in the comparison of subset fits, the data overlap
and change in analyzing power (notably the asymmetry) has to be
taken into account.

\begin{figure}
  \includegraphics*[width=0.75\textwidth]{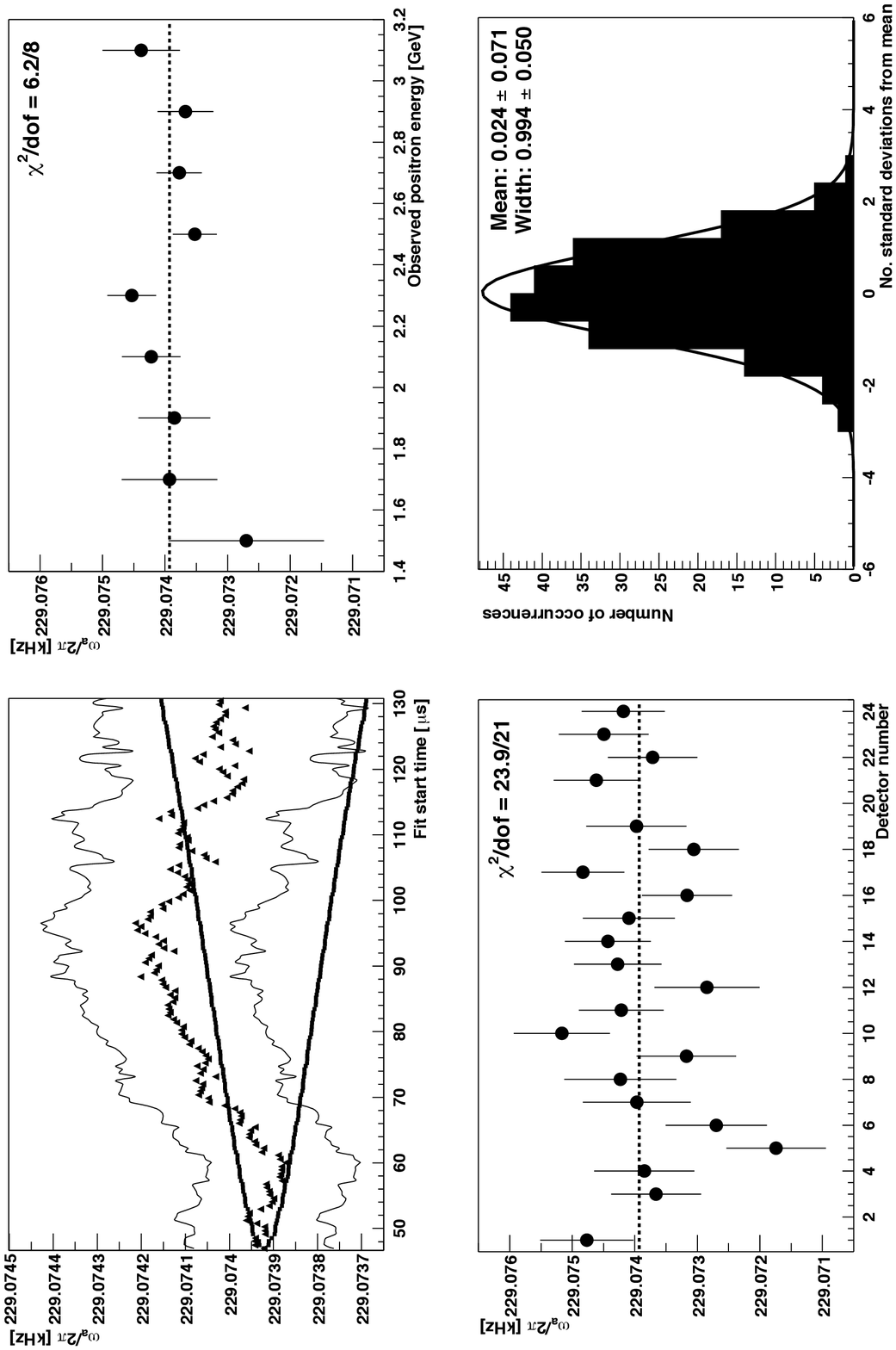}
  \caption{From a multi-parameter fitting of the R00 period data.
  Precession frequency: a) versus start time of the fit;
  b) versus detector; c) versus energy bin;  d) The difference between the
  precession frequency from each of the 198 individual fits of
  energy and detector subsets of the data compared to the mean.
  Unit width is expected for a statistical distribution. In a),
  the sideways ``parabolic'' band defines the
expected range of successive results while the outer envelope
represents the statistical uncertainty on any individual
fit.\label{fig:Fred00}}
%
\end{figure}

Data subset comparisons include:
\begin{itemize}
\item {\bf Start time:} As the fit start times increase,
the effects of short time-constant ($< \gamma\tau$)
perturbations---$V(t), B(t)$, fast rotation, pileup, gain
changes---disappear. Therefore the stability of results versus start
time is one of the most sensitive measures of a good fitting
function.
\item {\bf Run conditions:} Fits are carried out on individual runs
or sets of runs having special experimental conditions.  Examples
include variations in $n$ values, quadrupole scraping voltage or
timing, and radial magnetic field. Data from each of the 12
bunches in an AGS cycle were considered separately and compared.
\item {\bf Electron energy:} Data are sorted into discrete energy
bins (typically 200~MeV wide) or into sets having different
low-energy thresholds.  Fits to these data sets are naturally
quite sensitive to the details of energy scale and pileup
corrections.
\item {\bf Detector:} Fits are carried out on data from individual
detectors, on detector groups, and on the sum of all detectors. This
division is particularly sensitive to a correct treatment of CBO and
of the early-time background, which are highly dependent on position
around the ring.

\end{itemize}
Figure~\ref{fig:Fred00} is a representative set of important fit
subset tests for one of the R00 multi-parameter analyses. Individual
fits all had acceptable reduced $\chi^2$. The top left panel shows
the stability of the precession frequency versus start time of the
fit. The sideways ``parabolic'' band defines the expected range of
successive results, given the initial fit value and the steady
diminution of the data set. The outer envelope represents the
statistical uncertainty on any individual fit. The top right panel
shows the \wa\ results for data binned in 200~MeV energy bands, for
a fixed start time. This distribution is flat, as expected. The
bottom left panel shows the result as a function of detector station
around the ring~\cite{detector_remark}. Data were divided into
subsets indexed by both detector number and energy band. The
consistency of the \wa\ values can be verified by calculating the
mean of the ensemble and plotting the deviation of each individual
fit, normalized by the respective statistical error. The result,
shown in the bottom right panel, is a Gaussian of width 1, as
expected.

The expected spread in fitting results for data sets that partially
overlap depends on the size of the data samples, as well as on the
difference in analyzing power---the asymmetry and the phase of the
\gm\ modulation---giving
\begin{equation}
  \sigma_{diff} =
      \sqrt{
          \sigma_2^2 -
          \sigma_1^2\,
          \left(2\frac{A_1}{A_2}\cos\left(\phi_1-\phi_2\right)-1\right)}.
          \label{eq:spread}
\end{equation}
Here, the larger data set is indicated by subscript 1 and the
smaller, fully-contained subset, by subscript 2.  Data sets having
different energy thresholds will have a different analyzing power.
In the case that the analyzing powers are the same ($A_1\simeq A_2$
and $\phi_1\simeq\phi_2$), Eq.~\ref{eq:spread} reduces to
\begin{equation}
  \sigma_{diff} \simeq \sqrt{ \sigma_2^2 - \sigma_1^2},
\end{equation}
a result that is applicable to nearly all of the consistency
tests, and is the basis of the envelope shown in
Fig.~\ref{fig:Fred00}.

Results from the individual analyses were required to be
statistically compatible. When comparing different analyses of the
same nominal data set, a correct assessment of the expected
deviations must include a careful analysis of the data overlap. Each
analysis included the same accepted runs but the data overlap
differed at the few percent level, as explained in
Sec.~\ref{ssec:production}, because of the individual techniques
employed. For example, some of the more aggressive analyses used
very early start times or included a broader energy range of
accepted electrons.

Table~\ref{tab:comparisonresults} gives the final, offset-removed,
relative precession frequency results for the R99, R00 and R01
running periods based on the 14 analyses. The parameter \wa\ is
encoded in all the fitting programs as
\begin{equation}
\wa = 2\pi \cdot 0.2291~{\rm MHz} \cdot \left[1-(R-\Delta R)\times
10^{-6}\right]
\end{equation}
where $R$ is the actual free parameter and $\Delta R$ is the secret
offset for the given running period.  In this convenient form, the
precession results listed in Table~\ref{tab:comparisonresults} are
in ppm. The $R$ values cannot be compared {\em across} running
periods without further computation because the mean magnetic field
changed from year to year.

\begin{table}
  \caption{Comparison of the fitted relative precession frequency $R$
           from each of the analyses.  The units are ppm. The uncertainties are
           statistical and systematic, respectively.  Note that the
           variation in the magnetic field from year to year has {\em
           not} been corrected for. The labels A - E represent individual,
           independent analysis efforts for any given running period.
            \label{tab:comparisonresults}}
  \begin{tabular}{lccc}
  \toprule
  Analysis & R99 & R00 & R01 \\
  \colrule
  A   & ~~~ $119.60 \pm 1.23 \pm 0.08$ ~~~ & ~~~ $113.97 \pm 0.70 \pm 0.26$ ~~~  & ~~~ $108.63 \pm 0.63 \pm 0.24$ ~~~ \\
  B   & $119.33 \pm 1.28 \pm 0.19$ & $113.74 \pm 0.63 \pm 0.34$ & $107.98 \pm 0.69 \pm 0.28$ \\
  C   & $119.38 \pm 1.24 \pm 0.22$ & $113.57 \pm 0.64 \pm 0.36$ & $108.36 \pm 0.69 \pm 0.22$ \\
  D   & $119.67 \pm 1.28 \pm 0.17$ & $113.83 \pm 0.64 \pm 0.35$ & $108.31 \pm 0.71 \pm 0.23$ \\
  E   & $119.55 \pm 1.24 \pm 0.22$ &                            & $107.96 \pm 0.72 \pm 0.19$ \\
  \botrule
  \end{tabular}
\end{table}

\begin{figure}
  \begin{center}
    \includegraphics*[width=0.75\textwidth]{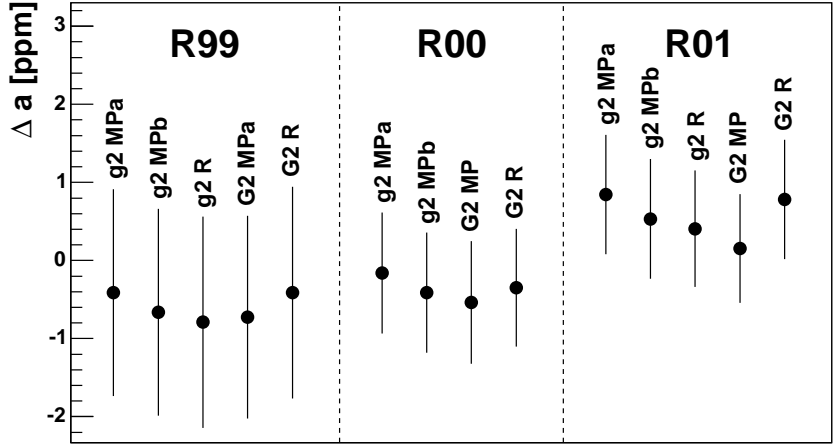}
    \caption{Results for the anomalous magnetic moment from the 14 different \wa\
    analyses performed in the R99, R00 and R01 running periods.  The analysis methods are coarsely distinguished by raw-event production,
    {\tt g2Off} ({\tt g2}) or {\tt G2Too} ({\tt G2}),
    and by the multi-parameter ({\tt MP}) or ratio-method ({\tt R}) of fitting.
    The results in a given period are highly correlated: the
    event samples largely overlap.
    \label{fig:AllDataOverview}}
  \end{center}
\end{figure}

The individual \amu\ results for the R99, R00, and R01 running
periods are shown in Fig.~\ref{fig:AllDataOverview}. The results
within a period are highly correlated. The analysis methods are
coarsely distinguished by raw-event production, {\tt g2Off} ({\tt
g2}) or {\tt G2Too} ({\tt G2}), and by the multi-parameter ({\tt
MP}) or ratio-method ({\tt R}) of fitting. The {\tt G2-MP} result
from the R01 period used the asymmetry-weighting method and the
{\tt G2-MP} result from the R00 period was based on an
energy-binned method (an indirect asymmetry weighting).


\subsubsection{Systematic errors in \wa}

The systematic uncertainties have been described in the relevant
sections.  Table~\ref{tab:syster} lists the numerical values, with
appropriate combinations of uncertainties taken into account for
the different analyses for each period.

For each running period, the final quoted result is obtained from a
statistical average of the individual results from each analysis.
This method of combination is robust, given that each analysis is
individually believed to be correct.  Each analysis included a
complete and independent evaluation of all fit-related systematic
uncertainties.  The methods, being different, feature somewhat
different sensitivities to individual uncertainties.  For example,
{\tt g2Too} and {\tt G2Off}-based productions have pulse-finding
algorithm differences and pileup sensitivities. Multi-parameter
versus ratio fitting implies different sensitivity to slow terms
such as muon loss.  The CBO sensitivity depends on whether the
analysis sums over all detectors or treats them individually. Gain
changes depend on the pileup subtraction treatment and the start
time selected. The published values take into account these
differences in the assignment of a systematic uncertainty for the
final result.

\begin{table}
\caption{Systematic errors for \wa\ in the R99, R00 and R01 data
periods. $\ddag$ In R01, the AGS background, timing shifts, E field
and vertical oscillations, beam debunching/randomization, binning
and fitting procedure together equaled 0.11 ppm.} \label{tab:syster}
\begin{center}
\begin{tabular}{lccc}
\toprule
$\sigma_{\rm syst}$ $\omega_a$  &~~~R99~~~  &  ~~~R00~~~ &~~~R01~~~ \\
 & (ppm) & (ppm) & (ppm) \\
\colrule
Pileup & 0.13 & 0.13  & 0.08    \\
AGS background~~~ & 0.10 & 0.01  &  $\ddag$   \\
Lost Muons & 0.10 & 0.10 & 0.09  \\
Timing Shifts & 0.10 & 0.02  & $\ddag$  \\
E-field and pitch & 0.08 & 0.03  & $\ddag$  \\
Fitting/Binning & 0.07 & 0.06  & $\ddag$ \\
CBO & 0.05 & 0.21 & 0.07  \\
Gain Changes & 0.02 & 0.13 & 0.12 \\ \hline
Total for $\omega_a$ & 0.3 & 0.31  &  0.21  \\
\botrule
\end{tabular}
\end{center}

\end{table}

\subsubsection{Consideration of a muon EDM}

Before presenting the final result, we remark on the possible
effect on the precession frequency from a non-zero muon electric
dipole moment (EDM). Equation~\ref{eq:amuR}, which gives the
dependence of \amu\ on the measured difference frequency (\wa),
tacitly assumes that the muon EDM is zero ($d_{\mu} = 0$).  A
non-zero EDM requires a modification of Eq.~\ref{eq:omega}:
\begin{equation}
\vec{\omega} = \vec{\omega}_{a} + \vec{\omega}_{EDM} =
\vec{\omega}_{a} - \frac{q\eta}{2m} \left( \vec{\beta} \times
\vec{B} \right), \label {eq:omegaedm}
\end{equation}
where $\eta$ is a unitless constant proportional to the EDM:
\begin{equation}
        \vec d_{\mu} = \frac{\eta q}{2mc} \vec S.
\end{equation}

The interaction of the EDM with the motional electric field (in
the muon rest frame) induces a radial component in the spin
precession vector, which is otherwise purely vertical. The spin
precession plane is tilted radially by angle $\delta$:

\begin{equation}
\delta \approx  \frac{\eta}{2a_{\mu}} =
\frac{d_{\mu}}{1.1\times10^{-16}~{\rm e\,cm}}. \label{eq:edmtodel}
\end{equation}
Independent of its sign, an EDM increases the measured precession
frequency,
\begin{equation}
\omega_{meas} = \omega_{a}\sqrt{1+\delta^{2}}. \label{eq:omega_mod}
\end{equation}
A precession plane tilt causes an oscillation at $\omega_{meas}$ in
the vertical direction of the decay electrons and thus in the mean
vertical position at the calorimeters. The vertical oscillation
reaches extrema when the spin points inward or outward radially,
$90^{\circ}$ out of phase with the usual \gm\ oscillation.

The vertical electron hit distribution on the FSDs was examined
for oscillations having this phase relation.  Additionally, the
traceback wire chambers were used to look for oscillations in the
vertical decay angle of the electrons. Both methods find null
results.  The final analysis of the EDM studies from our
experiment is being completed and we expect to achieve a limit of
a few times $10^{-19}~{\rm e\,cm}$.  The details will be described
in a separate paper~\cite{E821-EDMpaper}. However, to set the
scale of the potential effect, a non-zero EDM at $2.0 \times
10^{-19} {\rm e\,cm}$ would cause a systematic increase to \amu\
by 1.6~ppm. A more sensitive limit is obtained by invoking
muon-electron universality and the linear scaling relation
expected for many standard model extensions: $d_{\mu} \approx
(m_{\mu}/m_e)d_e$. With the current electron EDM limit, $d_e < 1.6
\times 10^{-27}$~e\,cm, linear scaling implies a muon upper limit
below $3.2 \times 10^{-25}$~e\,cm. An EDM of this magnitude is too
small to affect the \wa\ measured in this experiment. Therefore,
we adopt the assumption that the measured anomalous precession
frequency alone determines \amu.

\subsection{Final \amu\ result}
\begin{figure}
\begin{center}
\includegraphics*[angle=-90,width=.75\textwidth]{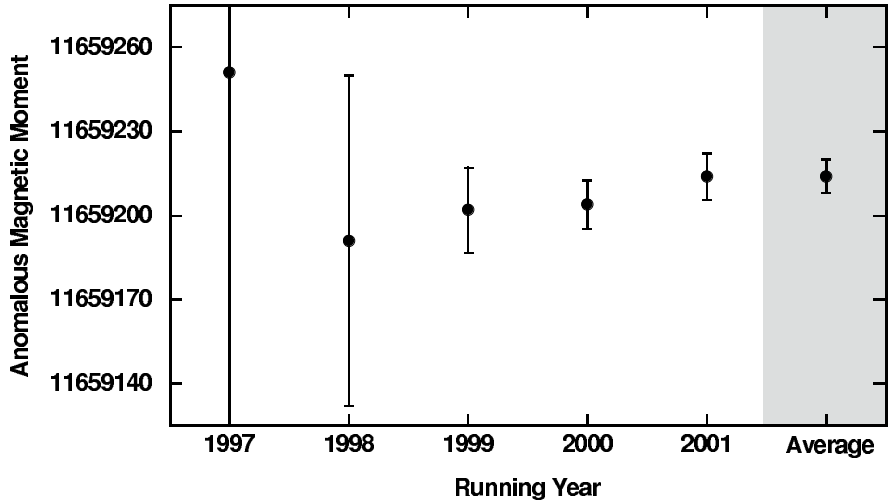}
\caption{Results for the E821 individual measurements of \amu\ by
running year, together with the final average.
\label{fig:summaryplot}}
\end{center}
\end{figure}

The final \amu\ result is obtained by combining the individual
\amu\ results from all running periods, which are listed in
Table~\ref{tab:aMuSummary}. All E821 results are plotted in
Figure~\ref{fig:summaryplot} together with the final average.

Recall that the relation
$$ a_{\mu} = \frac{\mathcal{R}}{\lambda - \mathcal{R}} $$
is used to determine \amu, where the ratio $\mathcal{R} =
\wa/\wpt$ is the experimental measurement: the anomalous
precession frequency divided by the event-weighted magnetic field.
The muon-to-proton magnetic moment ratio $\lambda = \mu_\mu/\mu_p
= 3.183\,345\,39(10)$ is obtained independently~\cite{Liu:1999}.
The appropriate comparison of results from E821 is made in terms
of $\mathcal{R}$. The precession frequencies, radial electric
field and pitch corrections, average magnetic field and
$\mathcal{R}$ ratios are given in
Table~\ref{tab:frequency-results} for the R99, R00 and R01
periods. Total uncertainties are given for each quantity. While
the magnetic field strength differed slightly from year to year,
and consequently the precession frequency changed, the agreement
in $\mathcal{R}$ between periods is excellent.

Correlations in certain systematic uncertainties exist across
running periods. These include the use of a common absolute
calibration probe, perturbations to the storage ring field from
kicker eddy currents, uncertainty in the lost muon population
phase, and the $E$/pitch correction. Their combined uncertainty is
less than 0.15~ppm. Other systematic uncertainties are
uncorrelated, as are the statistical uncertainties. The errors
quoted on combined results reflect a slight increase compared to a
direct weighted error because of correlations across periods.

The two positive muon values for $\mathcal{R}$ can be combined and
compared to the negative muon result:
\begin{eqnarray}
\mathcal{R}_{\mu^+} & = & 0.003\,707\,204\,7(2\,6)   \\
\mathcal{R}_{\mu^-} & = & 0.003\,707\,208\,3(2\,6),
\end{eqnarray}
giving $\Delta \mathcal{R} = \mathcal{R}_{\mu^-} -
\mathcal{R}_{\mu^+} = (3.6 \pm 3.7)\times 10^{-9}$, which is in
good agreement with the expectation from CPT invariance. Assuming
CPT invariance, we obtain the average value
\begin{equation}
\mathcal{R}_{\mu}({\rm E821}) = 0.003\,707\,206\,4 (2\,0),
\end{equation}
giving the anomalous magnetic moment
\begin{equation}
  a_\mu(\mathrm{Expt}) = 11\,659\,208.0(6.3) \times
  10^{-10}~~\mbox{(0.54\,ppm)}.\label{eq:e821}
\end{equation}
The total uncertainty includes a 0.46~ppm statistical uncertainty
and a 0.28~ppm systematic uncertainty, combined in quadrature.

\begin{table} [h]
\caption{Individual \wa\ and \wpt\ results, and the ratio for the
three high-statistics running periods. Column 3 gives the relative
electric field and pitch corrections, which have been applied to the
\wa\ values quoted in column~2. The total uncertainties for each
quantity are given.  The error on the average takes into account
correlations between the inter-period systematic uncertainties.}
\label{tab:frequency-results}
\begin{center}
\begin{tabular}{lllll}
\toprule
 Period~~~~~~~~~ & $\wa/(2\pi)$ [Hz]~~~~~~ & $E$/pitch [ppm]~~~& $\wpt/(2\pi)$ [Hz]~~~~~~ & $\mathcal{R} =
\wa/\wpt$ \\
 \colrule
  R99 (\mup) & $229\,072.8(3)$ & +0.81(8) & $61\,791\,256(25)$ & $0.003\,707\,204\,1(5\,1)$ \\
  R00 (\mup) & $229\,074.11(16)$ & +0.76(3) & $61\,791\,595(15)$ & $0.003\,707\,205\,0(2\,5)$ \\
  R01 (\mum) & $229\,073.59(16)$ & +0.77(6) & $61\,791\,400(11)$ & $0.003\,707\,208\,3(2\,6)$
  \\ \hline
  Average & -- & -- & -- & 0.003\,707\,206\,3(2\,0) \\
  \botrule
  \end{tabular}
\end{center}

\end{table}

\section{The Standard Model Value of the Anomaly}
%
%
%
%
Three classes of radiative processes contribute to $a_{\mu}$: QED
loops containing photons and leptons ($e,\mu,\tau$); hadronic
loops (Had) containing hadrons in vacuum polarization loops; and
weak (Weak) loops involving the $W,Z,$ and Higgs bosons. The
standard model anomaly is represented by the expression
\begin{equation} a_{\mu}{\rm (SM)} = a_{\mu}({\rm QED}) +
a_{\mu}({\rm Had}) + a_{\mu}({\rm Weak}).
\end{equation}
While the QED and electroweak contributions are well understood,
the hadronic terms remain the subject of intensive study. Recent
reviews of the complete standard model calculation are given by
Davier and Marciano~\cite{davmar} and by Passera~\cite{passera}.
Their accounting of the individual contributions is summarized
below; however, we include only those results that have been
published in refereed journals.

The QED contributions to $a_{\mu}$ are calculated through four
loops and an estimate is made of the leading five-loop
term~\cite{kinqed}. The total QED value is
\be a_{\mu}({\rm QED}) =
11\,658\,471.958(0.002)(0.115)(0.085)\times 10^{-10} \ee
where the first two uncertainties are from the $\alpha^4$ and
$\alpha^5$ terms, respectively, and the third is from the
uncertainty on $\alpha$.  The value and uncertainty on $\alpha$
are obtained from atom interferometry~\cite{kinqed}.

The electroweak contribution from one and two loops is
\be a_{\mu}({\rm Weak}) = 15.4 (0.1)(0.2) \times 10^{-10}
\label{eq:ew} \ee
where the first error comes from two-loop electroweak hadronic
effects in the quark triangle diagrams and the second comes from the
uncertainty on the Higgs mass~\cite{davmar,ew}.

\begin{figure}
  \includegraphics[width=0.85\textwidth]{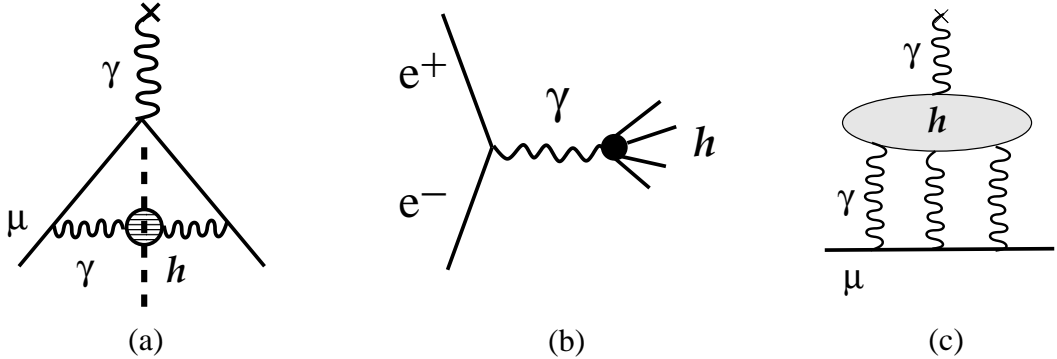}
  \caption{(a)The lowest order ``cut'' hadronic vacuum polarization
diagram and (b) the electroproduction of hadrons, which is related
to (a) through a dispersion relation. (c) The hadronic
light-by-light contribution.} \label{fig:had}
\end{figure}

Establishing an accurate and precise value for the hadronic
contributions to $a_{\mu}$ is the source of much theoretical and
experimental work worldwide. The lowest-order hadronic vacuum
polarization loop (Had; LO), shown in Fig.~\ref{fig:had}(a),
contributes approximately 60~ppm to \amu. This diagram can be
evaluated using the dispersion relation shown pictorially in
Fig.~\ref{fig:had}(a-b), which connects the bare cross section for
electroproduction of hadrons to the hadronic vacuum polarization
contribution to \amu:
\begin{equation}
a_{\mu}({\rm Had;LO})=\left({\alpha m_{\mu}\over 3\pi}\right)^2
\int^{\infty} _{4m_{\pi}^2} {ds \over s^2}K(s)R(s),
\end{equation}
where
\begin{equation}
R(s) \equiv{ {\sigma_{\rm tot}(e^+e^-\to{\rm hadrons})} \over
\sigma_{\rm tot}(e^+e^-\to\mu^+\mu^-)},
\end{equation}
and $K(s)$ is a kinematic factor. The measured cross section ratio
$R(s)$ is the critical input to the evaluation. The $s^{-2}$
dependence of the kernel weights preferentially the values of $R(s)$
at low energies (e.g., near the $\rho$ resonance) and, consequently,
the low-energy region dominates the determination of $a_{\mu}{\rm
(Had;LO)}$.  The higher-energy region is less critical~\cite{dehz2}.

The most precise data entering the dispersion relation at low
energies are from the CMD2 experiment at Novosibirsk~\cite{cmd}. The
CMD2 experiment measures  $R(s)$ by sweeping the center-of-mass
collision energy of the positron and electron beams in the VEPP-II
ring.   An alternate way of obtaining $R(s)$ is exploited by the
KLOE collaboration at Frascati~\cite{KLOE}.  They operate at a fixed
$e^+e^-$ collision energy corresponding to resonant $\phi$
production.  The hadronic cross sections at lower energies are
obtained from events having an initial-state radiated photon, which
reduces the actual center-of-mass collision energy.



The contribution to the dispersion integral from threshold to the
$\tau$ mass can also be derived from hadronic $\tau$ decays by
invoking the conserved vector current hypothesis and by making
necessary isospin corrections. For example, the decay rate for
$\tau^- \rightarrow \pi^- \pi^0 \nu_{\tau} $ can be related to the
$e^+e^-$ annihilation cross section into $\pi^+
\pi^-$~\cite{adh,dh98,dehz,dehz2}. Because the $\tau$ data only
contain an isovector component, the isoscalar piece present in
$e^+e^-$ annihilation has to be put in ``by hand'' to evaluate
$a_{\mu}{\rm (Had;LO)}$.  The $\tau$-data approach is attractive
because a large body of high-precision data exists from the LEP
experiments and from CLEO.  Unfortunately, there are significant
inconsistencies between these data and those obtained in direct
$e^+e^-$ annihilation~\cite{dehz2}. For example, the $\tau$
branching ratios predicted from the $e^+e^-$ data do not agree, nor
do the shapes of $F_{\pi}$ obtained from either the $e^+e^-$ or
$\tau$ data. Because of this inconsistency, we compared only to the
direct $e^+e^-$ annihilation data using the two recently published
analyses~\cite{dehz2,HMNT} for the $a_{\mu}{\rm (Had;LO)}$
contribution (see Table~\ref{tb:theoryvalue}). Use of the $\tau$
data leads to a higher dispersion integral.

Higher-order hadronic contributions fall into two classes. The
first class represents modifications of Fig.~\ref{fig:had}(a) with
an additional vacuum polarization loop (hadronic or leptonic), or
with a photon loop along with the hadronic vacuum polarization
loop.  These contributions, (Had; HO), can be calculated from a
dispersion relation with a different kernel function and
experimental data. Using the kernel function of
Krause~\cite{krause}, the evaluations reported in
Refs.~\cite{dehz2,HMNT} find $(-10.0 \pm 0.6)\times 10^{-10}$ and
$(-9.8 \pm 0.1)\times 10^{-10}$, respectively, which are in good
agreement and of sufficient precision compared to the experimental
uncertainty on \amu.

The hadronic light-by-light (Had; LbL) contribution shown in
Fig.~\ref{fig:had}(c) must be calculated using a theoretical
model.  Its evaluation has been the focus of considerable
theoretical activity~\cite{bpp,hk,kn,knpdr,bcm,mv}. In recent
work, Melnikov and Vainshtein (MV) report $13.6(2.5) \times
10^{-10}$ (0.22~ppm)~\cite{mv}, roughly 50 percent larger than
that obtained in earlier efforts by others~\cite{bpp,hk,kn,knpdr}.
This value is found by ignoring several small negative
contributions. For this reason, in their review~\cite{davmar},
Davier and Marciano assign $12 (3.5) \times 10^{-10}$ for this
contribution, the central value being an alternate result reported
by MV~\cite{mv}. The conservative uncertainty in the DM review
expands the range to include the earlier results. We use the DM
recommendation in our summary.

\begin{table}[ht]
\begin{center}
\caption{Contributions to the standard model value for $a_{\mu}$.
The error in ppm refers to the full value of  $a_{\mu}$. The
errors listed for (Had; LO) are the quadrature of those from the
data and from radiative corrections. The higher-order hadronic
contribution does not include the hadronic light-by-light term,
which is listed separately. In computing the total, the
higher-order hadronic is taken from the same reference as the
lowest-order hadronic. \label{tb:theoryvalue}}
\begin{tabular}{ c c c c c }
\toprule
Contribution & Value & Error & Error &\ \  Reference \\
             &  [$10^{-10}$]  &       & [ppm] &              \\
\colrule
QED & \ \ 11\,658\,471.958\ \ \  &\ \  0.143\ \  & \ \ 0.012 \ \  & \cite{kinqed} \\
Had; LO & 696.3 & 7.2 & 0.62 & \cite{dehz2} \\
Had; LO & 692.4 & 6.4 & 0.55 & \cite{HMNT}\\
Had; HO & -10.0 & 0.6 & 0.05 & \cite{dehz2} \\
Had; HO & -9.8 & 0.1 & 0.01 & \cite{HMNT}\\
Had; LBL & 12  & 3.5 & 0.3 & \cite{davmar}\\
Weak & 15.4 & 0.22 & 0.02 & \cite{davmar}\\
\hline
Total &  11\,659\,185.7 &8.0 & 0.69 &\cite{dehz2} \\
      &  11\,659\,182.0 & 7.3 & 0.62 & \cite{HMNT}\\
\botrule
\end{tabular}
\end{center}
\end{table}

The standard model theoretical summary is given in
Table~\ref{tb:theoryvalue}.  Two results are presented,
representing the two slightly different $e^+e^-$-based evaluations
of the leading-order hadronic vacuum polarization contribution.
The theoretical expectation should be compared to our experimental
result (Eq.~\ref{eq:e821}):
$$a_\mu({\rm Expt}) = 11\,659\,208.0 (6.3) \times 10^{-10}\quad
{\rm (0.54\,ppm)}.$$
The difference
\be \Delta a_{\mu}({\rm Expt-SM}) = (22.4 \pm 10\ {\rm to}\ 26.1 \pm
9.4  )\times 10^{-10}, \label{eq:delta} \ee
has a significance of 2.2 to 2.7 standard deviations.  Use of the
$\tau$-data gives a smaller discrepancy.

To show the sensitivity of the measured muon \gm\ value to the
electroweak gauge bosons, the electroweak contribution given in
Eq.~\ref{eq:ew} is subtracted from the standard model values in
Table \ref{tb:theoryvalue}. The resulting difference
with theory is \be \Delta a_{\mu} =
( 38 \ {\rm to}\ 41 \pm 10) \times 10^{-10}, \label{eq:deltanw}
\ee
or a 3.7 to 4.3 standard deviation discrepancy when the electroweak
contribution is left out.

The standard model value of the muon's magnetic anomaly is
entirely the result of radiative corrections from intermediate
states formed from a wide range of known particles. It is also
sensitive to speculative effects beyond the standard model such as
additional gauge bosons, muon or gauge boson substructure or the
existence of extra dimensions. Its value is potentially quite
sensitive to the presence of as yet undiscovered particles
associated with many generic manifestations of
supersymmetry~\cite{davmar,czmar}.

\section{Discussion and Conclusions \label{sec:summary}}
Experiment E821 at Brookhaven is formally complete. All of the data are analyzed for $a_{\mu}$ and
these independent muon anomalous magnetic moment evaluations have been
reported~\cite{Brown:2001mg,Bennett:2002jb,Bennett:2004xx}. Our combined result---based on nearly
equal samples of positive and negative muons---represents a 14-fold improvement on the CERN-III
experiment~\cite{Bailey:1977mm} of the mid-1970's. Our final measurement uncertainty is 0.54~ppm,
corresponding to 0.46~ppm statistical and 0.28~ppm systematic uncertainties, respectively. The
magnetic field and muon spin precession systematics are combined.  The experimental promise and
progress since the proposal submission in 1984 motivated a significant theoretical effort to
accurately predict the standard model expectation for the muon anomaly. The theoretical uncertainty,
now $0.62 - 0.69$~ppm, has been reduced by more than an order of magnitude over the same period. The
difference between the measured and theoretical values, $(22 - 26) \times 10^{-10}$, lies within the
expected range for many standard model extensions.

Our \gm\ Collaboration published individual results from the yearly
running periods and these were compared to the most up-to-date
theory expectations.
Improvements in the theoretical calculations, new input data for the
hadronic vacuum polarization analysis, and corrections to the theory
changed the standard model central value appreciably over this time
period. For example, the theoretical values quoted in
Table~\ref{tb:theoryvalue} are higher by $\sim24 \times 10^{-10}$
than the theory value quoted for our R99 result~\cite{Brown:2001mg}.
The dominant theoretical uncertainty is associated with the
leading-order hadronic vacuum polarization. Further work is in
progress at Novosibirsk, and additional data from both the CMD2 and
SND experiments there will be published in the near future. The $B$
factories at SLAC and KEK are also using initial-state radiation to
measure $R(s)$, and results should be forthcoming. Continued
theoretical modelling of the hadronic light-by-light contribution
can also be expected, and initial studies using the lattice have
begun~\cite{lattice}. We are confident that the precision on the
standard model value will be improved, enabling a more sensitive
comparison to experiment.

Because our measurement precision was ultimately limited by
statistics, the question naturally arises whether the current
technique can be extended using a more intense muon source. We have
studied this question and have outlined a plan~\cite{E969} that can
reduce the present uncertainty on \amu\ by a factor of 2.5 (or
more), to a relative precision of $\pm 0.2$~ppm. The effort requires
straight-forward improvements in the magnetic field uniformity and
mapping system and a five-fold increase in the muon production and
storage rate. An important feature of the design is the use of a
``backward'' decay beam to eliminate the hadronic-induced flash.
Other improvements are associated with increasing the muon
transmission fraction, optimizing the kicker efficiency, and
replacing the detectors with segmented calorimeters having
independent readout digitizers.

Historically, precision tests of the standard model have led to both
discoveries and refinements in the predictive power of the theory.
The series of CERN and BNL muon \gm\ experiments---spanning more
than 40 years---has methodically progressed such that the muon
anomaly is now measured to sub-ppm precision. Over the same time,
the standard model theoretical development has progressed, with QED
loops evaluated through fourth order and estimated through fifth,
weak loops through second order order, and hadronic loops through
second order. Many standard model extensions---SUSY is just one
example---suggest leading-order loops that will affect \amu\ at the
$\sim 1~$ppm (or slightly smaller) range. The present sensitivity of
the muon anomaly test of the standard model is $\pm$0.9~ppm, with
roughly equal contributions from theory and experiment.  Theory
uncertainty improvement can be expected from new experimental input
for the hadronic contribution and from new calculational approaches
for the hadronic light-by-light term. We have described an approach
to improve the experiment uncertainty on the anomaly to
$\pm$0.2~ppm. Thus, we may expect a significantly improved
sensitivity for the anomaly test in the future.   In the era of the
LHC and direct searches for specific standard model extensions,
precision measurements, such as that of the muon anomaly, represent
a continually improving sum rule of known physics and provide
independent insight into physics at high energies and short-distance
scales.

\section{Acknowledgments}
Many people have made important contributions towards the success of
E821 over the past twenty years when this experiment was proposed,
built and completed.  We thank N.P. Samios, R. Palmer, R.K. Adair,
T.L. Trueman, M. Schwartz, T.B.W. Kirk, P.D. Bond, M.
Murtagh$^{\dag}$, S. Aronson, D.I. Lowenstein, P. Pile and the staff
of the BNL AGS for the strong support they have given; H.
Hirabayashi and S. Kurokawa for their support of the magnet
technology provided by KEK; J.M. Bailey, J.R. Blackburn, W. Earle,
M.A. Green, E. Hazen, K. Ishida, J. Jackson, L. Jia, D. Loomba, W.P.
Lysenko, J. Ouyang, W.B. Sampson, R.T. Sanders, and K. Woodle for
their contributions to parts of the preparation and running of the
experiment; J. Cullen$^{\dag}$, C. Pai, C.~Pearson, I. Polk; L.
Snydstrup, J. Benante, D. von Lintig and S. Kochis, who played major
roles in the design and construction of the experiment.  We
gratefully acknowledge M. Tanaka$^{\dag}$ for his development of the
AGS extraction system required for our experiment. We thank U.
Haeberlen for his contributions to the development of the NMR
magnetometer system. We thank our colleagues J.~Bijnens, A.
Czarnecki, M.~Davier, E.~de~Rafael, S.I.~Eidelman, A.~H\"{o}cker,
F.~Jegerlehner, T. Kinoshita and W. Marciano for very useful
discussions on the theory.

We gratefully acknowledge L.M. Barkov for establishing the
collaboration between the Budker Institute and E821, both at BNL
and on the $R(s)$ measurements at  BINP. W.W. Williams$^{\dag}$
(U. Michigan) made important contributions in the early phases of
this work and in the development of the collaboration.

This work was supported in part by the U.S. Department of Energy, the U.S. National Science
Foundation, the German Bundesminister f\"ur Bildung und Forschung, the Russian Ministry of Science,
the US-Japan Agreement in High Energy Physics, the NATO Office of Scientific Research, and the U.S.
National Computational Science Alliance. M.~Deile, R. Prigl and A. Steinmetz acknowledge support by
the Alexander von Humboldt Foundation. F.E.~Gray was supported in part by a General Electric
Fellowship, and C.S.~\"{O}zben by the Scientific and Research Council of Turkey (TUBITAK).
D.W.~Hertzog acknowledges partial support from the John Simon Guggenheim Foundation during the writing
of this manuscript.



\begin{thebibliography}{99}
\bibitem{detector_remark}Detector $\#2$ is missing due to a hardware malfunction and
station $\#20$ is the traceback station.
\end{thebibliography}
\end{document}